\documentclass[prr,aps,twocolumn,10pt,superscriptaddress,notitlepage,longbibliography,nofootinbib,floatfix]{revtex4-2}

\usepackage[dvipsnames]{xcolor}
\usepackage[english]{babel} 
\usepackage{graphicx}
\usepackage{float}
\usepackage{braket}
\usepackage{epstopdf}
\usepackage{mathtools}
\usepackage{amsmath}
\usepackage{amssymb}
\usepackage{bbm,bm}
\usepackage[caption=false]{subfig}
\usepackage{pythonhighlight}
\usepackage{hyperref}
\usepackage{float}
\usepackage{bbold}
\usepackage[T1]{fontenc}

\usepackage{dsfont}
\usepackage{bm}

\graphicspath{{./Figures/}{./Figures_thesis/}}

\usepackage[markup=underlined]{changes}
\makeatletter
\@namedef{Changes@AuthorColor}{magenta}
\colorlet{Changes@Color}{magenta}
\makeatother
\usepackage{hyperref}
 \hypersetup{
     colorlinks=true,
     linkcolor=blue,
     filecolor=blue,
     citecolor = magenta,      
     urlcolor=red,
     }

\usepackage{braket}
\newcommand{\red}[1]{{\color{black} #1}}

\usepackage{xargs}
\newcommandx{\greencom}[2][1=]
{\todo[inline, color=green!40,#1]{#2}}
\newcommandx{\bluecom}[2][1=]
{\todo[inline, color=blue!40,#1]{#2}}
\newcommandx{\bluemargin}[2][1=]
{\todo[color=blue!40,#1]{#2}}

\usepackage{letltxmacro}
\LetLtxMacro{\ORIGselectlanguage}{\selectlanguage}
\makeatletter
\DeclareRobustCommand{\selectlanguage}[1]{%
  \@ifundefined{alias@\string#1}
    {\ORIGselectlanguage{#1}}
    {\begingroup\edef\x{\endgroup
       \noexpand\ORIGselectlanguage{\@nameuse{alias@#1}}}\x}%
}
\newcommand{\definelanguagealias}[2]{%
  \@namedef{alias@#1}{#2}%
}

\makeatother

\definelanguagealias{en}{english}
\definelanguagealias{EN}{english}
\definelanguagealias{eng}{english}
\definelanguagealias{de}{ngerman}

\newcommand{\ii}{\mathrm{i}}

\begin{document}

\title{
Generalized Dicke model and gauge-invariant master equations for two atoms in  ultrastrongly-coupled cavity
 quantum electrodynamics}
\author{Kamran~Akbari}
\thanks{These two authors contributed equally to this work: kamran.akbari@queensu.ca, will.salmon@queensu.ca}
\affiliation{Department of Physics, Engineering Physics and Astronomy, Queen's University, Kingston ON K7L 3N6, Canada}
\author{Will Salmon}
\thanks{These two authors contributed equally to this work: kamran.akbari@queensu.ca, will.salmon@queensu.ca}
\affiliation{Department of Physics, Engineering Physics and Astronomy, Queen's University, Kingston ON K7L 3N6, Canada}

\author{Franco Nori}
\affiliation{Theoretical Quantum Physics Laboratory, Cluster for Pioneering Research, RIKEN,  Wakoshi, Saitama 351-0198, Japan}
\affiliation{Quantum Computing Center, RiKEN, Wakoshi, Saitama, 351-0198, Japan}
\affiliation{Physics Department, The University of Michigan, Ann Arbor, Michigan 48109-1040, USA}

\author{Stephen Hughes}
\affiliation{Department of Physics, Engineering Physics and Astronomy, Queen's University, Kingston ON K7L 3N6, Canada}

\date{\today}

\begin{abstract} 
We  study a generalization of the well-known Dicke model,
using two dissimilar atoms in the regime of ultrastrongly coupled cavity
 quantum electrodynamics.
 Our theory uses gauge invariant master equations, which yields consistent results in either of the standard multipolar and 
 Coulomb gauges, including system-bath interactions for open cavity systems.
 We first show how a second atom can be 
 treated as a sensor atom to measure 
 the output spectrum from a single
 atom in the ultrastrong-coupling regime, and compare results with  the quantum regression theorem, explaining when they can be different. We then focus on the case where the second atom
 is also ultrastrongly coupled to the cavity, but with different parameters from those of the first atom, which introduces complex coupling effects and additional  resonances and spectral features.
 In particular, we show
 multiple resonances 
 in the cavity spectra that are visible off-resonance,  which cannot be seen when the second atom is on-resonance with the rest of the system. \red{We also observe clear anti-crossing features which are particularly pronounced when the second atom is tuned through its resonance.}

\end{abstract}

\maketitle


\section{Introduction}
\label{Sec:Intro}

Recent progress in the strong and ultrastrong (USC) regimes of light-matter interaction has opened up significant advances in theoretical and experimental research in quantum optical systems~\cite{niemczyk_circuit_2010,buluta2011natural,georgescu2012quantum,frisk_kockum_ultrastrong_2019,forn-diaz_ultrastrong_2019,mueller_deep_2020,leboite_theoretical_2020,leroux_enhancing_2018,scalari_ultrastrong_2012}.
These strong coupling regimes allow
one to coherently exchange excitations between matter and light, enabling 
breakthroughs in fundamental quantum experiments and technologies~\cite{frisk_kockum_ultrastrong_2019,forn-diaz_ultrastrong_2019,niemczyk_circuit_2010,PhysRevLett.89.067901,Volz2012}.

In particular, 
 USC  exploits the nature of counter-rotating wave physics and pondermotive forces~\cite{frisk_kockum_ultrastrong_2019,forn-diaz_ultrastrong_2019}, and 
pushes one toward a non-perturbative regime where the light and matter excitations must be treated on an equal footing, i.e., as joint/dressed states~\cite{de_bernardis_breakdown_2018}, where even the ground state can contain {\it virtual photons}.
These features make the USC regime responsible for many intriguing phenomena including the formation of 
 quasi-particle (e.g., exciton-polariton or plasmon-photon) collective modes \red{with finite lifetimes}, as well as hybrid and entangled states with higher degrees of controllablity~\cite{qin_exponentially_2018,gambino_exploring_2014,calvo_ultrastrong_2020,baranov_ultrastrong_2020,thomas_cavity-free_2021,leroux_enhancing_2018,scalari_ultrastrong_2012,todorov_ultrastrong_2010,anappara_signatures_2009}.

The intricate interactions between quantized cavity modes and quantum emitters can be  modeled in the framework of cavity
quantum electrodynamics (cavity-QED), where atoms and atom-like structures (e.g., quantum dots, molecules, superconducting circuits) interact with a (dominant) single quantized cavity mode~\cite{walther_cavity_2006,haroche_cavity_1989,miller_trapped_2005}. 
Traditionally, 
strong coupling occurs when the
cavity-emitter rate, $g$, exceeds any dissipation rate $\kappa$ (of the cavity) or $\gamma$ (decay of the emitter)~\cite{schuster_nonlinear_2008,Bishop2008,PhysRevLett.76.1800,flick_atoms_2017,yoshie_vacuum_2004,reithmaier_strong_2004,PhysRevLett.95.067401}, while
 the USC regime is characterized 
 not 
 only 
 by the lower rates of decoherence, but also when the atom-cavity coupling strength,
 $g$, becomes a significant fraction of the bare energies, $\omega_0$, of the system,
 commonly quantified as $g>0.1\omega_0$~\cite{frisk_kockum_ultrastrong_2019,forn-diaz_ultrastrong_2019}.
 Additionally, the hybridization of quantum states with different numbers of excitations in the USC regime results in a population of virtual photons in the collective system's  ground state, also with significant dissipation (and thus not 
even requiring the strong coupling regime) ~\cite{de_liberato_virtual_2017}.

The profound applicability of cavity-QED and its ease of modeling are derived from 
truncating the full emitter problem to a two-level system (TLS), which is typically coupled to a single quantized  cavity mode.
However,  the truncation of the Hilbert space, in either the material and/or photonic part, causes problems for gauge invariance when working in the USC regime~\cite{de_bernardis_breakdown_2018,di_stefano_resolution_2019,PhysRevLett.125.123602,gustin_gauge-invariant_2022}. 
Recently, many of these issues have been partly fixed for the standard quantum Rabi model (QRM) Hamiltonian~\cite{di_stefano_resolution_2019,salmon2021gauge,PhysRevResearch.4.023048}, and extended recently to ensure that dissipation
and input/output is also included in a gauge invariant way~\cite{salmon2021gauge}.
\red{While we focus our work in the context
of natural atoms (including molecules and quantum dots) in cavities, the general ideas also extend to other cavity-QED systems including circuit QED,
e.g., see 
Ref.~\cite{Settineri2021Apr}. More general quantization for arbitrary media (dealing with matter and field truncation), and the USC regime, is discussed in Ref.~\cite{gustin_gauge-invariant_2022}.}

\subsection{Gauge invariance}
In the dipole gauge (specifically, the dipole approximation in the multipolar gauge), the QRM describes the TLS-cavity system via the Hamiltonian
\cite{de_bernardis_breakdown_2018,di_stefano_resolution_2019}
(in units of $\hbar=1$):
\begin{equation}
\begin{split}
\mathcal{\hat{H}}^{\rm D}_{\rm QR}&=\omega_\mathrm{c}{a}^\dagger{a} + \frac{\omega_{a}}{2}\sigma_z+ \ii g^{\rm D}({a}^\dagger-{a}){\sigma}_x,
\end{split}
\label{eqn:HD_QR}
\end{equation}
up to a constant (${\bf 1}\omega_c\eta^2$),
where $\omega_c$ is the cavity transition frequency, ${a}$ (${a}^\dagger$) is the cavity photon annihilation (creation) operator, $\omega_a$ is the TLS transition frequency,  ${\sigma}_z={\sigma}^+{\sigma}^--{\sigma}^-{\sigma}^+$ 
and ${\sigma}_x={\sigma}^++{\sigma}^-$, with ${\sigma}^+=\vert{e}\rangle\langle{g}\vert$ (${\sigma}^-=\vert{g}\rangle\langle{e}\vert$)  the atomic raising (lowering) operator; also, $g^{\rm D}$ is the atom-cavity coupling in the dipole gauge ($g^{\rm D} \propto \sqrt{\omega_c}$), and $\eta=g^{\rm D}/\omega_c$ is the normalized coupling parameter.
We can neglect terms proportional to the identity as these do not affect the system dynamics; they simply introduce an offset in the ground state energy, which we can normalize to any value.
Equation~\eqref{eqn:HD_QR}
 reduces to the Jaynes-Cummings model (JCM) in the rotating wave approximation (RWA),  yielding~\cite{jaynes_comparison_1963,cummings_reminiscing_2013} 
\begin{equation}
\begin{split}
\mathcal{\hat{H}}^{\rm D}_{\rm JC}&=\omega_\mathrm{c}{a}^\dagger{a} + \frac{\omega_{a}}{2}\sigma_z+ \ii g^{\rm D}({a}^\dagger{\sigma}^--{a}{\sigma}^+).
\end{split}
\label{eqn:HD_JC}
\end{equation}

When the system is subjected to matter truncation,
$\hat{\mathcal{H}}_{\rm QR}^{\rm D}$ produces
the correct eigenenergies~\red{\cite{braak_integrability_2011}}, but the 
electric field operator
\cite{Settineri2021Apr,salmon2021gauge}
$\hat{E} \propto -\ii({a}'^\dagger-{a}')$,
where ${a}'={a}+\ii\eta{\sigma}_x$, which can be derived from several different viewpoints
\cite{Settineri2021Apr,salmon2021gauge,di_stefano_resolution_2019}. 
For example,
in the restricted TLS subspace,
one can transform the
Coulomb gauge operators to the dipole gauge operators, through the {\it projected} unitary transform \cite{di_stefano_resolution_2019} ${\cal U}=\exp[-\ii\eta({a}+{a}^\dagger){\sigma}_x]$,
so that ${a}'\rightarrow {\cal U} {a} {\cal U}^\dagger = {a}+\ii\eta{\sigma}_x$~\cite{Settineri2021Apr}.
These transformed operators then must be used when computing cavity field observables and for deriving  master equations.
\red{We note that in the Coulomb gauge, the electric field operator   can be directly written in terms of the cavity photon operators. However, this is not the case for the electric field operator described in the multipolar gauge, where it takes on a matter component from the TLS~\cite{salmon2021gauge,gustin_gauge-invariant_2022}.}

In the Coulomb gauge,
the standard system Hamiltonian
for the QRM is~\cite{de_bernardis_breakdown_2018,adam_stokes_gauge_2019,di_stefano_resolution_2019}
\begin{equation}
\hat{\mathcal{H}}^{\rm C,naive}_{\rm QR}=\omega_\mathrm{c}{a}^\dagger{a} + (\omega_a/2)\sigma_z+ g^{\rm C}({a} + {a}^\dagger){\sigma}_y + D({a} + {a}^\dagger)^2,
\end{equation}
where ${\sigma}_y=\ii({\sigma}^--{\sigma}^+)$, 
 $g^{\rm C}=g^{\rm D}\omega_a/\omega_c$, 
 and $D=(g^{\rm C})^2/\omega_a$ is the ponderomotive coupling strength~\cite{savasta_thomasreichekuhn_2021}.
 Unfortunately, 
this ``naive'' system Hamiltonian
is wrong (which is why we use this name in the superscript) as it does not 
 produce the
correct eigenenergies in the USC regime~\cite{de_bernardis_breakdown_2018}, and  breaks
gauge invariance.
The breakdown of gauge invariance here
can be seen as a formation of a potential nonlocality due to the truncation of the matter Hilbert space~\cite{starace_length_1971,di_stefano_resolution_2019}. 
Instead, by applying an appropriate unitary gauge transformation 
to the dipole gauge-independent QRM model, the correct 
\emph{gauge-fixed} Coulomb QRM Hamiltonian is~\cite{di_stefano_resolution_2019} 
\begin{equation}
\begin{split}
 \hat{\mathcal{H}}^{\rm C}_\mathrm{QR}&=\omega_\mathrm{c}{a}^\dagger{a} 
 \\
 &\hspace{0.3cm}+ \frac{\omega_{a}}{2}\left(\sigma_z\cos[2({a}+{a}^\dagger)\eta ]+\sigma_y\sin[2({a}+{a}^\dagger)\eta]\right),  
\end{split}
\end{equation} 
which  produces identical
eigenenergies to $\hat{\mathcal{H}}^{\rm D}_{\rm QR}$. 

\subsection{Gauge-invariant generalized master equation}
For  realistic cavities, one must also  account for dissipation/losses and photon input-output
channels.
Generally, open-system cavity-QED problems  are modelled by considering the atom and the cavity are interacting with general baths, as an open quantum system. 
In such situations, a master equation  description is widely used, leading to 
a detailed understanding of the cavity spectra and other desired observables~\cite{carmichael_statistical_2013,carmichael_dissipation_1999,settineri_dissipation_2018,ciuti_input-output_2006}. 

Commonly, the bare-state master equation formalism, 
where the joint basis states are constructed from the bare light states and the bare matter states before light-matter interaction, 
is often used
in open system cavity-QED, yielding the
standard Lindblad master equation.
However,  the bare-state master equation formalism uses
  the wrong states in the USC regime (including the ground state, which can now be an entangled state of photons and matter) and it has been shown that one needs a dressed-state approach to avoid unphysical transitions~\cite{beaudoin_dissipation_2011}.
Moreover, one also needs a ``generalized'' master equation (GME) approach to account for frequency-dependent baths and non-secular effects~\cite{settineri_dissipation_2018}.
\red{A secular 
 approximation,
 which relies on, e.g., $|\omega-\omega'|\gg \kappa$ for $\omega\neq\omega'$
 in the GME (described later),
 is not
able to describe dissipation or decoherence in open quantum systems with mixed harmonic-anharmonic or quasi-harmonic spectra~\cite{LeBoite2020Jul,settineri_dissipation_2018}. 
This is especially important 
in the USC regime, where $g$ dominates the 
eigenenergies. 
Specific examples include cavity QED in the dispersive regime~\cite{beaudoin_dissipation_2011},
cavity optomechanics~\cite{PhysRevA.91.013812},
resonant Raman interactions in cavity-QED~\cite{PhysRevB.104.045431},
electron-phonon interactions~\cite{PhysRevB.65.235311,PhysRevB.92.205406}, as well as frequency-dependent pure dephasing bath~\cite{Neuman2018} and/or  radiative
decay~\cite{salmon_master_2021}. Thus, our model below with an Ohmic bath ($\kappa(\omega) \propto \omega$)  is also an example application of the 
 GME approach, since a secular approximation leads to the same assumption as a spectrally flat bath (i.e.,
$\kappa(\omega) = \kappa$), and we show explicitly how these differ.}

Beyond these details, 
 in the USC regime,
such approaches are typically gauge relative, and again one must use
a corrected ${a}'$ for cavity mode  operators with the dipole gauge or
use the corrected Coulomb gauge Hamiltonian. Although such studies
have so far assumed very  simple models for the system-bath interactions, these approaches do produce gauge-independent results~\cite{salmon2021gauge,PhysRevResearch.4.023048}.

An advantage of
using a GME approach is that realistic observables can be computed, such as 
 the cavity-emitted spectra, typically using the quantum regression theorem~\cite{lax_formal_1963,salmon2021gauge}. However, 
gauge-independent GMEs have so far  only been applied to the case of one atom/TLS,
and we can also expect a significant 
impact when applied to multiple atoms.
In this regard,  the Dicke model is a fundamental model of quantum optics describing the light-matter interaction where a cavity mode is coupled \red{to}
a set of {\it identical} TLSs~\cite{dicke_coherence_1954,hepp_superradiant_1973}.
The model is known to be an
established description for a class of intriguing phenomena in cavity-QED  such as superradiant phase transitions and quantum chaos~\cite{hepp_superradiant_1973,wang_phase_1973,emary_quantum_2003,baumann_dicke_2010,garraway_dicke_2011,baden_realization_2014,bastarrachea-magnani_chaos_2015,klinder_dynamical_2015,larson_remarks_2017,kirton_introduction_2019,lambert_superradiance_2016,shammah_superradiance_2017,shammah_open_2018}.

\subsection{Dicke model in the USC regime}
The Dicke model
has also been investigated in the USC regime~\cite{lolli_ancillary_2015,jaako_ultrastrong-coupling_2016,dimer_proposed_2007,garbe_superradiant_2017,chen_finite-size_2018,garziano_gauge_2020,bhaseen_dynamics_2012,aedo_analog_2018}.
In the study of  effective light-matter interactions in a circuit QED system, coupled symmetrically to multiple superconducting qubits,  Ref.~\cite{jaako_ultrastrong-coupling_2016} studied a microscopic model Hamiltonian that not only describes the usual collective qubit-photon coupling but also the effect of direct qubit-qubit interactions.
Various other works in the USC regime have been presented
on related coupling effects, mainly at the thermodynamic limit, using simple system Hamiltonians~\cite{de_bernardis_cavity_2018,stokes_uniqueness_2020}.
In these extended Dicke models, similar to the previous studies on the Dicke model or even the Hopfield model in the USC regime~\cite{dimer_proposed_2007,garbe_superradiant_2017,chen_finite-size_2018,garziano_gauge_2020}, the  atoms are degenerate (i.e., they share the same coupling coefficient and resonant frequency).
Recent studies also include 
gauge-invariant system Hamiltonian models~\cite{garziano_gauge_2020}, or  discuss more exotic schemes of the Dicke model, such as the {\it anisotropic} or {\it nonequilibrium} models in which the counter- and co-rotating terms have different coupling strengths, but the two atoms are still identical~\cite{bhaseen_dynamics_2012,aedo_analog_2018}.

\subsection{Generalized {Dicke} model}
It is  desirable to explore a more general two-atom case where the TLS parameters can be different, and a natural extension to investigate is a system of two dissimilar atoms, which we term a {\it generalized Dicke model} (GDM), in the limit of two atoms. Extensions to multiple dissimilar atoms could be the subject of future work.
From a practical viewpoint, one must also include
realistic dissipation and input-output channels to the system.
In this paper, we present such a study, using 
 {\it gauge-independent} master equations valid for exploring USC dynamics.
Ultimately,
this GDM is a more realistic scenario
for studying how atoms interact in the USC regime, as it is practically impossible to experimentally produce two identical  TLSs for  experimental systems~\cite{reitzenstein_polarization-dependent_2010,kim_strong_2011,majumdar_phonon-mediated_2012,maragkou_bichromatic_2013}.
Coupling with two different atoms
also leads to new coupling regimes that are not accessible  with identical atoms.

\red{
With the recent technological and experimental advances, promising opportunities are emerging to study  GDM~\cite{bourassa_ultrastrong_2009,niemczyk_circuit_2010,leroux_enhancing_2018,Yahiaoui2022,sanchez-barquilla_theoretical_2022,benz_single-molecule_2016,chikkaraddy_single-molecule_2016,calvo_ultrastrong_2020,reitzenstein_polarization-dependent_2010,al-ani_recent_2022,albrechtsen_nanometer-scale_2022}.
While most of these are in the strong coupling regime, new designs and systems are emerging in the USC regime as well. References
\cite{frisk_kockum_ultrastrong_2019,forn-diaz_ultrastrong_2019} give a review of the various experiments systems, including molecular as well as solid state 
systems.
Moreover, it has been extensively shown that the efficiency of light-matter interaction in bound systems can be enhanced in nanostructures,
especially using 
metallic nanostructures,
where experiments have even demonstrated the deep USC regime ($g>1$) \cite{mueller_deep_2020}.  
}

\begin{figure*}[!htb]
    \centering   
    \includegraphics[width=.485\linewidth]{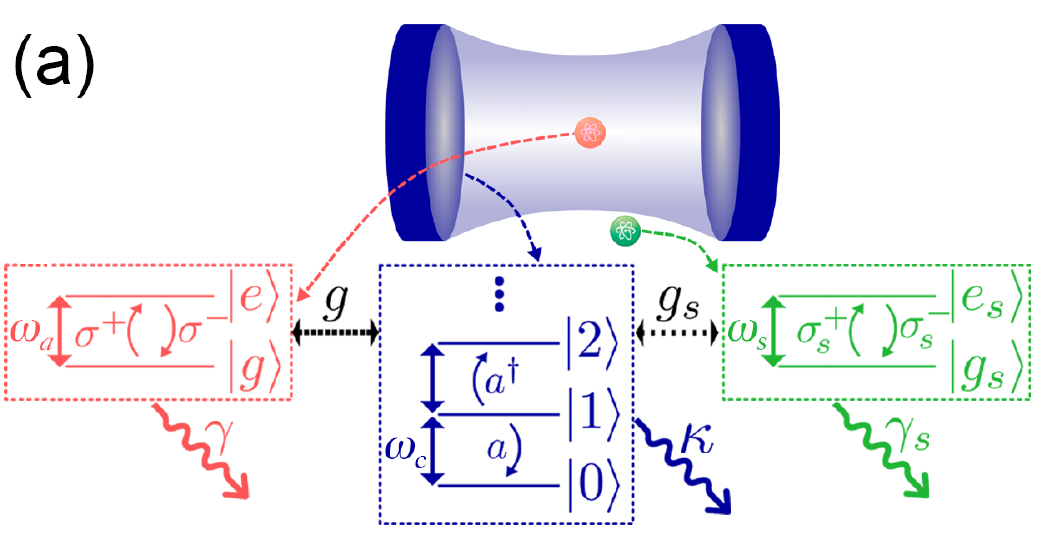}\hspace{0.3cm}
    \includegraphics[width=.485\linewidth]{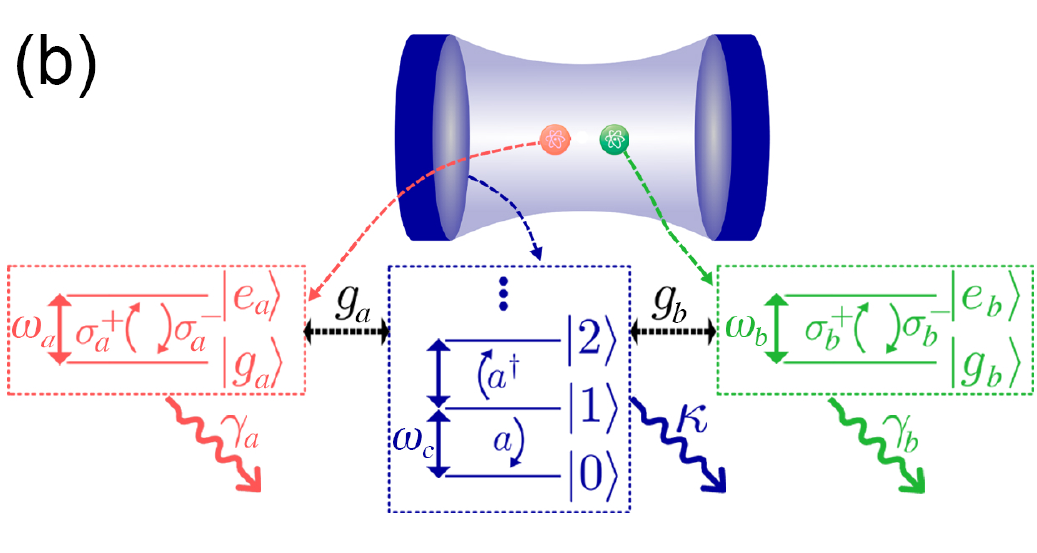}
    \caption{\textbf{Cavity-QED schemes with two atoms.} Schematics of the cavity-QED model with a second atom, including: (a) the sensor atom approach and (b) the generalized Dicke model in the USC regime. In the sensor atom approach (a), the addition of a second TLS shown as a sensor atom is  weekly coupled to the cavity (hence shown outside of the cavity). In the generalized Dicke model (b), the second atom is also  considered to be ultrastrongly coupled to the cavity (similar to the first atom, but it can have different coupling parameters).
    }
    \label{Fig1}
\end{figure*}

We study the two-atom GDM by introducing a \red{dissimilar} second atom to a general one-atom-cavity USC problem, using a gauge-invariant GME description. We exploit this model in two different ways:
($i$) we first introduce a second TLS as a weakly coupled 
{\it sensor atom} for the cavity-emitted spectrum [sensor atom approach, Fig.~\ref{Fig1}(a)], and show that it produces  qualitatively similar spectra to that computed with the quantum regression theorem, though only with certain types of bath coupling; we also confirm that these sensing atom results are identical in both the dipole gauge and Coulomb gauge, as they must be; ($ii$) we then focus on the main topic where the second atom is now also treated as an ultrastrongly coupled atom, distinct from the first atom [GDM, Fig.~\ref{Fig1}(b)], and demonstrate several new spectral features that emerge as we change the coupling parameters of the second TLS.

The rest of our paper is organized as follows:
In Sec.~\ref{sec:theory}, we present the main theory, which includes
a description of the GME, our excitation 
scheme, as well  the various system Hamiltonians, bath interactions, and observables, including the 
cavity-emitted spectra.

In Sec.~\ref{sec:results_SAA}, we  present the
main calculations and results for the sensing atom approach, and show how the sensing atom coupling can be used to model the detection of light. We also show how these results compare to calculations with the quantum regression theorem and explore the more general case of different bath couplings (for the atoms as well as the cavity).
Next, in Sec.~\ref{sec:results_GDM}, we consider the case of two atoms in the USC regime, where we change the parameters of the second atom, and study the effect
that this has on both the system eigenenergies as well as the cavity observables. We first show explicitly how our GME produces 
gauge-independent results when using the correct gauge-fixed approaches as described in the main text. Subsequently, we then present 
a series of investigations
using the dipole gauge.
Finally, we conclude in Sec.~\ref{sec:conclusions}.

\section{Theory}
\label{sec:theory}
In this section, we present the GME, 
as well as the different bath models and system Hamiltonians that we will use. We also 
show how these can be used to compute
the cavity spectra, using either the quantum regression theorem, or a sensing atom approach.

\subsection{Generalized master equation}
We first introduce the main GME that we use to compute the key observables of interest:
\begin{equation}
\begin{split}
    \frac{\partial}{\partial t}{\rho}_\mathrm{} &= -\ii\left[\hat{\cal H},{\rho}_\mathrm{}\right] +\sum_{\Lambda}\mathcal{L}^{\Lambda}{\rho}+\mathcal{L}^{\rm pump}{\rho},
\end{split}
\label{GME}
\end{equation}
where ${\rho}_\mathrm{}$ is the composite system (composed of the cavity and the atom, or atoms) density matrix, and $\hat{\mathcal{H}}\equiv\hat{\mathcal{H}}^\mathrm{D/C}$ is the 
system Hamiltonian in either gauge (dipole, `${\rm D}$', or Coulomb, `${\rm C}$'). 

The Lindbladian 
 for each dissipation channel is of the same form, so we write it generally as~\cite{settineri_dissipation_2018}
\begin{gather}
\begin{split}
    \mathcal{L}^{\Lambda}{\rho}_\mathrm{} &= \frac{1}{2} \sum\limits_{\omega,\omega'>0}
    \Gamma^{\Lambda}(\omega) \left[ {X}^{+}_{\Lambda}(\omega)\,{\rho}_\mathrm{}\,{X}^{-}_{\Lambda}(\omega')\right.     
    \\
    &\hspace{3cm}\left.- {X}^{-}_{\Lambda}(\omega') {X}^{+}_{\Lambda}(\omega)\,{\rho}_\mathrm{}\right] 
        \\
    &\hspace{1.4cm}
    + \Gamma^{\Lambda}(\omega')\left[ {X}^{+}_{\Lambda}(\omega)\,{\rho}\, {X}^{-}_{\Lambda}(\omega')\right.
    \\
    &\hspace{3.2cm}\left.- {\rho}\,{X}^{-}_{\Lambda}(\omega') {X}^{+}_{\Lambda}(\omega)\right].
\end{split}
\end{gather}
Since we now have several possible dissipation channels for the cavity, and the atoms, $\Lambda$  indexes the cavity and the atom, or atoms.

The dressed operators are defined 
from
\begin{equation}
\begin{alignedat}{2}
    {X}^{+}_{\rm cav}(\omega) &= \braket{j|\hat{\Pi}|k}\ket{j}\bra{k}, \\
    {X}^{+}_{\rm atom}(\omega) &= \braket{j|\sigma_x|k}\ket{j}\bra{k}, 
\end{alignedat}
\label{eqn:Xp}
\end{equation}
with $\omega=\omega_k-\omega_j > 0$, ${X}^{-}_{\Lambda}(\omega)=[{X}^{+}_{\Lambda}(\omega)]^\dagger$, and we assume that $\hat{\Pi}$ 
has electric field coupling, such that
$\hat{\Pi}^{\rm C}=\ii({a}^\dagger-{a})$ in the Coulomb gauge,
and   $\hat{\Pi}^{\rm D}=\ii({a}'^\dagger-{a}')$ in the dipole gauge~\cite{salmon2021gauge}.
We note that the dressed eigenstates $\{\vert j\rangle\}$ are required to construct the correct dressed operators utilized in the GME; 
these
are the 
eigenstates of the full light-matter system Hamiltonian \emph{including} the interaction term~\cite{settineri_dissipation_2018,salmon_master_2021,salmon2021gauge}.
The dressed states are naturally gauge-dependent, but the observables are not. 

Modeled by a (continuous) superposition of damped bosonic harmonic oscillators, baths are generally described by their correlation functions and, in turn, their spectral densities of states which contain information on the frequencies of the baths' modes and their coupling to the system~\cite{settineri_dissipation_2018}. 
For our purpose, 
the frequency-dependence of the baths is modeled as either a flat bath,
\begin{equation}
\Gamma^{\rm cav}(\omega)=\kappa, \ \ \Gamma^{\rm atom}(\omega)=\gamma
\end{equation}
or an Ohmic bath,
\begin{equation}
\Gamma^{\rm cav}(\omega)=\frac{\kappa\omega}{\omega_c}, \ \ \Gamma_{a,b}^{\rm atom}(\omega)=\frac{\gamma_{a,b}\,\omega}{\omega_{a,b}}.
\end{equation}
However,
in the case of a sensor atom, we use
\begin{equation}
\Gamma^{\rm sen}(\omega)=\frac{\gamma_{s}\omega}{\omega_{c}},
\end{equation}
since in reality the sensor
will also have a center frequency at the main detection frequency of interest, while we assume is at the cavity resonance frequency.

If the open system also includes a sensing element, special considerations for the sensing atom's bath must be taken into account. 
Essentially, we must add the  dissipation channel for this sensor atom in an analogous way to the primary atom. However, in principle, we require that the inclusion of the sensor should act as a noninvasive measurement.
Therefore, we must ensure that $\gamma_{s}\ll\kappa$ in either the flat or Ohmic shape of 
$\Gamma^\mathrm{sen}(\omega)$, where $\gamma_{s}$ is the sensor atom decay rate, or else the sensor atom introduces additional broadening to the existing peaks in the spectra. Careful attention is also needed as the dissipation rate of the sensor puts a limit on the coupling strength between itself and the cavity. We will cautiously take into account these considerations in our results. 

For a cavity-QED system in the USC regime, the
$\gamma \ll \kappa$ process is usually negligible; however, $\gamma$
plays an important role in the sensing atom approach (for its light detection), so we keep the bath functions general for such a  study. However, in the case of two atoms in the USC regime, we will use Ohmic baths throughout, where only
$\Gamma^\mathrm{cav}(\omega)$ is generally important.

For the excitation process,
 we also include the incoherent driving through the pump Lindbladian, with
\begin{equation}
    \mathcal{L}^{\rm pump}=\frac{1}{2}P_{\rm inc}\,\mathcal{D}[X^{-}_{\rm cav}]\,{\rho},
\end{equation}
where ${\cal D}[\hat{O}]{\rho} = \frac{1}{2}(2\hat{O}{\rho}\hat{O}^\dagger - {\rho}\hat{O}^\dagger\hat{O} - \hat{O}^\dagger\hat{O}{\rho})$,  and $P_{\rm inc}$ is the incoherent driving strength. 

\subsection{Observables}

Now that our main master equation model is established, we next present the key observables with which to explore the dynamics of the system.
These can also be used to ensure we have properly enforced gauge invariance. 
We will focus on 
the cavity-emitted spectrum.

The cavity spectrum is typically computed from the Fourier transform of the two-time cavity correlation function, which exploits the quantum regression theorem. In such an approach,
the cavity spectrum is defined
from~\cite{cui_emission_2006}
\begin{align}\label{eqn:spectra}
    &S_{\rm{cav}}(\omega) & \\ \nonumber&\propto \text{Re} \left[\int_{0}^{\infty} d\tau e^{i\omega\tau} \int_{0}^{\infty} \left\langle {X}^{-}_{\rm cav}(t)\,{X}^{+}_{\rm cav}(t+\tau) \right\rangle dt\right],
\end{align}
where $\omega$  is the emission frequency. With incoherent steady-state driving, this simplifies to a single time integration,
\begin{align}\label{eqn:spectra_SS}
    &S_{\rm{cav}}(\omega) \propto
    \text{Re} \left[\int_{0}^{\infty} d\tau e^{i\omega\tau}  \left\langle {X}^{-}_{\rm cav}(0)\,{X}^{+}_{\rm cav}(\tau) \right\rangle \right],
\end{align}
carried out after the system dynamics 
has reached  steady state.

An alternative method for computing the spectra
is to include a sensing atom, and compute its excitation flux.
Reference \cite{del_valle_theory_2012}
showed how 
 normal-order correlation functions,
 used to compute the spectrum and other observables, 
 can be computed from a frequency-tunable sensing atom in the limit of small coupling 
 with the field. In the USC regime, such methods have been discussed
 at the system Hamiltonian level~\cite{Settineri2021Apr}, and here we test how well such an approach recovers the same sort of spectra as the quantum regression theorem. If such a model is correct
 and is gauge invariant, it naturally extends to allow
 us to explore two atoms in the USC regime. In the latter case, we will use the quantum regression theorem again, primarily for convenience. However, we remark that the sensing atom approach has several potential advantages:
 ($i$) it does not require a Born-Markov approximation  to be valid, and ($ii$) it can easily be used to model pulsed excitations, without the need for a double-time integral.

 \subsection{Spectrum detected from the sensing atom approach}
 The detected spectrum from the sensing atom approach (SAA)
 is defined through 
\begin{equation}
    S_{\rm cav}^{\rm SAA}(\omega_s) = \int_0^\infty dt
    \braket{{X}^{-}_{\rm sen}(t)\,{X}^{+}_{\rm sen}(t)},
\end{equation}
where $\omega_s$ is the sensing atom frequency and, according to Eq.~\eqref{eqn:Xp}, $X^+_{\rm sen}=\braket{j|\sigma_{x,s}|k}\ket{j}\bra{k}$, with $\sigma_{x,s}$ being the Pauli $x$-matrix for the sensor. In contrast to the quantum regression theorem, such an approach 
does not require any two-time correlation functions.
Moreover, 
since we are exciting the system with a steady-state drive, then
\begin{equation}
    S_{\rm cav}^{\rm SAA}(\omega_s) = 
    \braket{{X}^{-}_{\rm sen}(0)\,{X}^{+}_{\rm sen}(0)},
\end{equation}
which is (again) computed when the system reaches a steady state. This method has a simple physical interpretation; the sensing atom excitation number is proportional to the photon flux of the cavity-QED system
which contains another atom in the USC regime.
The sensing atom then ``detects''
the cavity output flux.

For the SAA to be valid, the sensing parameters should generally be noninvasive
 to avoid affecting the detected spectrum. To be specific, the sensor atom should have a
vanishing coupling strength, $g_s \ll g$. 
 In order to determine acceptable parameters for the ``sensor atom'' (non-perturbative coupling), we use
 parameters that provide constant results
 over a range of frequencies, with acceptable run times, which are guided by
 the criteria $g_s\ll\sqrt{\gamma_s R/2}$, where $R$ is any rate in the system~\cite{del_valle_theory_2012}.

\subsection{Photodetection rate of cavity photons}
Another useful quantity to calculate is the 
 photodetection rate of cavity photons, 
 emitted from the $\ket{j}\rightarrow\ket{k}$ transition~\cite{salmon_master_2021},
 which is proportional to the ${\cal P}$ 
 ($\hat \Pi/\sqrt{2}$) quadrature matrix elements squared, namely:
\begin{equation}\label{eqn:TR}
    \lvert{\cal P}_{jk}\rvert^2 = \frac{1}{2}\lvert \braket{j|\hat{\Pi}|k}\rvert^2.
\end{equation}
This is the main system-level quantity 
that affects the spectral transition rates; however, the transition rates, $T_{ij}$, also must be multiplied  by a factor  $D^2(\omega_{jk})$, where $D(\omega)$ is the density of states of the relevant bath~\cite{savasta_thomasreichekuhn_2021},
so that, in the case of cavity emission, $T_{jk} =2 \pi   D_{\rm cav}^2(\omega_{jk})
\lvert{\cal P}_{jk}\rvert^2$;
this can be derived from Fermi's golden rule.
Moreover,
in the presence of a sensing atom, there is additional filtering through the sensing atom's density of states,
as  will show below.

For all our numerical calculations, 
we will use \verb|Python| and we also exploit the QuTiP module for quantum objects and operations~\cite{johansson2012qutip,johansson_qutip_2013}. 

\subsection{System Hamiltonians and gauge-fixing for the sensor atom approach}
\label{Sec:SAA}
We  first introduce a second atom (TLS) as a sensor for the cavity-emitted spectrum.
While the sensor atom approach does not introduce any further observables to probe or any vastly new physics, it provides  a check for gauge invariance and verifies that the second atom is included correctly in the model, before elevating it to a second atom also in the USC regime. 
It also demonstrates the influence
of an additional bath coupling, which is relevant for other types of detection, including cavity detection.
As mentioned earlier, this method
for simulating spectra 
also holds some potential advantages over
the quantum regression theorem, which requires the calculation of the two-time
correlation function.
In particular, the sensing atom approach 
is potentially more powerful when used to compute various multi-time correlation functions~\cite{del_valle_theory_2012}
and to model short-pulse excitation.
The sensor atom approach is also a valid physical model for the detection of photons emitted from the cavity, and a similar approach
could be used to describe a sensing cavity as well.

In order to not affect the spectrum, this sensor atom should have a vanishing coupling strength, $g_{s}\ll g$. In practice, however, it may have a minor 
influence on the computed spectrum, and a qualitatively different one if it also has a non-trivial bath function.

The  system Hamiltonian in such a case may be naively constructed by the addition of two terms: $\left({\omega_{s}}/{2}\right){\sigma}_{z,s}$   and $\ii g^{\rm D}_{s}({a}^\dagger - {a}){\sigma}_{x,s}$, namely the bare Hamiltonian of the sensor, and the interaction Hamiltonian between the sensor and the cavity, respectively~\cite{del_valle_theory_2012}, to Eq.~\eqref{eqn:HD_QR}; this is analogous to that of the primary atom, in the dipole gauge. 
Perhaps counter-intuitively, after our discussions on the dipole gauge Hamiltonian, here the gauge correction, including only the main atom, also needs to be applied at the Hamiltonian level for the sensor interaction~\cite{Settineri2021Apr}. This is because the sensor atom couples to the electric field of the cavity with its coupling to the primary atom already included, which explicitly contains the corrected $\hat{a}'$ operators (similar to the cavity bath operators terms in the GME).
Therefore, the naive choice is incorrect, and gauge fixing must be applied so that the interaction Hamiltonian between the sensor atom and the cavity becomes
\begin{align}
    \ii g^{\rm D}_{s}({a}^\dagger - {a}){\sigma}_{x,s}
    &\rightarrow
    g^{\rm D}_{s}\left[\ii({a}^\dagger - {a})+2\eta\sigma_x+2\eta_s\sigma_{x,s}\right]{\sigma}_{x,s} \nonumber \\
    & \approx g^{\rm D}_{s}\left[\ii({a}^\dagger - {a})+2\eta\sigma_x\right]{\sigma}_{x,s},
\end{align}
 since $\eta_s\ll \eta$
 and also $\sigma_{x,s}^2={\bf 1}$ (which only gives an energy offset).

Thus, in the dipole gauge, the
gauge-corrected full system Hamiltonian reads
\cite{Settineri2021Apr,salmon_master_2021,macri_spontaneous_2022}
\begin{equation}\label{eqn:HD_SAA}
\begin{split}\displaystyle
\hat{\cal H}^{\rm D}_{\rm SAA} 
& = \hat{\cal H}^{\rm D}_{\rm QR} +\frac{\omega_{s}}{2}{\sigma}_{z,s} +  g_{s}^{\rm D}\left[\ii({a}^\dagger - {a})+2\eta{\sigma}_x\right]{\sigma}_{x,s}.
    \end{split}
\end{equation}
Applying a RWA, then we have 
\begin{equation}\label{eqn:HD_SAA_RWA}
\begin{split}
\hat{\cal H}^{\rm D}_{\rm SAA}|_{\rm RWA} 
& = \hat{\mathcal{H}}^{\rm D}_{\rm JC} + \frac{\omega_{s}}{2}\sigma_{z,s}
\\
&+ g^{\rm D}_s\left[\ii({a}^\dagger\sigma_s^--a\sigma_s^+)+2\eta(\sigma^+\sigma_s^-+\sigma^-{\sigma}_s^+)\right],
\end{split}
\end{equation}
where  $\eta$ is the normalized coupling for the primary atom.
Clearly, in the sensor atom approach, the sensor atom does not need to modify the principal atom-cavity coupling and related observable operators, though its bath 
interactions can play a qualitatively important role. In the USC regime, of course, the RWA does not work, but it is useful to highlight the effects of counter-RWA terms, at least at the level of how these affect the system eigenfrequencies.

In the Coulomb gauge, the Hamiltonian  can be also obtained 
from the dipole-gauge one,
by conducting the unitary transform $\hat{\cal H}^{\rm C}_{\rm SAA} = {\cal U}^\dagger_{1,s}\hat{\cal H}^{\rm D}_{\rm SAA}{\cal U}_{1,s}$, 
with ${\cal U}_{1,s}=\exp[\ii({a}+{a}^\dagger)(\eta\sigma_x+\eta_s{\sigma}_{x,s})]$, 
yielding~\cite{garziano_gauge_2020}
\begin{equation}\label{eqn:HD+sen}
\begin{split}\displaystyle
\hat{\cal H}^{\rm C}_{\rm SAA} &= \hat{\cal H}^{\rm C}_{\rm QR}
    \\\displaystyle
    &\hspace{-0.5cm} + \frac{\omega_{s}}{2}\left\{{\sigma}_{z,s}\cos{} [2\eta_s({a}+{a}^\dagger)] +{\sigma}_{y,s}\sin{} [2\eta_{s}({a}+{a}^\dagger)]\right\},
    \end{split}
\end{equation}
where $g^{\rm C}_s=g^{\rm D}_s\omega_s/\omega_c$, $\eta_s = g^{\rm C}_s/\omega_s=g_s^{\rm D}/\omega_c$, and again we have neglected terms proportional to the identity operator.
Note here that, as in the Coulomb Hamiltonian without the sensor, there is no separable bare sensor Hamiltonian when including the gauge correction. However, without the gauge correction, we have, $(\omega_s/2){\sigma}_{z,s}+ g^{\rm D}_{s}({a}+{a}^\dagger){\sigma}_{y,s}$ for the bare sensor and cavity-sensor interaction Hamiltonian.

Naturally, we can apply this transformation in reverse, i.e., to the cavity operators in the Coulomb gauge, to find the two-atom version of the gauge correction in the dipole gauge. As expected, this results in the corrected operators in the dipole gauge, and yields
\begin{equation}
    {a}\,\rightarrow\,{a}' = {a} - \ii(\eta\sigma_x+\eta_{s}{\sigma}_{x,s}).
\end{equation}
As discussed before, computing field observables in the dipole gauge must then be done with these corrected electric-field operators.
The GME for the sensor atom approach has the same format as Eq.~\eqref{GME}. However, one must use the correct definition of Hamiltonian  $\hat{\mathcal{H}}\equiv\hat{\mathcal{H}}_\mathrm{SAA}^\mathrm{D/C}$, to ensure gauge-invariant results, and also one has additional bath coupling terms, which need not have the same spectral function (i.e., the density of states seen by the sensing atom could be different to the density of states seen by the cavity).

\subsection{System Hamiltonians and gauge-fixing for the generalized Dicke model}
\label{Sec:GDM}
The  Dicke model in the USC regime takes the QRM and adds a second identical atom to the system, also in the USC regime. This also has to be done with the gauge-corrections, to ensure the results are gauge invariant. The difference between the regular (or previously extended) Dicke model~\cite{PhysRevA.75.013804,PhysRevA.95.053854,PhysRevA.97.053821,garziano_gauge_2020} and the GDM is that we allow the two ultrastrongly coupled atoms to vary in their frequency and coupling strength.
We have already described how to include a second atom (TLS) into our system in both the dipole gauge and the Coulomb gauge, using  the sensor atom approach, where gauge invariance is also ensured. The only difference here is that we  use Eq.~\eqref{eqn:spectra} to compute the spectrum since the second atom is now participating in energy exchange with the cavity and its population cannot be relied on to obtain the spectrum, as it is no longer acting as a weak sensor.
Of course, one could bring in a third atom as a sensor, but in general, the quantum regression theorem is efficient and accurate for the problems we study below, especially with time-independent
incoherent driving.

Conveniently, the required Hamiltonians
for two atoms in the USC regime are similar to the above, except that for the operators and quantities associated with the original atom we assign a subscript $`a$' and for those of the sensor atom we assign a subscript $`b$',
and we no longer assume
$\eta_b\ll \eta_a$. 
Thus we have 
\begin{equation}\label{eqn:HD_GDM}
\begin{aligned}
    \hat{\cal H}^{\rm D}_{\rm GDM} &= \omega_c{a}^\dagger {a} 
    + \frac{\omega_a}{2}\sigma_{z,a} + 
    \ii g^{\rm D}_a({a}^\dagger-{a})\sigma_{x,a}\\
    &+ \frac{\omega_b}{2}\sigma_{z,b} + \ii g^{\rm D}_b({a}^\dagger-{a})\sigma_{x,b} 
    + 2\omega_c\eta_a\eta_b\sigma_{x,a}\sigma_{x,b},
\end{aligned}
\end{equation}
and for reference, 
with the application of the RWA, one has
\begin{equation}\label{eqn:HD_GDM_RWA}
\begin{aligned}
    \hat{\cal H}^{\rm D}_{\rm GDM}|_{\rm RWA}  = \omega_c{a}^\dagger {a} 
    &+ \frac{\omega_a}{2}\sigma_{z,a} + 
    \ii g^{\rm D}_a({a}^\dagger\sigma_{a}^--{a}\sigma_{a}^+)\\
    &+ \frac{\omega_b}{2}\sigma_{z,b} + \ii g^{\rm D}_b({a}^\dagger\sigma_{b}^- -{a}\sigma_{b}^+ )
    \\
    &\hspace{1.0cm}+ 2\omega_c\eta_a\eta_b(\sigma_{a}^-\sigma_{b}^++\sigma_{a}^+\sigma_{b}^-).
\end{aligned}
\end{equation}

As before, from the transformation $\hat{\cal H}_{\rm GDM}^{\rm C}={\cal U}^\dagger_{1,2}\hat{\cal H}^{\rm D}_{\rm GDM}{\cal U}_{1,2}$, 
with ${\cal U}_{1,2}=\exp[\ii({a}+{a}^\dagger)(\eta_a\sigma_{x,a}+\eta_b{\sigma}_{x,b})]$, then the correct Coulomb-gauge Hamiltonian is
\begin{equation}\label{eqn:HC_GDM}
\begin{aligned}
    &\hat{\cal H}_{\rm GDM}^{\rm C} = \omega_c {a}^\dagger {a} 
    \\
    &\ \ \ + \frac{\omega_a}{2}\left\{\sigma_{z,a}\cos{} [2\eta_a({a}+{a}^\dagger)] + \sigma_{y,a}\sin{} [2\eta_a({a}+{a}^\dagger)]\right\}
    \\
    &\ \ \ + \frac{\omega_b}{2}\left\{\sigma_{z,b}\cos{} [2\eta_b({a}+{a}^\dagger)] + \sigma_{y,b}\sin{} [2\eta_b({a}+{a}^\dagger)]\right\}.
\end{aligned}
\end{equation}

Simulations can then proceed as before,
e.g., if using the dipole gauge, then one must use the corrected cavity operator
\begin{equation}
\begin{split}
a\,\to\,{a}' = {a} - \ii(\eta_a\sigma_{x,a}+\eta_b\sigma_{x,b}), 
\end{split}
\end{equation}
to compute cavity observables in the dipole gauge,
and these are also used for deriving the dissipation terms in the GME.
The GME [Eq.~\eqref{GME}] then uses the appropriate Hamiltonians, $\hat{\mathcal{H}}\equiv\hat{\mathcal{H}}_\mathrm{GDM}^\mathrm{D/C}$, which are the total gauge-corrected Hamiltonians of the system in their respective gauge for the GDM.  In the Coulomb gauge, no modification is needed
 on the field operators.


\section{Results and discussions of the sensor atom approach}
\label{sec:results_SAA}

Thus far, we have presented a gauge-corrected model for the system Hamiltonian for the sensing atom approach; namely, one that yields the same eigenfrequencies for either the dipole gauge or the Coulomb gauge. This leads to the proper understanding of the eigenenergies and eigenstates of the closed system, which is basically an extension of the original QRM. 
We must also consider dissipation for this sensor atom, including it in an analogous way to the primary atom,
under the condition
 $\gamma_{s}\ll\kappa$. The chosen dissipation rate of the sensor also puts a limit on the coupling strength between itself and the cavity. In general, we require that the coupling must be small enough to ensure that losses from the cavity into the sensor and the sensor back-action into the main system are negligible. This crucial argument leads to an acceptable range of parameters discussed in Ref.~\cite{del_valle_theory_2012}, and also discussed in more details below.

For computing spectra
in either considered gauge (dipole and Coulomb), we then allow the system to evolve to  steady state, 
also including an
incoherent pump term. Once the steady state has been reached, we take the expectation value of the sensor excitation. We do this for a range of  frequencies  of interest to form the 
sensor atom spectra, performing a calculation for each scanned $\omega_s$. 

For this sensor atom section,
we  will focus our attention on 
the sensing atom interacting with 
the
primary atom
in the USC regime, using a fixed coupling parameter of $\eta=0.5$ to the primary atom. This USC regime has been studied recently using 
gauge-independent master equations,
and shown to yield identical results in the dipole and Coulomb gauges, and is thus an excellent test-bed to also compare with a sensing atom simulation~\cite{salmon2021gauge}. 

  We must first ensure an approximately vanishing coupling rate compared to the coupling between the cavity and the main atom. We take $g_s=0.001g$ to satisfy this condition. Then, to obtain a lower limit on $\gamma_{s}$, we find the \emph{smallest} transition rate in our system at $\eta=0.5$ to be $R\approx0.3g$, and we chose
  a slightly smaller value than this, to  satisfy the previously-mentioned criterion $g_s\ll\sqrt{\gamma_s R/2}$~\cite{del_valle_theory_2012}. If we then use $\gamma_{s}=0.0025g$, we obtain $\sqrt{\gamma_{s}R/2}\approx0.02\gg 0.001g=g_{s}$.
  Therefore, we 
  use $\kappa\gg\gamma_{s}\geq0.0025g$, as the acceptable range of values to choose from.

\subsection{Dressed eigenenergies/eigenstates and example transitions}

\begin{figure}[!htb]
    \centering
\includegraphics[width=.95\linewidth]{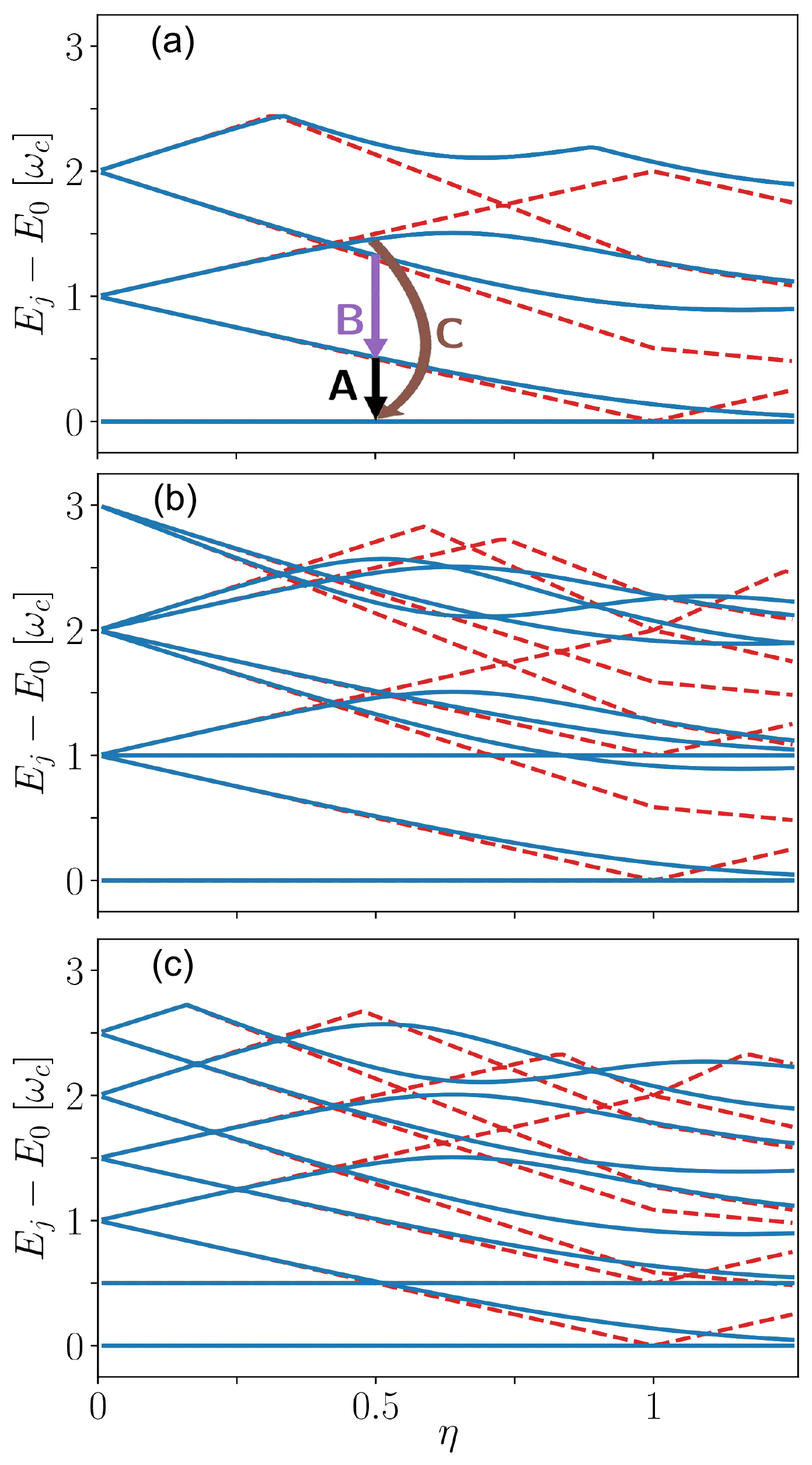}
    \caption{\textbf{Cavity-QED eigenenergies with and without a sensor atom.} Computed first ten eigenenergies of the QRM (blue solid lines) and the JCM (red dashed lines) for the cavity-QED system with (a) a single atom that is in resonance with the cavity, i.e., $\omega_a=\omega_c$, obtained from Eqs.~\eqref{eqn:HD_QR} and \eqref{eqn:HD_JC}, respectively, and (b) the sensor atom approach Hamiltonian, where both the primary and the sensor atoms are in resonance with the cavity, $\omega_a=\omega_s=\omega_c$, and the sensor atom has the coupling of $g_s^{\rm D}=0.001g^{\rm D}$; and (c) the sensor atom approach Hamiltonian, where the primary atom is in resonance with the cavity, $\omega_a=\omega_c$, and the sensor atom has $\omega_s= \omega_c/2$ and $g_s^{\rm D}=0.001g^{\rm D}$, obtained from Eqs.~\eqref{eqn:HD_SAA} and \eqref{eqn:HD_SAA_RWA}, respectively. In panel (a), the three significant transitions are identified with the downward arrows and the letters `A' ($\omega_{10}/\omega_c\approx0.5$), `B' ($\omega_{31}/\omega_c\approx8.2$), and `C' ($\omega_{20}/\omega_c\approx1.45$), for $\eta=0.5$. These resonances are highlighted for  reference when explaining the key features of the cavity spectrum.
    }
    \label{fig:cavityQED_Eigenenergies}
\end{figure}

In Fig.~\ref{fig:cavityQED_Eigenenergies}(a), we plot the 
eigenenergies of the single-atom cavity-QED without (blue solid curves) and with (red dashed curves) the RWA.
This helps to highlight the role
of the counter-rotating wave 
terms for increasing $\eta$.
The computed energies are gauge-independent (namely,
the Coulomb and dipole gauge results yield identical results),
as they should be.
In the sensor atom approach, one expects no difference between the main eigenenergies with a single TLS-cavity system, which we have confirmed to be the case; however, additional states naturally appear because of the sensing atom states, which depend on $\omega_s$. 

The three significant optical transitions are identified with the downward arrows and the letters `A' ($\omega_{10}/\omega_c\approx0.5$), `B' ($\omega_{31}/\omega_c\approx8.2$), and `C' ($\omega_{20}/\omega_c\approx1.45$), for $\eta=0.5$
(primary atom). These transitions are responsible for the significant peaks that will appear in the incoherent spectra shown in Fig.~\ref{fig:inc_sensor},
discussed below.

When the primary atom and sensor atom are both on resonance with the cavity, as expected, the eigenenergy lines start together,
at low $\eta$, and then split from the same initial points at multiples of the cavity transition energy, as shown in Fig.~\ref{fig:cavityQED_Eigenenergies}(b).
When the primary atom is on resonance but $\omega_b=0.5\omega_c$ in Fig.~\ref{fig:cavityQED_Eigenenergies}(c), the addition of extra eigenenergy lines starts at  half-multiples of the cavity transition energy.
In all three panels of Fig.~\ref{fig:cavityQED_Eigenenergies}, the deviation  of the JCM eigenenergies from the full QRM Hamiltonian eigenenergies is apparent as the normalized coupling parameter increases, a failure that is of course fully expected~\cite{frisk_kockum_ultrastrong_2019,forn-diaz_ultrastrong_2019}. 

\subsection{Cavity-emitted spectra via incoherent driving}
We next study the cavity-emitted spectra via incoherent driving,
and  consider weak drives so as not to perturb the system eigenstates too much.
We will compare our sensor results to those obtained using the  quantum regression theorem~\cite{salmon2021gauge}. Additionally, we investigate the effects of changing the 
various bath spectral functions. 

\begin{figure}[!htb]
    \centering
    \includegraphics[width=.95\linewidth]{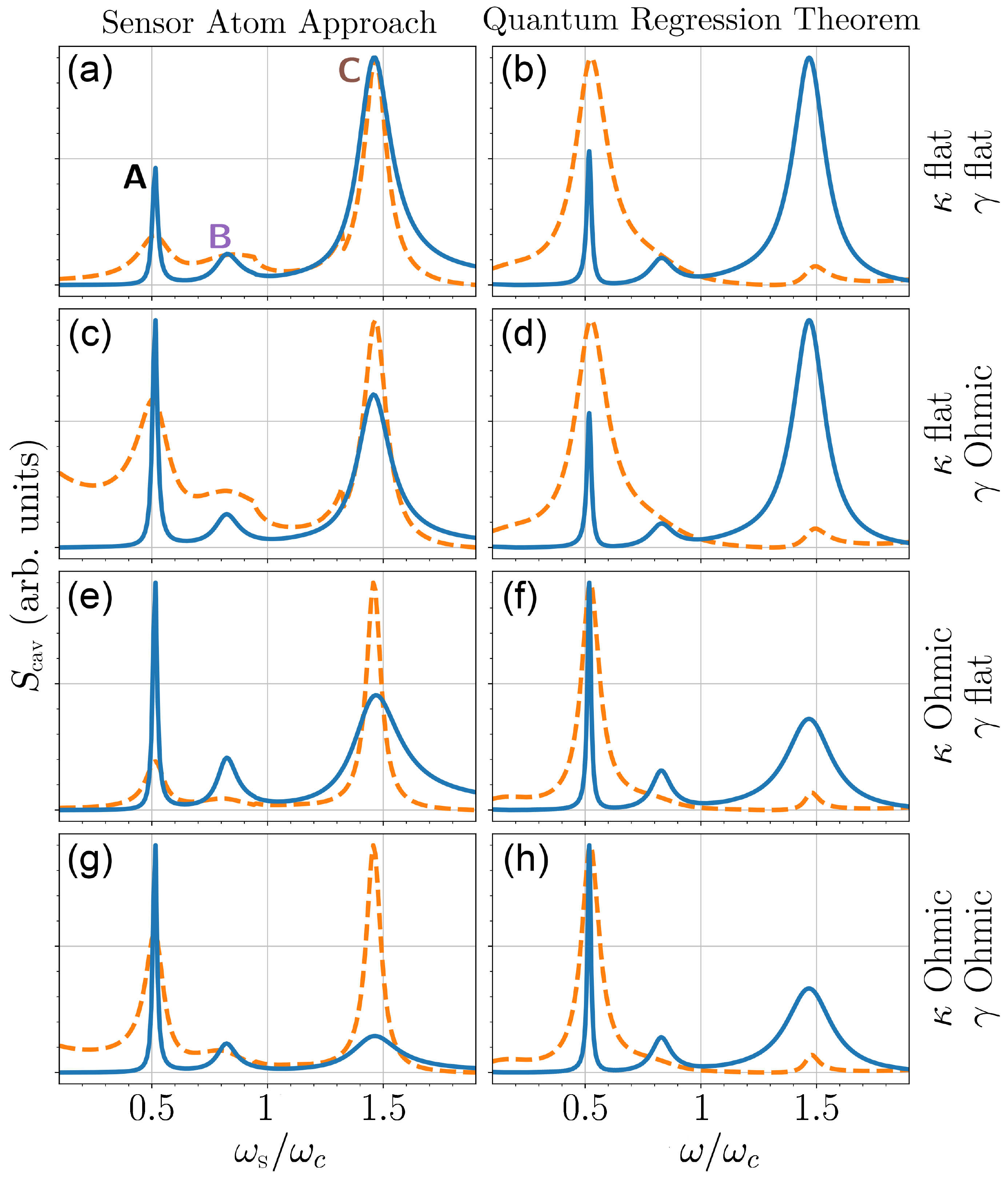}
    \caption[]{
    \textbf{Cavity-QED incoherent spectra for one atom in the ultrastrong coupling regime.} Cavity emitted spectra were computed using the sensor atom approach (left column) and the full quantum regression theorem (right column). On the right, we list the type of bath used for the cavity and the two atoms (which use the same bath type). Gauge-corrected (not corrected) results are shown with solid blue (dashed orange) curves. We use incoherent driving with $P_{\rm inc}=0.01g$. Other system parameters are $\kappa = 0.25g$, $\gamma = \gamma_{s} = 0.005g$, $g_s = 0.001g$, $\omega_c=\omega_a$, $\eta=0.5$. The three significant peaks A, B, and C represent the major transitions shown in Fig.~\ref{fig:cavityQED_Eigenenergies}. 
}
    \label{fig:inc_sensor}
\end{figure}

In the top two rows of Fig.~\ref{fig:inc_sensor}, panels~(a-d), we consider a flat bath for the cavity (i.e., $\Gamma^{\rm cav}(\omega)=\kappa$) and show the effect of changing the atomic baths from flat to Ohmic. Using the quantum regression theorem, changing the atomic bath has almost no 
visible effect on the spectrum when the cavity bath is flat. However, the spectrum detected by the  sensor atom is \emph{drastically} modified if the atomic bath of the sensor atom has a non-trivial frequency dependence. 
This can be viewed as an additional filtering process, e.g., in the case of 
an Ohmic sensor bath (i.e., $\Gamma^{\rm sen}(\omega)=\gamma_s \omega/\omega_c$), there is  increasing  dissipation at higher frequencies, thus reducing the strength of the peak on the right and increasing the relative strength of the peak on the left.

Next, as shown in the bottom half of Fig.~\ref{fig:inc_sensor}, panels~(e-h), we consider again an Ohmic cavity bath and  look at the effect of changing the atomic baths. The first result to note is that the Ohmic cavity bath produces the largest change of any of the models explored here. This is not too surprising, as the cavity dissipation is the largest by far to begin with ($\kappa = 0.25g$ vs. $\gamma = \gamma_{s} = 0.005g$), and we are modeling cavity emission. Thus, the dissipation is overwhelmingly dominated by $\kappa$, and any frequency dependence included with it will have a larger effect. Also, since the two models (quantum regression theorem and sensor atom) here have the same dependence on the single cavity, we see the change in the cavity bath having a similar effect on both spectra 
(specifically, reversing the asymmetry and modifying the relative peak heights to a similar extent). When we now also change the atomic baths to be Ohmic, we see similar effects to those above. The quantum regression theorem results are now slightly affected, and we again see a large effect on the sensor approach model.

\red{
Obviously, the gauge-uncorrected results are not expected to produce the correct physical results
in the USC regime, and thus one obtains different plots when computed by different approaches.
Beyond this, as shown and discussed above, the gauge-fixed results,
with the two different detection models,
may produce different results
depending on the model for the bath function of the detection.
Consequently, the various bath functions are generally important in
the USC regime (e.g., the cavity bath may alter the transition rates by its spectral density).
Depending on the context, one may
argue that in certain detection models, the results from the SAA would be more experimentally relevant.
In terms of highlighting the main physics for our 
two-atom USC below, either model is adequate, 
and we will use the quantum regression theorem approach as it is simpler
and more computationally efficient.
}

\section{Results and discussions of the generalized Dicke model}
\label{sec:results_GDM}

In the description of the GDM, we   extended our sensor atom approach to be a primarily part of the coupled system  (i.e., no longer weakly coupled but also in the USC regime).
Using this approach, 
we now allow the second atom's properties to vary relative to the first atom, but now when the second atom is also in the USC regime. Our two-atom Hamiltonian in the dipole gauge [Eq.~\eqref{eqn:HD_GDM}] is equivalent to the extended Dicke model in Ref.~\cite{jaako_ultrastrong-coupling_2016} in the case that the two atoms are degenerate (i.e., $g_a=g_b$ and $\omega_a=\omega_b$). 
However, 
our main focus will be on analyzing spectra obtained with {\it dissimilar} atoms, where the coupling parameters and resonant frequencies need not be the same.

Similar to the sensor atom approach, one has to first identify the dressed operators which are now found using the eigenstates of the full gauge-corrected GDM Hamiltonians, \emph{including} the second atom. We can also consider dissipation for this second atom, including it in the same way as for the primary atom, but these rates are basically negligible, as cavity decay is the main source of loss. 
In either the Coulomb gauge or the dipole gauge, we then allow the system to evolve to a steady state, again including an incoherent pump term. 
From now on, we
 use the quantum regression theorem to compute the spectra. Our first calculations will show explicitly the effect of gauge fixing and confirm gauge invariance for the spectra, and then we just choose the dipole gauge, since both gauge results yield identical results.

\subsection{System characterization: dressed eigenenergies/eigenstates and transitions}
\begin{figure}[!htb]
    \centering
\includegraphics[width=.95\linewidth]{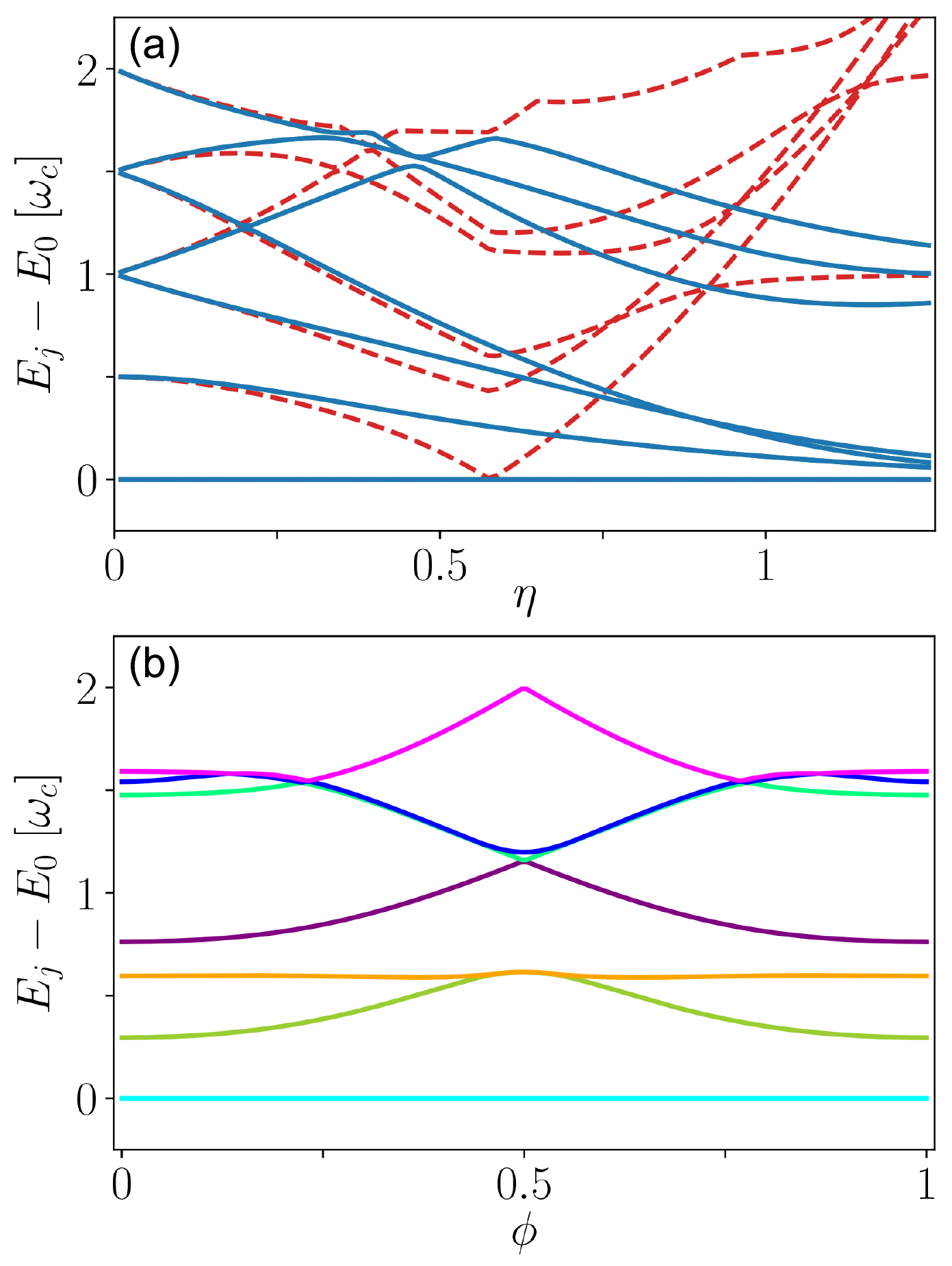}
    \caption{\textbf{Selected eigenenergies using the 
    generalized Dicke model.} We show  the first seven eigenenergies of the GDM. (a) Full quantum model without a RWA (blue solid lines) and with a RWA (red dashed lines), displaying eigenenergies with  $\omega_a=2\omega_b=\omega_c$ versus $\eta\equiv\eta_a=\eta_b$ are plotted. (b) Full model eigenenergies for  $\omega_a=2\omega_b=\omega_c$ and $|g_b|=g_a=0.5\omega_c$ ($\eta=0.5=|\eta_a|
    =|\eta_b|$, with $g_b = g_a\exp[\ii\pi\phi]$)  versus the variation of the relative phase between the coupling parameters are represented; here we do not show RWA results, as they are all clearly wrong and also gauge dependent. The random color coding in panel (b) is to help distinguish the different eigenenergies.
    }
    \label{fig:GDM_Eigenenergies}
\end{figure}

In Fig.~\ref{fig:GDM_Eigenenergies}, we show the 
first seven
eigenenergies of the GDM without (blue solid curves) and with (red dashed curves) the RWA. In panel (a), we display the eigenenergies with respect to the variation of the equal normalized coupling parameter of the two atoms, $\eta\equiv\eta_a=\eta_b$. In the GDM, one expects a considerable difference between the resulting eigenenergies compared to the single TLS-cavity system (or with the sensor atom approach) as compared to Fig.~\ref{fig:cavityQED_Eigenenergies}. In particular, one observes significant hybridization of the two TLSs in the system leading to the production of the splitting of the eigenstate curves.
As expected from our previous observation in the sensor atom approach, when $\omega_b=0.5\omega_c$ in Fig.~\ref{fig:GDM_Eigenenergies}(a), the extra eigenenergy lines start at half-multiples of the cavity transition energy. As opposed to a regular Dicke model with similar atoms, the different starting and splitting point of eigenenergies results a different
hybridization and crossing/anti-crossing. Therefore, the possibility of the existence of more exotic transitions and spectra,
in comparison to the one-atom spectra and regular Dicke model, appears.

In panel (b) of Fig.~\ref{fig:GDM_Eigenenergies}, the GDM eigenenergies are plotted when the coupling parameters  have a phase difference. Letting $g_b = g_a\exp[\ii\pi\phi]$, we equate their amplitude but vary their phase via the sweep of $\phi$ from 0 to 1. The eigenenergies in both models show symmetry  and crossing at $\phi=0.5$, as the sign of the real part of the coupling does not change the physics. 

\begin{figure}[!htb]
    \centering
\includegraphics[width=0.9\columnwidth]{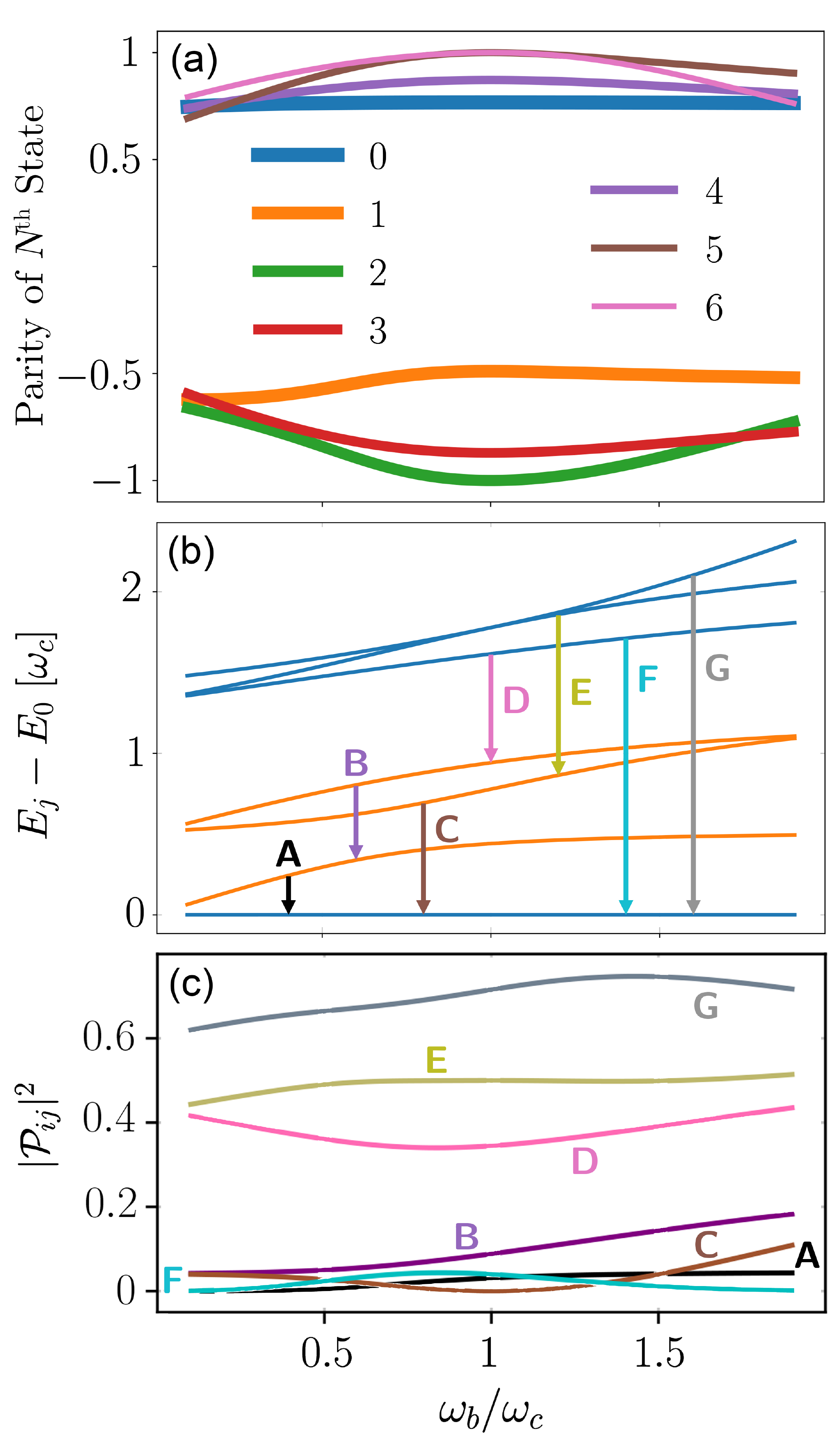}
    \caption[]{\textbf{Generalized Dicke model  state parities,  optical transitions
    and (normalized) transition rates.} (a) Parities of the first seven states of the GDM \emph{versus} the second atom's normalized frequency. (b) Eigenenergies of the first seven states of the GDM \emph{versus} the second atom's normalized frequency with positive (blue) and negative (orange) parity. (c) ${\cal P}$ quadrature matrix element squared of the selected transitions in panel (b) versus the second atom's normalized frequency obtained via Eq.~\eqref{eqn:TR}. The position of the arrows in panel (b) is irrelevant. The plots are for $\eta_a=\eta_b=0.5$.
   }    
    \label{fig:GDM_Eigenstates}
\end{figure}

In Fig.~\ref{fig:GDM_Eigenstates}, we show further details about the eigenstate properties and transitions. In particular, in Fig.~\ref{fig:GDM_Eigenstates}(a) the parities of the first seven eigenstates are shown for a range of interest in the second atom's normalized frequency, $\omega_b/\omega_c$. 
We define the parity of a state $\ket{j}$ as $\bra{j}\hat{P}\ket{j}$ where $\hat{P} = \exp[\ii\pi \hat{N}]$ and $\hat{N}={\sigma}^{+}_{a}{\sigma}^{-}_{a} + {\sigma}^{+}_{b}{\sigma}^{-}_{b} + {a}'^\dagger {a}'$ is the total excitation number (in the dipole gauge).
We label the states in Fig.~\ref{fig:GDM_Eigenstates}(a) even (odd) if their parity is positive (negative). 
We see that for the considered range of frequencies,  the first three excited states have odd parity while the ground state and the $4-6\,$th excited states have even parity.
Correspondingly, in panel (b), we plot the energy eigenvalues of these lowest seven states, where we distinguish their parity and label the main transitions
that will show up in the cavity spectra. Also, in Fig.~\ref{fig:GDM_Eigenstates}(c), $\lvert\mathcal{P}_{ij}\rvert^2$ for these transitions are shown which are related to the transition rates (which  also  depend on the  density of states of the cavity bath). All of these properties clearly explain the main spectral peaks that emerge in the computed spectrum.

As mentioned previously, and defined in Eq~\eqref{eqn:TR}, we relate the rate of a transition from state $\ket{j}$ to state $\ket{k}$~\cite{savasta_thomasreichekuhn_2021,salmon2021gauge}, as proportional to $\lvert{\cal P}_{jk}\rvert^2 = \frac{1}{2}\lvert \braket{j|\ii({a}'^\dagger-{a}')|k}\rvert^2$,
in the dipole gauge.

\subsection{Cavity emitted spectra via incoherent driving}
Our first set of GDM spectra  to study is with $\omega_a\neq\omega_b$, assuming that the two atoms have the same coupling strength.  We will first highlight that our current models do indeed ensure gauge invariance. 

\begin{figure}[!htb]
    \centering
\includegraphics[width=0.99\columnwidth]{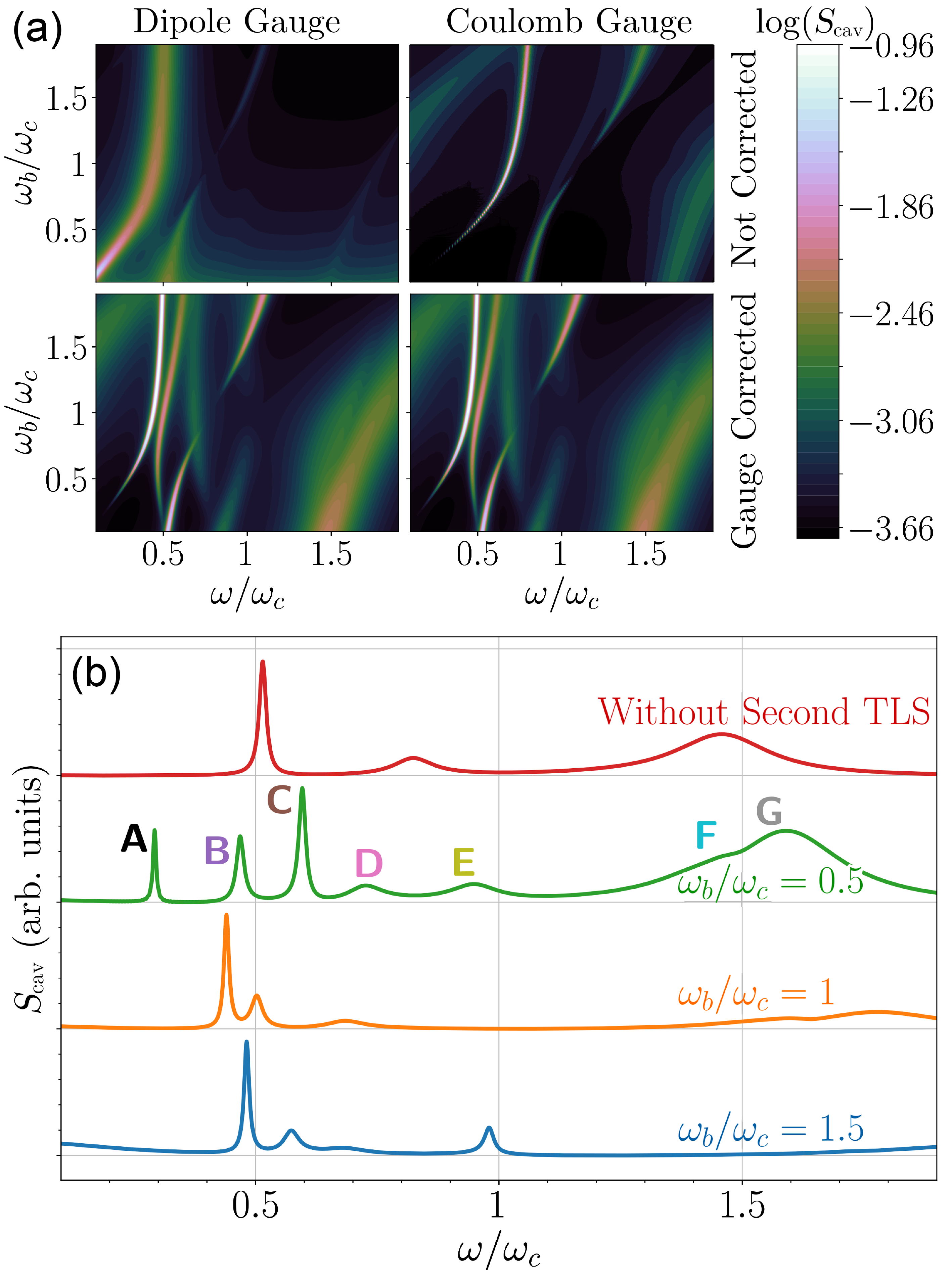}
    \caption{\textbf{Cavity spectra for the GDM, using two different atoms that are both ultrastrongly coupled to the cavity.} (a) Cavity spectra computed using the quantum regression theorem with two USC atoms in GDM, the first on resonance ($\omega_a=\omega_c$) and the second ($\omega_b$) we sweep through resonance. 
    (b) Selected cavity spectra as in the lower left panel of (a) for $\omega_b/\omega_c= \{0.5, 1, 1.5\}$. The labeled peaks in $\omega_b/\omega_c=0.5$ plot in (b) reflect on the corresponding transitions in Fig.~\ref{fig:GDM_Eigenstates}c.
    We also plot the spectra obtained in the absence of the second TLS.
    All baths are Ohmic and here $\eta_a=\eta_b=0.5$. We use the following parameters throughout the section: $\kappa=0.25g$, $\gamma_a=\gamma_b=0.005g$, and $P_{\rm inc} = 0.01 g$. }
    \label{fig:contour_eta=0.5}
\end{figure}

In Fig.~\ref{fig:contour_eta=0.5}(a), we compare the spectra obtained in the dipole gauge and Coulomb gauge and also show the
naive non-gauge-corrected counterparts. Throughout this section, we use 
Ohmic baths for the cavity and atoms. We display the spectra as a function of the second atom's frequency, while the first is held on resonance with the cavity mode, and both atoms are in the USC regime. It can easily be seen that, while the non-gauge-corrected spectra do pick up some of the correct features, they clearly do not satisfy gauge invariance. The corrected spectra are not only clearly gauge invariant but are also much richer, with additional features including a visible anti-crossing around $\omega_b/\omega_c \approx 1$ and the disappearance of a main peak in this regime as well.

Beyond the numerical success of our model and codes, we can once again identify key differences in the behavior of our system with and without gauge correction.
In all subsequent calculations, we will just choose 
the dipole gauge, since the results 
produce observables that are clearly gauge-invariant.

In Fig.~\ref{fig:contour_eta=0.5}(b), we plot selected spectra at a few selected $\omega_b$ values of interest, along with the spectra in the absence of the second TLS.
The anticrossing behavior in Fig.~\ref{fig:contour_eta=0.5}(a) is at its closest at $\omega_b/\omega_c \approx 1$; examining the 2D spectra at this frequency in Fig.~\ref{fig:contour_eta=0.5}(b), we can see that the splitting is about $0.063\omega_c = 0.126$, or about $g/8$. The location of this minimal splitting can be understood by looking at Fig.~\ref{fig:GDM_Eigenstates}(b) and noting that the eigenvalue of the third excited state is closest to twice that of the first excited state at this point, thereby making the frequencies of the \textbf{B} and \textbf{A} transitions closest. 

The change in 
$|\mathcal{P}_{ij}|^2$
(which is proportional to the 
 transition rates) is shown in Fig.~\ref{fig:GDM_Eigenstates}(c), which correlates with a change in the associated peak's height, e.g.,  peak \textbf{C}, which is absent from the spectra (peak height is zero) on resonance; this can be 
 explained by the transition rate going to zero in this regime, as shown in Fig.~\ref{fig:GDM_Eigenstates}(c). This is a strong indicator that there are features of the GDM that can only be accessed when the atoms are dissimilar (i.e., when the second atom is off-resonance). However, one cannot solely rely on the $\mathcal{P}$ quadrature matrix element squared to determine the relative heights of the peaks.
If this were the case, we would expect peak \textbf{G} to be the largest by far, and peak \textbf{A} to be very small, whereas it dominates above $\omega_b/\omega_c=0.5$. This is primarily due to the increased damping of higher states~\cite{PhysRevLett.96.127006}, and the fact that, as mentioned before, one must take into account the effect of the cavity Ohmic bath in the definition of the transition rate as $T_{jk}\propto \omega_{ij}^2
\lvert{\cal P}_{jk}\rvert^2$.
 Indeed, this clarifies that transitions involving higher states (\textbf{D}, \textbf{E}, \textbf{F}, \textbf{G}) are significantly more broadened than those involving just the lower states (\textbf{A}, \textbf{B}, \textbf{C}), which appear very sharp on the spectra. Furthermore, transition \textbf{C} has a lower $\lvert{\cal P}_{jk}\rvert^2$ value than \textbf{B}, but since \textbf{B} involves higher states, \textbf{C} dominates. Even peak \textbf{A}, with a far lower $\lvert{\cal P}_{jk}\rvert^2$ value than \textbf{B}, is larger due to the effect of damping. To sum up, this trend largely depends on the spectral bath function so that one can expect more broadening at higher energy levels, as well as the proper definition of the transition rate for a general (though Ohmic here) relevant bath.

From the eigenvalues in Fig.~\ref{fig:GDM_Eigenstates}(b), we can identify the peaks with their transitions. As  labelled in Fig.~\ref{fig:GDM_Eigenstates}(c), the visible peaks when $\eta=0.5$ are caused by transitions varying from $\ket{1}\rightarrow\ket{0}$ to $\ket{6}\rightarrow\ket{0}$. Hence, it is clear that the incoherent drive (even though weak) excites states up to at least $\ket{6}$. Moreover, it appears that most of the peaks are due to relaxation to the ground state.
This is partly because higher-order photons are already part of the lower hybrid states in the USC regime.
The key transitions are summarized in Table~\ref{tab:transitions}. Now that these peaks have been identified with specific transitions, below we next vary the coupling strength of the second atom to determine how the spectra are affected.

\begin{table}[htb]
    \centering
    \begin{tabular}{c|c}
        Peak & Transition\\ \hline
        \textbf{A} & $\ket{1}\rightarrow\ket{0}$ \\
        \textbf{B} & $\ket{3}\rightarrow\ket{1}$ \\
        \textbf{C} & $\ket{2}\rightarrow\ket{0}$ \\
        \textbf{D} & $\ket{4}\rightarrow\ket{3}$ \\
        \textbf{E} & $\ket{5}\rightarrow\ket{2}$ \\
        \textbf{F} & $\ket{4}\rightarrow\ket{0}$ \\
        \textbf{G} & $\ket{6}\rightarrow\ket{0}$ \\
    \end{tabular}
    \caption{Identification of the key transitions causing some of the peaks. Note that not all peaks are visible at all values of $\eta$. At $\eta=1$, there are other peaks present that we have not labeled.}
    \label{tab:transitions}
\end{table}

\begin{figure*}
    \centering
\includegraphics[width=.9\linewidth]{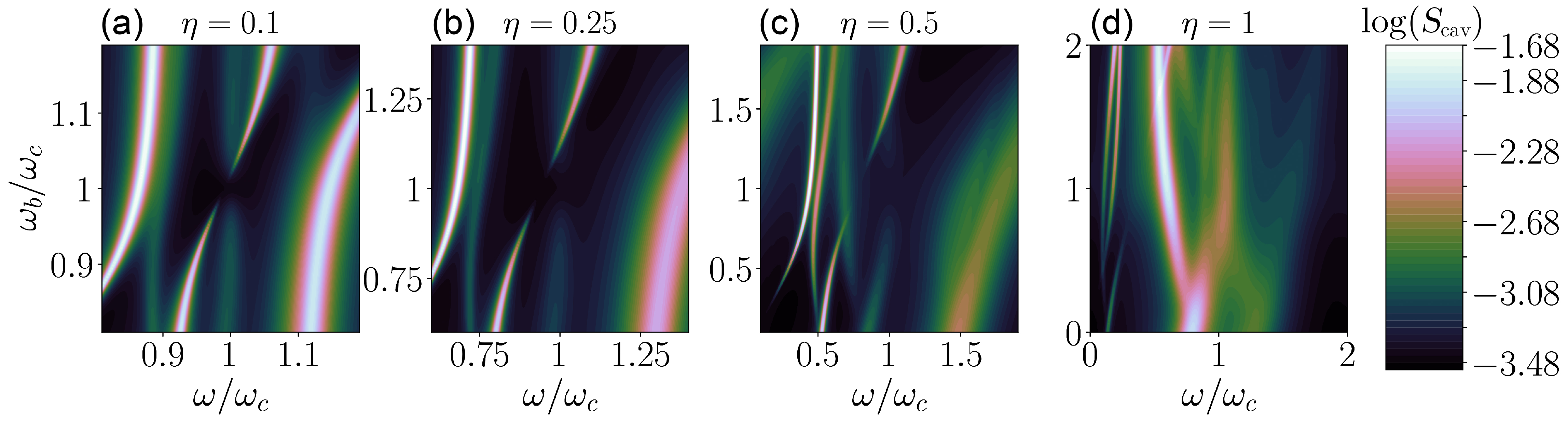}
    \caption{\textbf{Influence of coupling strength on the GDM spectra.} Cavity spectra as in Fig.~\ref{fig:contour_eta=0.5} for various $\eta$ values, with both atoms in the same coupling regime (i.e., $\eta=\eta_a=\eta_b$). As before, we keep the first atom on resonance with the cavity and sweep the second atom resonance. 
    }
    \label{fig:contour_eta=0.1-1}
\end{figure*}

Next, we vary $\eta$
from near the threshold of the USC regime ($\eta=0.1$), to the verge of the deep strong coupling regime ($\eta=1$), plotted in Fig.~\ref{fig:contour_eta=0.1-1}. At $\eta=0.1$, we see a reasonable level of symmetry around $\omega=\omega_c$, yet we also see the appearance of a new resonance which anticrosses with the lower polariton peak, near $\omega/\omega_c\approx0.9$. As we increase $\eta$, this symmetry significantly reduces and the anticrossing peaks shift to lower frequencies,
while the general Rabi splittings increase as expected, in addition to various Stark shifts.
We also see reduced broadening (sharper peaks) with increasing $\eta$ for the lower frequency peaks, as expected from the GME baths. At $\eta=1$, we discern some of the background peaks becoming the main peaks, and the apparent anti-crossing at lower $\eta$ appears to become a true crossing, i.e., near $\omega/\omega_c \approx 1$.  One of the peaks we can identify through the entire range is the one that appears forbidden (or highly reduced) when the second atom is near resonance. This peak can be identified as the \textbf{C} ($\ket{2}\rightarrow\ket{0}$) transition. However, states $\ket{2}$ and $\ket{3}$ cross in energy between $\eta=0.5$ and $\eta=1$, and are degenerate up to about $\omega_b/\omega_c=0.5$ at $\eta=1$. For simplicity, we retain the label $\ket{2}$ even after it crosses with $\ket{3}$, so that this feature is indeed due to the same transition throughout.

\subsection{Relative coupling strength variation}
\subsubsection{Influence of amplitude variation of {$g_b$}}  
\begin{figure}[!htb]
    \centering
    \includegraphics[width=1\linewidth]{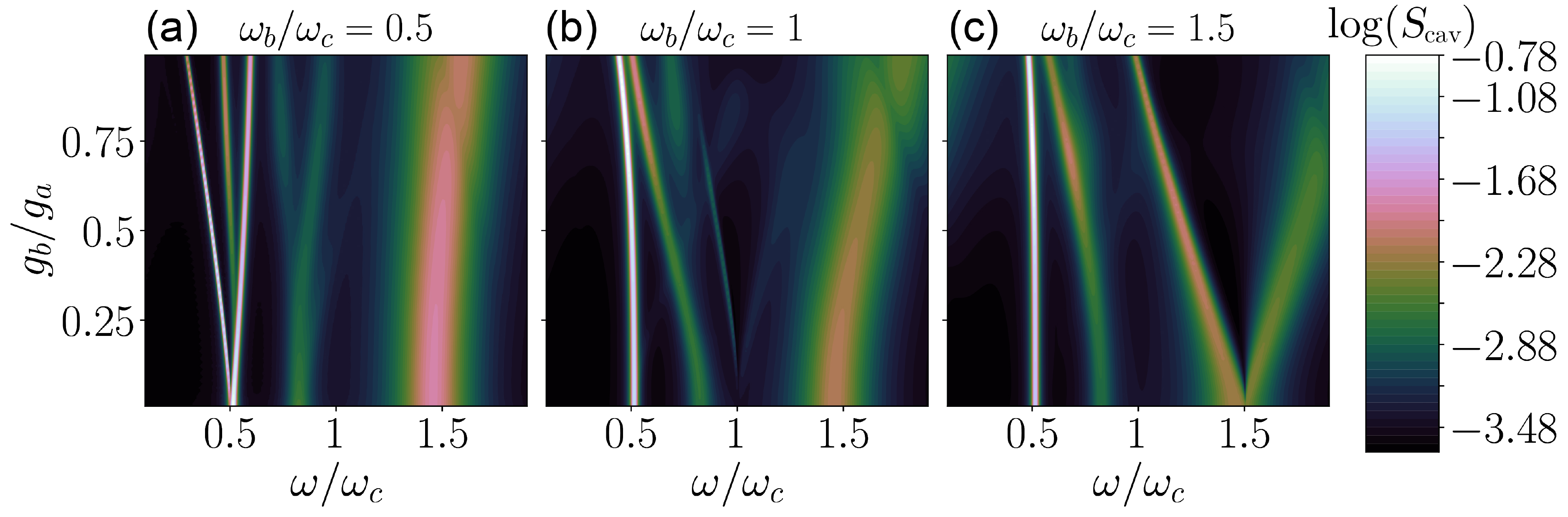}
    \caption{\textbf{Influence of relative coupling amplitude variation on the GDM cavity spectra.} Spectra at selected $\omega_b$, where we now sweep $g_b$ from negligible coupling to the same level as the first atom, at $\eta_b=\eta_a=0.5$.
}
    \label{fig:contour_wa2=1.0-3.0}
\end{figure}

In the above investigations, we chose a few values of $\omega_b$ to study in  detail. We now extend this study
further, by examining the role of
$g_b$ when it varies from zero to $g_a$. First, in Fig.~\ref{fig:contour_wa2=1.0-3.0}, we show how the spectra change when increasing the second atom's coupling strength at a few interesting values of the second atom's frequency. The main feature we can identify is a splitting of some peaks with increased $g_b$ and, interestingly, the merging of some other peaks. Some peaks also shift in frequency without any other behavior appearing (Stark shifts).
\begin{figure}[!htb]
    \centering
   \includegraphics[width=1\linewidth]{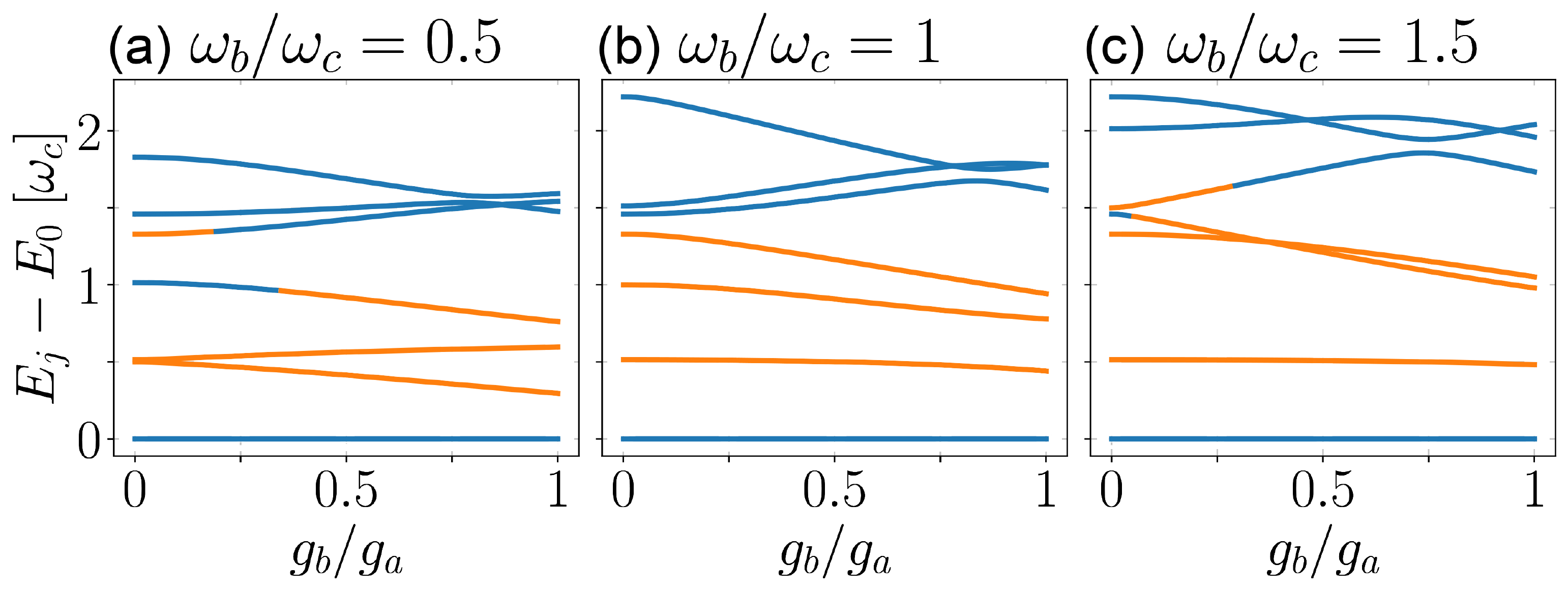}
    \caption{\textbf{Relative coupling amplitude variation in GDM eigenenergies.} Eigenvalues of the lowest seven states at selected $\omega_b$ values, as a function of $g_b$. The parity of the states is again given by the line color: blue (dark) represents even parity and orange (light) represents odd parity. 
    }
    \label{fig:evals_dicke_vs_g2}
\end{figure}

In the first example, at $\omega_b/\omega_c=0.5$ [panel (a) of Fig.~\ref{fig:contour_wa2=1.0-3.0}], we see one peak splitting into three at low frequency. Since we have already identified the origin of these peaks and given them labels at $g_b=g_a$, we can easily explain where this splitting  comes from by examining the change in energy eigenvalues as we increase $g_b$. In Fig.~\ref{fig:evals_dicke_vs_g2}, we can see that states $\ket{1}$ and $\ket{2}$, initially near degenerate at $g_b=0$, split in energy. Recalling from Table~\ref{tab:transitions} that peaks \textbf{A}, \textbf{C}, and \textbf{B} are due to transitions $\ket{1}\rightarrow\ket{0}$, $\ket{2}\rightarrow\ket{0}$, and $\ket{3}\rightarrow\ket{1}$, respectively, we can see why these peaks decrease, increase, or remain roughly unchanged in energy respectively over the range of $g_b$ considered here. Turning now to the peaks involving higher states, these are more complex due to the anticrossing of states $\ket{4}$ and $\ket{6}$ (labeled according to the order at $g_b=g_a$, to be consistent with the previous sections) around $g_b/g_a=0.8$. Considering $g_b=g_a$, the energy differences in Fig.~\ref{fig:evals_dicke_vs_g2} do explain the peaks with the same transitions as in Table~\ref{tab:transitions}. Below the anticrossing, however, the peaks can only be explained by different transitions, namely switching $\ket{4}$ with $\ket{6}$. 

Next, consider the case of the resonant second atom [panel (b) of Fig.~\ref{fig:contour_wa2=1.0-3.0}]. Once again the bright left-most peak can be trivially associated with the \textbf{A} transition. Similarly, the next peak, which almost merges with the first, is identified as transition \textbf{B}, as expected. Transition \textbf{C} is not visible in this regime, but transition \textbf{D} is seen as a broad peak at high $g_b$. Transition \textbf{E} is also not visible, but transition \textbf{F} is visible throughout and transition \textbf{G} is visible at high $g_b$.

Finally, at $\omega_b/\omega_c=1.5$ [panel (c) of Fig.~\ref{fig:contour_wa2=1.0-3.0}], we again see a significant dressing of the resonances as we change $g_b$, and all of the peaks can be identified as aligning with the transitions in Table~\ref{tab:transitions} throughout. The differences here are that transition \textbf{C} is strongly visible throughout and merges with \textbf{F} at low $g_b$ and that peaks \textbf{D}, \textbf{E}, and \textbf{G} are not visible, except \textbf{D}  at higher $g_b$. 


\subsubsection{Influence of phase variation of $g_b$}

The transition dipole of the two TLSs might not be necessarily in the same direction,
e.g., if the atomic dipoles are anisotropic and/or the field-polarization is different at the different atom locations. 
This will change the nature of the couplings in our GDM from pure real to generally complex quantities. 
Such effects have 
implications in real-world nanoengineered photonic systems,  to manipulate the quantum states and control quantum optical 
interference effects~\cite{hughes_anisotropy-induced_2017}.

\begin{figure}[!htb]
    \centering    \includegraphics[width=1\linewidth]{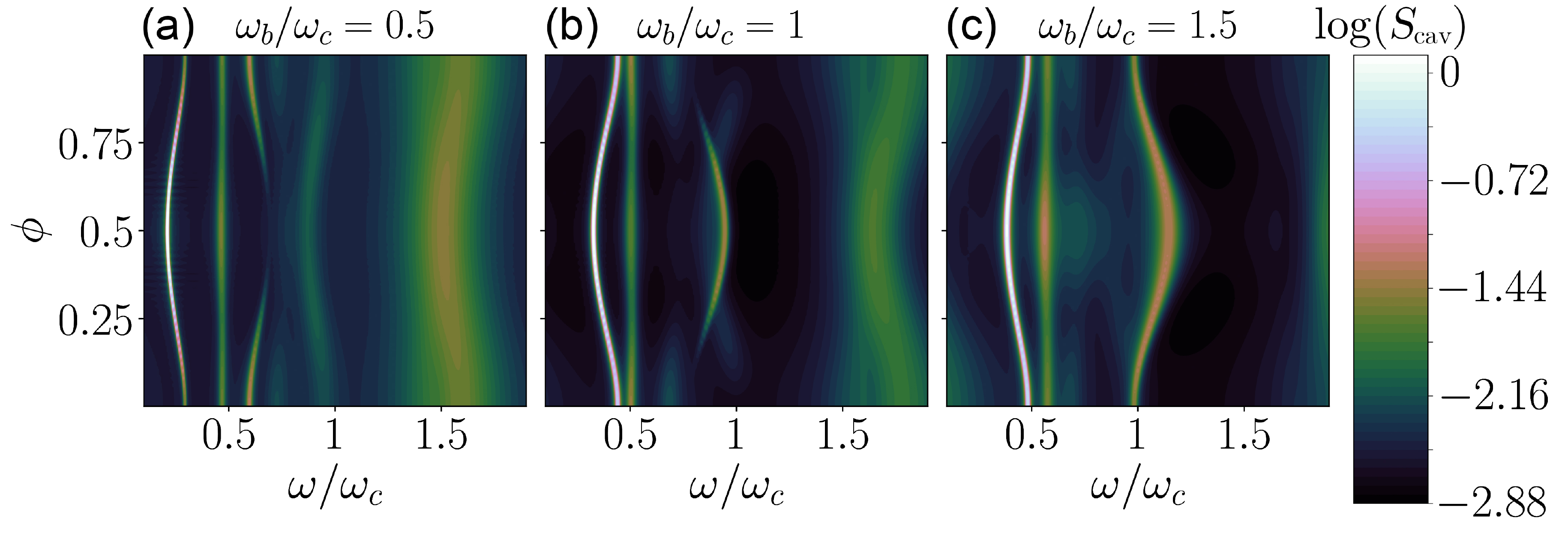}
    \caption{\textbf{Influence of relative coupling phase variation in GDM cavity spectra.} Spectra at selected $\omega_b$, where we now sweep the phase of the second atom. Notably,  when we lower the coupling strength below the USC regime $\eta_a \leq 0.1$, there is no dependence on the phase. 
    }
    \label{fig:phiSweep}
\end{figure}

To investigate the effects of a phase-dependent GDM, we next allow $g_b$ to be complex, and vary its phase. We take $g_b = g_a\exp[\ii\pi\phi]$ and sweep $\phi$ from 0 to 1, similar to Fig.~\ref{fig:GDM_Eigenenergies}(c). In Fig.~\ref{fig:phiSweep}, we show the spectra 
for $g_b$ ranging from $g_a$ to $\ii g_a$ to $-g_a$. 
Apart from being gauge independent for all results, we mention that all three of the 3D spectra (contours) 
in Fig.~\ref{fig:phiSweep} can be simulated in a matter of minutes on a standard desktop computer, where we use typically 200 bare photon states and 12 dressed states. Moreover, a single 2D spectra can be calculated at \emph{arbitrary} coupling strengths, including complex coupling from the second atom, in typically a few 10s of seconds. Thus the dressed-state truncation is not
only necessary for the GME, but it considerably simplifies the numerical Hilbert space from a bare state basis.

As observed above in Fig.~\ref{fig:contour_eta=0.5}(a), the gauge-fixed model produces peaks that are 
absent without the gauge correction, or vice versa. Indeed, at $\omega_b/\omega_c=1.5$, we can identify peak \textbf{C} as persisting throughout the change in $\phi$. Conversely, at $\omega_b/\omega_c=0.5$, peak \textbf{C} is completely absent at $\phi=0.5$ (but, persists throughout the range without the gauge-correction~\cite{salmon_master_2021}, not shown here). 
 Extending the second atom's coupling strength to a complex quantity increases the separation between the first three peaks 
 The broadening is quite low (comparatively) throughout, but the phase change does act to increase the strength of the \textbf{A} transition at $\omega_b/\omega_c=0.5$, and increases the broadening of peak \textbf{C} at $\omega_b/\omega_c=1.5$. We finally note that the spectra are symmetric about $\phi=0.5$, meaning that the spectra are invariant to changes in the sign of the real part of the coupling strength.
All these features are in accordance with the eigenenergy lines in Fig.~\ref{fig:GDM_Eigenenergies}(c).

\section{Conclusions}
\label{sec:conclusions}
We have presented a gauge-invariant GME approach
to model two atoms in ultrastrong coupling regimes of open system cavity-QED, where the atoms are modelled as Fermionic TLSs.
We first analyzed the applicability of a sensor atom approach for computing 
the detected spectra from the cavity-QED system. This is an alternative approach to using the quantum regression theorem, allowing for the computation of spectra even when driving with extremely short pulses or with multiple time-dependent fields, when the spectra solution from the quantum regression theorem may break down. This two-atom model also provides  confirmation of the gauge independence of the 
general theory for light-matter interaction in the USC regime~\cite{salmon2021gauge}, when more than one atom is included in the system, which is a non-trivial task, even without including dissipation and optical excitations.

Using incoherent driving, we demonstrated the ability of the sensor approach to produce spectra that match well with the quantum regression theorem results, when using spectrally flat baths. We also showed the influence
on the spectra when changing the bath function for both the cavity and atomic baths. We compared the Ohmic and flat baths for each case and demonstrated that the spectra only agree well when the atomic baths are flat. 
This is, however, not a realistic model for real-world detection over very large frequencies. 

For the main part of the article, we then presented results obtained using a generalized Dicke model, in the limit of two atoms. Previous studies on the Dicke model have used identical atoms, only varying properties of both at the same. However, it is practically impossible to produce this situation in a physical lab environment. Motivated by this fact, our studies presented results obtained with dissimilar atoms,
extending previous works. We first showed that our model produces gauge invariant results when including the gauge correction terms. We also showed that the gauge corrected spectra (correct spectra) are much richer than naive models and with more striking features. 

We then examined the effect of allowing the resonant frequency of the second atom to vary, and showed that there are significant peaks visible off-resonance that cannot be seen when the second atom is on-resonance with the rest of the system. We demonstrated that this effect holds for a large range of normalized coupling strengths even down to the verge of USC. This shows that this first extension, namely the ability to model two atoms with dissimilar resonant frequencies, has important implications not just in the usual USC regime. We also identified the main transitions for these visible spectral peaks.

Next, we  chose a few values of the second atom's frequency, including resonant with the first atom, to explore the second extension of our model, where we changed the coupling strength of the second atom relative to the first. We observed that some of the separate peaks can only be identified as separate peaks due to the coupling of the second atom. Indeed, a single peak without the second atom's coupling splits into three when the coupling is introduced in one of the regimes considered. Finally, we allowed the second atom's coupling to have a phase difference relative to the first,
and showed how the relative phase can substantially tune the spectral energy levels.

\red{

Qualitatively, the degree of tunability of the system to shift, produce or nullify the resonances in the emission spectra
is much richer in a GDM.
A clear observable to probe is the nullification of the main cavity resonant radiation when one of the dissimilar atoms is off-resonant with the cavity, as well as new resonances unique to the USC regime. With the naturally entangled states (even the ground state) in the USC regime, when a second
dissimilar atom is added to a single-atom cavity system, the transitions between the states are
highly modified so that some of the natural cavity-QED radiation modes are absent, and/or higher-order photons are already part of the lower hybrid states in that regime. In addition, the effects
of the separate atomic baths may be intensified in a GDM where, for example, they can massively
broaden the higher-order photons.
This can relate to emerging experimental
systems
for probing the USC regime in the near future
\cite{frisk_kockum_ultrastrong_2019,forn-diaz_ultrastrong_2019}.
}

\acknowledgements
This work was supported by the Natural Sciences and Engineering Research Council of Canada (NSERC),
the National Research Council of Canada (NRC),
the Canadian Foundation for Innovation (CFI), and Queen's University, Canada.
S.H.  acknowledges the
Japan Society for the Promotion of Science (JSPS) for funding support through an Invitational Fellowship.
F.N. is supported in part by: 
Nippon Telegraph and Telephone Corporation (NTT) Research, 
the Japan Science and Technology Agency (JST) 
[via the Quantum Leap Flagship Program (Q-LEAP), and the Moonshot R\&D Grant Number JPMJMS2061], 
the 
JSPS
[via the Grants-in-Aid for Scientific Research (KAKENHI) Grant No. JP20H00134],
the Asian Office of Aerospace Research and Development (AOARD) (via Grant No. FA2386-20-1-4069), and 
the Foundational Questions Institute Fund (FQXi) via Grant No. FQXi-IAF19-06.

\bibliography{main}

\begin{thebibliography}{100}%
\makeatletter
\providecommand \@ifxundefined [1]{%
 \@ifx{#1\undefined}
}%
\providecommand \@ifnum [1]{%
 \ifnum #1\expandafter \@firstoftwo
 \else \expandafter \@secondoftwo
 \fi
}%
\providecommand \@ifx [1]{%
 \ifx #1\expandafter \@firstoftwo
 \else \expandafter \@secondoftwo
 \fi
}%
\providecommand \natexlab [1]{#1}%
\providecommand \enquote  [1]{``#1''}%
\providecommand \bibnamefont  [1]{#1}%
\providecommand \bibfnamefont [1]{#1}%
\providecommand \citenamefont [1]{#1}%
\providecommand \href@noop [0]{\@secondoftwo}%
\providecommand \href [0]{\begingroup \@sanitize@url \@href}%
\providecommand \@href[1]{\@@startlink{#1}\@@href}%
\providecommand \@@href[1]{\endgroup#1\@@endlink}%
\providecommand \@sanitize@url [0]{\catcode `\\12\catcode `\$12\catcode
  `\&12\catcode `\#12\catcode `\^12\catcode `\_12\catcode `\%12\relax}%
\providecommand \@@startlink[1]{}%
\providecommand \@@endlink[0]{}%
\providecommand \url  [0]{\begingroup\@sanitize@url \@url }%
\providecommand \@url [1]{\endgroup\@href {#1}{\urlprefix }}%
\providecommand \urlprefix  [0]{URL }%
\providecommand \Eprint [0]{\href }%
\providecommand \doibase [0]{https://doi.org/}%
\providecommand \selectlanguage [0]{\@gobble}%
\providecommand \bibinfo  [0]{\@secondoftwo}%
\providecommand \bibfield  [0]{\@secondoftwo}%
\providecommand \translation [1]{[#1]}%
\providecommand \BibitemOpen [0]{}%
\providecommand \bibitemStop [0]{}%
\providecommand \bibitemNoStop [0]{.\EOS\space}%
\providecommand \EOS [0]{\spacefactor3000\relax}%
\providecommand \BibitemShut  [1]{\csname bibitem#1\endcsname}%
\let\auto@bib@innerbib\@empty
\bibitem [{\citenamefont {Niemczyk}\ \emph {et~al.}(2010)\citenamefont
  {Niemczyk}, \citenamefont {Deppe}, \citenamefont {Huebl}, \citenamefont
  {Menzel}, \citenamefont {Hocke}, \citenamefont {Schwarz}, \citenamefont
  {Garcia-Ripoll}, \citenamefont {Zueco}, \citenamefont {H{\"u}mmer},
  \citenamefont {Solano} \emph {et~al.}}]{niemczyk_circuit_2010}%
  \BibitemOpen
  \bibfield  {author} {\bibinfo {author} {\bibfnamefont {T.}~\bibnamefont
  {Niemczyk}}, \bibinfo {author} {\bibfnamefont {F.}~\bibnamefont {Deppe}},
  \bibinfo {author} {\bibfnamefont {H.}~\bibnamefont {Huebl}}, \bibinfo
  {author} {\bibfnamefont {E.~P.}\ \bibnamefont {Menzel}}, \bibinfo {author}
  {\bibfnamefont {F.}~\bibnamefont {Hocke}}, \bibinfo {author} {\bibfnamefont
  {M.~J.}\ \bibnamefont {Schwarz}}, \bibinfo {author} {\bibfnamefont {J.~J.}\
  \bibnamefont {Garcia-Ripoll}}, \bibinfo {author} {\bibfnamefont
  {D.}~\bibnamefont {Zueco}}, \bibinfo {author} {\bibfnamefont
  {T.}~\bibnamefont {H{\"u}mmer}}, \bibinfo {author} {\bibfnamefont
  {E.}~\bibnamefont {Solano}}, \emph {et~al.},\ }\bibfield  {title} {\bibinfo
  {title} {Circuit quantum electrodynamics in the ultrastrong-coupling
  regime},\ }\href {https://doi.org/https://doi.org/10.1038/nphys1730}
  {\bibfield  {journal} {\bibinfo  {journal} {Nature Physics}\ }\textbf
  {\bibinfo {volume} {6}},\ \bibinfo {pages} {772} (\bibinfo {year}
  {2010})}\BibitemShut {NoStop}%
\bibitem [{\citenamefont {Buluta}\ \emph {et~al.}(2011)\citenamefont {Buluta},
  \citenamefont {Ashhab},\ and\ \citenamefont {Nori}}]{buluta2011natural}%
  \BibitemOpen
  \bibfield  {author} {\bibinfo {author} {\bibfnamefont {I.}~\bibnamefont
  {Buluta}}, \bibinfo {author} {\bibfnamefont {S.}~\bibnamefont {Ashhab}},\
  and\ \bibinfo {author} {\bibfnamefont {F.}~\bibnamefont {Nori}},\ }\bibfield
  {title} {\bibinfo {title} {Natural and artificial atoms for quantum
  computation},\ }\href {https://doi.org/10.1088/0034-4885/74/10/104401}
  {\bibfield  {journal} {\bibinfo  {journal} {Reports on Progress in Physics}\
  }\textbf {\bibinfo {volume} {74}},\ \bibinfo {pages} {104401} (\bibinfo
  {year} {2011})}\BibitemShut {NoStop}%
\bibitem [{\citenamefont {Georgescu}\ and\ \citenamefont
  {Nori}(2012)}]{georgescu2012quantum}%
  \BibitemOpen
  \bibfield  {author} {\bibinfo {author} {\bibfnamefont {I.}~\bibnamefont
  {Georgescu}}\ and\ \bibinfo {author} {\bibfnamefont {F.}~\bibnamefont
  {Nori}},\ }\bibfield  {title} {\bibinfo {title} {Quantum technologies: an old
  new story},\ }\href {https://doi.org/10.1088/2058-7058/25/05/28} {\bibfield
  {journal} {\bibinfo  {journal} {Physics World}\ }\textbf {\bibinfo {volume}
  {25}},\ \bibinfo {pages} {16} (\bibinfo {year} {2012})}\BibitemShut {NoStop}%
\bibitem [{\citenamefont {Frisk~Kockum}\ \emph {et~al.}(2019)\citenamefont
  {Frisk~Kockum}, \citenamefont {Miranowicz}, \citenamefont {De~Liberato},
  \citenamefont {Savasta},\ and\ \citenamefont
  {Nori}}]{frisk_kockum_ultrastrong_2019}%
  \BibitemOpen
  \bibfield  {author} {\bibinfo {author} {\bibfnamefont {A.}~\bibnamefont
  {Frisk~Kockum}}, \bibinfo {author} {\bibfnamefont {A.}~\bibnamefont
  {Miranowicz}}, \bibinfo {author} {\bibfnamefont {S.}~\bibnamefont
  {De~Liberato}}, \bibinfo {author} {\bibfnamefont {S.}~\bibnamefont
  {Savasta}},\ and\ \bibinfo {author} {\bibfnamefont {F.}~\bibnamefont
  {Nori}},\ }\bibfield  {title} {\bibinfo {title} {Ultrastrong coupling between
  light and matter},\ }\href {https://doi.org/10.1038/s42254-018-0006-2}
  {\bibfield  {journal} {\bibinfo  {journal} {Nature Reviews Physics}\ }\textbf
  {\bibinfo {volume} {1}},\ \bibinfo {pages} {19} (\bibinfo {year}
  {2019})}\BibitemShut {NoStop}%
\bibitem [{\citenamefont {Forn-D{\'i}az}\ \emph {et~al.}(2019)\citenamefont
  {Forn-D{\'i}az}, \citenamefont {Lamata}, \citenamefont {Rico}, \citenamefont
  {Kono},\ and\ \citenamefont {Solano}}]{forn-diaz_ultrastrong_2019}%
  \BibitemOpen
  \bibfield  {author} {\bibinfo {author} {\bibfnamefont {P.}~\bibnamefont
  {Forn-D{\'i}az}}, \bibinfo {author} {\bibfnamefont {L.}~\bibnamefont
  {Lamata}}, \bibinfo {author} {\bibfnamefont {E.}~\bibnamefont {Rico}},
  \bibinfo {author} {\bibfnamefont {J.}~\bibnamefont {Kono}},\ and\ \bibinfo
  {author} {\bibfnamefont {E.}~\bibnamefont {Solano}},\ }\bibfield  {title}
  {\bibinfo {title} {Ultrastrong coupling regimes of light-matter
  interaction},\ }\href {https://doi.org/10.1103/RevModPhys.91.025005}
  {\bibfield  {journal} {\bibinfo  {journal} {Reviews of Modern Physics}\
  }\textbf {\bibinfo {volume} {91}},\ \bibinfo {pages} {025005} (\bibinfo
  {year} {2019})}\BibitemShut {NoStop}%
\bibitem [{\citenamefont {Mueller}\ \emph {et~al.}(2020)\citenamefont
  {Mueller}, \citenamefont {Okamura}, \citenamefont {Vieira}, \citenamefont
  {Juergensen}, \citenamefont {Lange}, \citenamefont {Barros}, \citenamefont
  {Schulz},\ and\ \citenamefont {Reich}}]{mueller_deep_2020}%
  \BibitemOpen
  \bibfield  {author} {\bibinfo {author} {\bibfnamefont {N.~S.}\ \bibnamefont
  {Mueller}}, \bibinfo {author} {\bibfnamefont {Y.}~\bibnamefont {Okamura}},
  \bibinfo {author} {\bibfnamefont {B.~G.~M.}\ \bibnamefont {Vieira}}, \bibinfo
  {author} {\bibfnamefont {S.}~\bibnamefont {Juergensen}}, \bibinfo {author}
  {\bibfnamefont {H.}~\bibnamefont {Lange}}, \bibinfo {author} {\bibfnamefont
  {E.~B.}\ \bibnamefont {Barros}}, \bibinfo {author} {\bibfnamefont
  {F.}~\bibnamefont {Schulz}},\ and\ \bibinfo {author} {\bibfnamefont
  {S.}~\bibnamefont {Reich}},\ }\bibfield  {title} {\bibinfo {title} {Deep
  strong light–matter coupling in plasmonic nanoparticle crystals},\ }\href
  {https://doi.org/10.1038/s41586-020-2508-1} {\bibfield  {journal} {\bibinfo
  {journal} {Nature}\ }\textbf {\bibinfo {volume} {583}},\ \bibinfo {pages}
  {780} (\bibinfo {year} {2020})}\BibitemShut {NoStop}%
\bibitem [{\citenamefont {Le~Boit{\'e}}(2020)}]{leboite_theoretical_2020}%
  \BibitemOpen
  \bibfield  {author} {\bibinfo {author} {\bibfnamefont {A.}~\bibnamefont
  {Le~Boit{\'e}}},\ }\bibfield  {title} {\bibinfo {title} {Theoretical methods
  for ultrastrong light--matter interactions},\ }\href
  {https://doi.org/https://doi.org/10.1002/qute.201900140} {\bibfield
  {journal} {\bibinfo  {journal} {Advanced Quantum Technologies}\ }\textbf
  {\bibinfo {volume} {3}},\ \bibinfo {pages} {1900140} (\bibinfo {year}
  {2020})}\BibitemShut {NoStop}%
\bibitem [{\citenamefont {Leroux}\ \emph {et~al.}(2018)\citenamefont {Leroux},
  \citenamefont {Govia},\ and\ \citenamefont {Clerk}}]{leroux_enhancing_2018}%
  \BibitemOpen
  \bibfield  {author} {\bibinfo {author} {\bibfnamefont {C.}~\bibnamefont
  {Leroux}}, \bibinfo {author} {\bibfnamefont {L.~C.~G.}\ \bibnamefont
  {Govia}},\ and\ \bibinfo {author} {\bibfnamefont {A.~A.}\ \bibnamefont
  {Clerk}},\ }\bibfield  {title} {\bibinfo {title} {Enhancing cavity quantum
  electrodynamics via antisqueezing: Synthetic ultrastrong coupling},\ }\href
  {https://doi.org/10.1103/PhysRevLett.120.093602} {\bibfield  {journal}
  {\bibinfo  {journal} {Phys. Rev. Lett.}\ }\textbf {\bibinfo {volume} {120}},\
  \bibinfo {pages} {093602} (\bibinfo {year} {2018})}\BibitemShut {NoStop}%
\bibitem [{\citenamefont {Scalari}\ \emph {et~al.}(2012)\citenamefont
  {Scalari}, \citenamefont {Maissen}, \citenamefont {Turcinková},
  \citenamefont {Hagenmüller}, \citenamefont {De~Liberato}, \citenamefont
  {Ciuti}, \citenamefont {Reichl}, \citenamefont {Schuh}, \citenamefont
  {Wegscheider}, \citenamefont {Beck},\ and\ \citenamefont
  {Faist}}]{scalari_ultrastrong_2012}%
  \BibitemOpen
  \bibfield  {author} {\bibinfo {author} {\bibfnamefont {G.}~\bibnamefont
  {Scalari}}, \bibinfo {author} {\bibfnamefont {C.}~\bibnamefont {Maissen}},
  \bibinfo {author} {\bibfnamefont {D.}~\bibnamefont {Turcinková}}, \bibinfo
  {author} {\bibfnamefont {D.}~\bibnamefont {Hagenmüller}}, \bibinfo {author}
  {\bibfnamefont {S.}~\bibnamefont {De~Liberato}}, \bibinfo {author}
  {\bibfnamefont {C.}~\bibnamefont {Ciuti}}, \bibinfo {author} {\bibfnamefont
  {C.}~\bibnamefont {Reichl}}, \bibinfo {author} {\bibfnamefont
  {D.}~\bibnamefont {Schuh}}, \bibinfo {author} {\bibfnamefont
  {W.}~\bibnamefont {Wegscheider}}, \bibinfo {author} {\bibfnamefont
  {M.}~\bibnamefont {Beck}},\ and\ \bibinfo {author} {\bibfnamefont
  {J.}~\bibnamefont {Faist}},\ }\bibfield  {title} {{\selectlanguage
  {eng}\bibinfo {title} {Ultrastrong coupling of the cyclotron transition of a
  {2D} electron gas to a {THz} metamaterial}},\ }\href
  {https://doi.org/10.1126/science.1216022} {\bibfield  {journal} {\bibinfo
  {journal} {Science}\ }\textbf {\bibinfo {volume} {335}},\ \bibinfo {pages}
  {1323} (\bibinfo {year} {2012})}\BibitemShut {NoStop}%
\bibitem [{\citenamefont {Kuhn}\ \emph {et~al.}(2002)\citenamefont {Kuhn},
  \citenamefont {Hennrich},\ and\ \citenamefont
  {Rempe}}]{PhysRevLett.89.067901}%
  \BibitemOpen
  \bibfield  {author} {\bibinfo {author} {\bibfnamefont {A.}~\bibnamefont
  {Kuhn}}, \bibinfo {author} {\bibfnamefont {M.}~\bibnamefont {Hennrich}},\
  and\ \bibinfo {author} {\bibfnamefont {G.}~\bibnamefont {Rempe}},\ }\bibfield
   {title} {\bibinfo {title} {Deterministic single-photon source for
  distributed quantum networking},\ }\href
  {https://doi.org/10.1103/PhysRevLett.89.067901} {\bibfield  {journal}
  {\bibinfo  {journal} {Phys. Rev. Lett.}\ }\textbf {\bibinfo {volume} {89}},\
  \bibinfo {pages} {067901} (\bibinfo {year} {2002})}\BibitemShut {NoStop}%
\bibitem [{\citenamefont {Volz}\ \emph {et~al.}(2012)\citenamefont {Volz},
  \citenamefont {Reinhard}, \citenamefont {Winger}, \citenamefont {Badolato},
  \citenamefont {Hennessy}, \citenamefont {Hu},\ and\ \citenamefont
  {Imamo{\u{g}}lu}}]{Volz2012}%
  \BibitemOpen
  \bibfield  {author} {\bibinfo {author} {\bibfnamefont {T.}~\bibnamefont
  {Volz}}, \bibinfo {author} {\bibfnamefont {A.}~\bibnamefont {Reinhard}},
  \bibinfo {author} {\bibfnamefont {M.}~\bibnamefont {Winger}}, \bibinfo
  {author} {\bibfnamefont {A.}~\bibnamefont {Badolato}}, \bibinfo {author}
  {\bibfnamefont {K.~J.}\ \bibnamefont {Hennessy}}, \bibinfo {author}
  {\bibfnamefont {E.~L.}\ \bibnamefont {Hu}},\ and\ \bibinfo {author}
  {\bibfnamefont {A.}~\bibnamefont {Imamo{\u{g}}lu}},\ }\bibfield  {title}
  {\bibinfo {title} {Ultrafast all-optical switching by single photons},\
  }\href {https://doi.org/10.1038/nphoton.2012.181} {\bibfield  {journal}
  {\bibinfo  {journal} {Nature Photonics}\ }\textbf {\bibinfo {volume} {6}},\
  \bibinfo {pages} {605} (\bibinfo {year} {2012})}\BibitemShut {NoStop}%
\bibitem [{\citenamefont {De~Bernardis}\ \emph
  {et~al.}(2018{\natexlab{a}})\citenamefont {De~Bernardis}, \citenamefont
  {Pilar}, \citenamefont {Jaako}, \citenamefont {De~Liberato},\ and\
  \citenamefont {Rabl}}]{de_bernardis_breakdown_2018}%
  \BibitemOpen
  \bibfield  {author} {\bibinfo {author} {\bibfnamefont {D.}~\bibnamefont
  {De~Bernardis}}, \bibinfo {author} {\bibfnamefont {P.}~\bibnamefont {Pilar}},
  \bibinfo {author} {\bibfnamefont {T.}~\bibnamefont {Jaako}}, \bibinfo
  {author} {\bibfnamefont {S.}~\bibnamefont {De~Liberato}},\ and\ \bibinfo
  {author} {\bibfnamefont {P.}~\bibnamefont {Rabl}},\ }\bibfield  {title}
  {\bibinfo {title} {Breakdown of gauge invariance in ultrastrong-coupling
  cavity {QED}},\ }\href {https://doi.org/10.1103/PhysRevA.98.053819}
  {\bibfield  {journal} {\bibinfo  {journal} {Physical Review A}\ }\textbf
  {\bibinfo {volume} {98}},\ \bibinfo {pages} {053819} (\bibinfo {year}
  {2018}{\natexlab{a}})}\BibitemShut {NoStop}%
\bibitem [{\citenamefont {Qin}\ \emph {et~al.}(2018)\citenamefont {Qin},
  \citenamefont {Miranowicz}, \citenamefont {Li}, \citenamefont {L\"u},
  \citenamefont {You},\ and\ \citenamefont {Nori}}]{qin_exponentially_2018}%
  \BibitemOpen
  \bibfield  {author} {\bibinfo {author} {\bibfnamefont {W.}~\bibnamefont
  {Qin}}, \bibinfo {author} {\bibfnamefont {A.}~\bibnamefont {Miranowicz}},
  \bibinfo {author} {\bibfnamefont {P.-B.}\ \bibnamefont {Li}}, \bibinfo
  {author} {\bibfnamefont {X.-Y.}\ \bibnamefont {L\"u}}, \bibinfo {author}
  {\bibfnamefont {J.~Q.}\ \bibnamefont {You}},\ and\ \bibinfo {author}
  {\bibfnamefont {F.}~\bibnamefont {Nori}},\ }\bibfield  {title} {\bibinfo
  {title} {Exponentially enhanced light-matter interaction, cooperativities,
  and steady-state entanglement using parametric amplification},\ }\href
  {https://doi.org/10.1103/PhysRevLett.120.093601} {\bibfield  {journal}
  {\bibinfo  {journal} {Phys. Rev. Lett.}\ }\textbf {\bibinfo {volume} {120}},\
  \bibinfo {pages} {093601} (\bibinfo {year} {2018})}\BibitemShut {NoStop}%
\bibitem [{\citenamefont {Gambino}\ \emph {et~al.}(2014)\citenamefont
  {Gambino}, \citenamefont {Mazzeo}, \citenamefont {Genco}, \citenamefont
  {Di~Stefano}, \citenamefont {Savasta}, \citenamefont {Patanè}, \citenamefont
  {Ballarini}, \citenamefont {Mangione}, \citenamefont {Lerario}, \citenamefont
  {Sanvitto},\ and\ \citenamefont {Gigli}}]{gambino_exploring_2014}%
  \BibitemOpen
  \bibfield  {author} {\bibinfo {author} {\bibfnamefont {S.}~\bibnamefont
  {Gambino}}, \bibinfo {author} {\bibfnamefont {M.}~\bibnamefont {Mazzeo}},
  \bibinfo {author} {\bibfnamefont {A.}~\bibnamefont {Genco}}, \bibinfo
  {author} {\bibfnamefont {O.}~\bibnamefont {Di~Stefano}}, \bibinfo {author}
  {\bibfnamefont {S.}~\bibnamefont {Savasta}}, \bibinfo {author} {\bibfnamefont
  {S.}~\bibnamefont {Patanè}}, \bibinfo {author} {\bibfnamefont
  {D.}~\bibnamefont {Ballarini}}, \bibinfo {author} {\bibfnamefont
  {F.}~\bibnamefont {Mangione}}, \bibinfo {author} {\bibfnamefont
  {G.}~\bibnamefont {Lerario}}, \bibinfo {author} {\bibfnamefont
  {D.}~\bibnamefont {Sanvitto}},\ and\ \bibinfo {author} {\bibfnamefont
  {G.}~\bibnamefont {Gigli}},\ }\bibfield  {title} {\bibinfo {title} {Exploring
  {Light}–{Matter} {Interaction} {Phenomena} under {Ultrastrong} {Coupling}
  {Regime}},\ }\href {https://doi.org/10.1021/ph500266d} {\bibfield  {journal}
  {\bibinfo  {journal} {ACS Photonics}\ }\textbf {\bibinfo {volume} {1}},\
  \bibinfo {pages} {1042} (\bibinfo {year} {2014})}\BibitemShut {NoStop}%
\bibitem [{\citenamefont {Calvo}\ \emph {et~al.}(2020)\citenamefont {Calvo},
  \citenamefont {Zueco},\ and\ \citenamefont
  {Martin-Moreno}}]{calvo_ultrastrong_2020}%
  \BibitemOpen
  \bibfield  {author} {\bibinfo {author} {\bibfnamefont {J.}~\bibnamefont
  {Calvo}}, \bibinfo {author} {\bibfnamefont {D.}~\bibnamefont {Zueco}},\ and\
  \bibinfo {author} {\bibfnamefont {L.}~\bibnamefont {Martin-Moreno}},\
  }\bibfield  {title} {\bibinfo {title} {Ultrastrong coupling effects in
  molecular cavity qed},\ }\href {https://doi.org/doi:10.1515/nanoph-2019-0403}
  {\bibfield  {journal} {\bibinfo  {journal} {Nanophotonics}\ }\textbf
  {\bibinfo {volume} {9}},\ \bibinfo {pages} {277} (\bibinfo {year}
  {2020})}\BibitemShut {NoStop}%
\bibitem [{\citenamefont {Baranov}\ \emph {et~al.}(2020)\citenamefont
  {Baranov}, \citenamefont {Munkhbat}, \citenamefont {Zhukova}, \citenamefont
  {Bisht}, \citenamefont {Canales}, \citenamefont {Rousseaux}, \citenamefont
  {Johansson}, \citenamefont {Antosiewicz},\ and\ \citenamefont
  {Shegai}}]{baranov_ultrastrong_2020}%
  \BibitemOpen
  \bibfield  {author} {\bibinfo {author} {\bibfnamefont {D.~G.}\ \bibnamefont
  {Baranov}}, \bibinfo {author} {\bibfnamefont {B.}~\bibnamefont {Munkhbat}},
  \bibinfo {author} {\bibfnamefont {E.}~\bibnamefont {Zhukova}}, \bibinfo
  {author} {\bibfnamefont {A.}~\bibnamefont {Bisht}}, \bibinfo {author}
  {\bibfnamefont {A.}~\bibnamefont {Canales}}, \bibinfo {author} {\bibfnamefont
  {B.}~\bibnamefont {Rousseaux}}, \bibinfo {author} {\bibfnamefont
  {G.}~\bibnamefont {Johansson}}, \bibinfo {author} {\bibfnamefont {T.~J.}\
  \bibnamefont {Antosiewicz}},\ and\ \bibinfo {author} {\bibfnamefont
  {T.}~\bibnamefont {Shegai}},\ }\bibfield  {title} {{\selectlanguage
  {en}\bibinfo {title} {Ultrastrong coupling between nanoparticle plasmons and
  cavity photons at ambient conditions}},\ }\href
  {https://doi.org/10.1038/s41467-020-16524-x} {\bibfield  {journal} {\bibinfo
  {journal} {Nat Commun}\ }\textbf {\bibinfo {volume} {11}},\ \bibinfo {pages}
  {2715} (\bibinfo {year} {2020})}\BibitemShut {NoStop}%
\bibitem [{\citenamefont {Thomas}\ \emph {et~al.}(2021)\citenamefont {Thomas},
  \citenamefont {Menghrajani},\ and\ \citenamefont
  {Barnes}}]{thomas_cavity-free_2021}%
  \BibitemOpen
  \bibfield  {author} {\bibinfo {author} {\bibfnamefont {P.~A.}\ \bibnamefont
  {Thomas}}, \bibinfo {author} {\bibfnamefont {K.~S.}\ \bibnamefont
  {Menghrajani}},\ and\ \bibinfo {author} {\bibfnamefont {W.~L.}\ \bibnamefont
  {Barnes}},\ }\bibfield  {title} {\bibinfo {title} {Cavity-{Free}
  {Ultrastrong} {Light}-{Matter} {Coupling}},\ }\href
  {https://doi.org/10.1021/acs.jpclett.1c01695} {\bibfield  {journal} {\bibinfo
   {journal} {J. Phys. Chem. Lett.}\ }\textbf {\bibinfo {volume} {12}},\
  \bibinfo {pages} {6914} (\bibinfo {year} {2021})}\BibitemShut {NoStop}%
\bibitem [{\citenamefont {Todorov}\ \emph {et~al.}(2010)\citenamefont
  {Todorov}, \citenamefont {Andrews}, \citenamefont {Colombelli}, \citenamefont
  {De~Liberato}, \citenamefont {Ciuti}, \citenamefont {Klang}, \citenamefont
  {Strasser},\ and\ \citenamefont {Sirtori}}]{todorov_ultrastrong_2010}%
  \BibitemOpen
  \bibfield  {author} {\bibinfo {author} {\bibfnamefont {Y.}~\bibnamefont
  {Todorov}}, \bibinfo {author} {\bibfnamefont {A.~M.}\ \bibnamefont
  {Andrews}}, \bibinfo {author} {\bibfnamefont {R.}~\bibnamefont {Colombelli}},
  \bibinfo {author} {\bibfnamefont {S.}~\bibnamefont {De~Liberato}}, \bibinfo
  {author} {\bibfnamefont {C.}~\bibnamefont {Ciuti}}, \bibinfo {author}
  {\bibfnamefont {P.}~\bibnamefont {Klang}}, \bibinfo {author} {\bibfnamefont
  {G.}~\bibnamefont {Strasser}},\ and\ \bibinfo {author} {\bibfnamefont
  {C.}~\bibnamefont {Sirtori}},\ }\bibfield  {title} {\bibinfo {title}
  {Ultrastrong {Light}-{Matter} {Coupling} {Regime} with {Polariton} {Dots}},\
  }\href {https://doi.org/10.1103/PhysRevLett.105.196402} {\bibfield  {journal}
  {\bibinfo  {journal} {Phys. Rev. Lett.}\ }\textbf {\bibinfo {volume} {105}},\
  \bibinfo {pages} {196402} (\bibinfo {year} {2010})}\BibitemShut {NoStop}%
\bibitem [{\citenamefont {Anappara}\ \emph {et~al.}(2009)\citenamefont
  {Anappara}, \citenamefont {De~Liberato}, \citenamefont {Tredicucci},
  \citenamefont {Ciuti}, \citenamefont {Biasiol}, \citenamefont {Sorba},\ and\
  \citenamefont {Beltram}}]{anappara_signatures_2009}%
  \BibitemOpen
  \bibfield  {author} {\bibinfo {author} {\bibfnamefont {A.~A.}\ \bibnamefont
  {Anappara}}, \bibinfo {author} {\bibfnamefont {S.}~\bibnamefont
  {De~Liberato}}, \bibinfo {author} {\bibfnamefont {A.}~\bibnamefont
  {Tredicucci}}, \bibinfo {author} {\bibfnamefont {C.}~\bibnamefont {Ciuti}},
  \bibinfo {author} {\bibfnamefont {G.}~\bibnamefont {Biasiol}}, \bibinfo
  {author} {\bibfnamefont {L.}~\bibnamefont {Sorba}},\ and\ \bibinfo {author}
  {\bibfnamefont {F.}~\bibnamefont {Beltram}},\ }\bibfield  {title} {\bibinfo
  {title} {Signatures of the ultrastrong light-matter coupling regime},\ }\href
  {https://doi.org/10.1103/PhysRevB.79.201303} {\bibfield  {journal} {\bibinfo
  {journal} {Physical Review B}\ }\textbf {\bibinfo {volume} {79}},\ \bibinfo
  {pages} {201303(R)} (\bibinfo {year} {2009})}\BibitemShut {NoStop}%
\bibitem [{\citenamefont {Walther}\ \emph {et~al.}(2006)\citenamefont
  {Walther}, \citenamefont {Varcoe}, \citenamefont {Englert},\ and\
  \citenamefont {Becker}}]{walther_cavity_2006}%
  \BibitemOpen
  \bibfield  {author} {\bibinfo {author} {\bibfnamefont {H.}~\bibnamefont
  {Walther}}, \bibinfo {author} {\bibfnamefont {B.~T.~H.}\ \bibnamefont
  {Varcoe}}, \bibinfo {author} {\bibfnamefont {B.-G.}\ \bibnamefont
  {Englert}},\ and\ \bibinfo {author} {\bibfnamefont {T.}~\bibnamefont
  {Becker}},\ }\bibfield  {title} {{\selectlanguage {en}\bibinfo {title}
  {Cavity quantum electrodynamics}},\ }\href
  {https://doi.org/10.1088/0034-4885/69/5/R02} {\bibfield  {journal} {\bibinfo
  {journal} {Rep. Prog. Phys.}\ }\textbf {\bibinfo {volume} {69}},\ \bibinfo
  {pages} {1325} (\bibinfo {year} {2006})}\BibitemShut {NoStop}%
\bibitem [{\citenamefont {Haroche}\ and\ \citenamefont
  {Kleppner}(1989)}]{haroche_cavity_1989}%
  \BibitemOpen
  \bibfield  {author} {\bibinfo {author} {\bibfnamefont {S.}~\bibnamefont
  {Haroche}}\ and\ \bibinfo {author} {\bibfnamefont {D.}~\bibnamefont
  {Kleppner}},\ }\bibfield  {title} {{\selectlanguage {en}\bibinfo {title}
  {Cavity {Quantum} {Electrodynamics}}},\ }\href
  {https://doi.org/10.1063/1.881201} {\bibfield  {journal} {\bibinfo  {journal}
  {Physics Today}\ }\textbf {\bibinfo {volume} {42}},\ \bibinfo {pages} {24}
  (\bibinfo {year} {1989})}\BibitemShut {NoStop}%
\bibitem [{\citenamefont {Miller}\ \emph {et~al.}(2005)\citenamefont {Miller},
  \citenamefont {Northup}, \citenamefont {Birnbaum}, \citenamefont {Boca},
  \citenamefont {Boozer},\ and\ \citenamefont {Kimble}}]{miller_trapped_2005}%
  \BibitemOpen
  \bibfield  {author} {\bibinfo {author} {\bibfnamefont {R.}~\bibnamefont
  {Miller}}, \bibinfo {author} {\bibfnamefont {T.~E.}\ \bibnamefont {Northup}},
  \bibinfo {author} {\bibfnamefont {K.~M.}\ \bibnamefont {Birnbaum}}, \bibinfo
  {author} {\bibfnamefont {A.}~\bibnamefont {Boca}}, \bibinfo {author}
  {\bibfnamefont {A.~D.}\ \bibnamefont {Boozer}},\ and\ \bibinfo {author}
  {\bibfnamefont {H.~J.}\ \bibnamefont {Kimble}},\ }\bibfield  {title}
  {\bibinfo {title} {Trapped atoms in cavity {QED}: coupling quantized light
  and matter},\ }\href {https://doi.org/10.1088/0953-4075/38/9/007} {\bibfield
  {journal} {\bibinfo  {journal} {Journal of Physics B: Atomic, Molecular and
  Optical Physics}\ }\textbf {\bibinfo {volume} {38}},\ \bibinfo {pages} {S551}
  (\bibinfo {year} {2005})}\BibitemShut {NoStop}%
\bibitem [{\citenamefont {Schuster}\ \emph {et~al.}(2008)\citenamefont
  {Schuster}, \citenamefont {Kubanek}, \citenamefont {Fuhrmanek}, \citenamefont
  {Puppe}, \citenamefont {Pinkse}, \citenamefont {Murr},\ and\ \citenamefont
  {Rempe}}]{schuster_nonlinear_2008}%
  \BibitemOpen
  \bibfield  {author} {\bibinfo {author} {\bibfnamefont {I.}~\bibnamefont
  {Schuster}}, \bibinfo {author} {\bibfnamefont {A.}~\bibnamefont {Kubanek}},
  \bibinfo {author} {\bibfnamefont {A.}~\bibnamefont {Fuhrmanek}}, \bibinfo
  {author} {\bibfnamefont {T.}~\bibnamefont {Puppe}}, \bibinfo {author}
  {\bibfnamefont {P.~W.~H.}\ \bibnamefont {Pinkse}}, \bibinfo {author}
  {\bibfnamefont {K.}~\bibnamefont {Murr}},\ and\ \bibinfo {author}
  {\bibfnamefont {G.}~\bibnamefont {Rempe}},\ }\bibfield  {title} {\bibinfo
  {title} {Nonlinear spectroscopy of photons bound to one atom},\ }\href
  {https://doi.org/10.1038/nphys940} {\bibfield  {journal} {\bibinfo  {journal}
  {Nature Physics}\ }\textbf {\bibinfo {volume} {4}},\ \bibinfo {pages} {382}
  (\bibinfo {year} {2008})}\BibitemShut {NoStop}%
\bibitem [{\citenamefont {Bishop}\ \emph {et~al.}(2008)\citenamefont {Bishop},
  \citenamefont {Chow}, \citenamefont {Koch}, \citenamefont {Houck},
  \citenamefont {Devoret}, \citenamefont {Thuneberg}, \citenamefont {Girvin},\
  and\ \citenamefont {Schoelkopf}}]{Bishop2008}%
  \BibitemOpen
  \bibfield  {author} {\bibinfo {author} {\bibfnamefont {L.~S.}\ \bibnamefont
  {Bishop}}, \bibinfo {author} {\bibfnamefont {J.~M.}\ \bibnamefont {Chow}},
  \bibinfo {author} {\bibfnamefont {J.}~\bibnamefont {Koch}}, \bibinfo {author}
  {\bibfnamefont {A.~A.}\ \bibnamefont {Houck}}, \bibinfo {author}
  {\bibfnamefont {M.~H.}\ \bibnamefont {Devoret}}, \bibinfo {author}
  {\bibfnamefont {E.}~\bibnamefont {Thuneberg}}, \bibinfo {author}
  {\bibfnamefont {S.~M.}\ \bibnamefont {Girvin}},\ and\ \bibinfo {author}
  {\bibfnamefont {R.~J.}\ \bibnamefont {Schoelkopf}},\ }\bibfield  {title}
  {\bibinfo {title} {Nonlinear response of the vacuum {Rabi} resonance},\
  }\href {https://doi.org/10.1038/nphys1154} {\bibfield  {journal} {\bibinfo
  {journal} {Nature Physics}\ }\textbf {\bibinfo {volume} {5}},\ \bibinfo
  {pages} {105} (\bibinfo {year} {2008})}\BibitemShut {NoStop}%
\bibitem [{\citenamefont {Brune}\ \emph {et~al.}(1996)\citenamefont {Brune},
  \citenamefont {Schmidt-Kaler}, \citenamefont {Maali}, \citenamefont {Dreyer},
  \citenamefont {Hagley}, \citenamefont {Raimond},\ and\ \citenamefont
  {Haroche}}]{PhysRevLett.76.1800}%
  \BibitemOpen
  \bibfield  {author} {\bibinfo {author} {\bibfnamefont {M.}~\bibnamefont
  {Brune}}, \bibinfo {author} {\bibfnamefont {F.}~\bibnamefont
  {Schmidt-Kaler}}, \bibinfo {author} {\bibfnamefont {A.}~\bibnamefont
  {Maali}}, \bibinfo {author} {\bibfnamefont {J.}~\bibnamefont {Dreyer}},
  \bibinfo {author} {\bibfnamefont {E.}~\bibnamefont {Hagley}}, \bibinfo
  {author} {\bibfnamefont {J.~M.}\ \bibnamefont {Raimond}},\ and\ \bibinfo
  {author} {\bibfnamefont {S.}~\bibnamefont {Haroche}},\ }\bibfield  {title}
  {\bibinfo {title} {Quantum {Rabi} oscillation: A direct test of field
  quantization in a cavity},\ }\href
  {https://doi.org/10.1103/PhysRevLett.76.1800} {\bibfield  {journal} {\bibinfo
   {journal} {Phys. Rev. Lett.}\ }\textbf {\bibinfo {volume} {76}},\ \bibinfo
  {pages} {1800} (\bibinfo {year} {1996})}\BibitemShut {NoStop}%
\bibitem [{\citenamefont {Flick}\ \emph {et~al.}(2017)\citenamefont {Flick},
  \citenamefont {Ruggenthaler}, \citenamefont {Appel},\ and\ \citenamefont
  {Rubio}}]{flick_atoms_2017}%
  \BibitemOpen
  \bibfield  {author} {\bibinfo {author} {\bibfnamefont {J.}~\bibnamefont
  {Flick}}, \bibinfo {author} {\bibfnamefont {M.}~\bibnamefont {Ruggenthaler}},
  \bibinfo {author} {\bibfnamefont {H.}~\bibnamefont {Appel}},\ and\ \bibinfo
  {author} {\bibfnamefont {A.}~\bibnamefont {Rubio}},\ }\bibfield  {title}
  {\bibinfo {title} {Atoms and molecules in cavities, from weak to strong
  coupling in quantum-electrodynamics ({QED}) chemistry},\ }\href
  {https://doi.org/10.1073/pnas.1615509114} {\bibfield  {journal} {\bibinfo
  {journal} {Proceedings of the National Academy of Sciences}\ }\textbf
  {\bibinfo {volume} {114}},\ \bibinfo {pages} {3026} (\bibinfo {year}
  {2017})}\BibitemShut {NoStop}%
\bibitem [{\citenamefont {Yoshie}\ \emph {et~al.}(2004)\citenamefont {Yoshie},
  \citenamefont {Scherer}, \citenamefont {Hendrickson}, \citenamefont
  {Khitrova}, \citenamefont {Gibbs}, \citenamefont {Rupper}, \citenamefont
  {Ell}, \citenamefont {Shchekin},\ and\ \citenamefont
  {Deppe}}]{yoshie_vacuum_2004}%
  \BibitemOpen
  \bibfield  {author} {\bibinfo {author} {\bibfnamefont {T.}~\bibnamefont
  {Yoshie}}, \bibinfo {author} {\bibfnamefont {A.}~\bibnamefont {Scherer}},
  \bibinfo {author} {\bibfnamefont {J.}~\bibnamefont {Hendrickson}}, \bibinfo
  {author} {\bibfnamefont {G.}~\bibnamefont {Khitrova}}, \bibinfo {author}
  {\bibfnamefont {H.~M.}\ \bibnamefont {Gibbs}}, \bibinfo {author}
  {\bibfnamefont {G.}~\bibnamefont {Rupper}}, \bibinfo {author} {\bibfnamefont
  {C.}~\bibnamefont {Ell}}, \bibinfo {author} {\bibfnamefont {O.~B.}\
  \bibnamefont {Shchekin}},\ and\ \bibinfo {author} {\bibfnamefont {D.~G.}\
  \bibnamefont {Deppe}},\ }\bibfield  {title} {\bibinfo {title} {Vacuum {Rabi}
  splitting with a single quantum dot in a photonic crystal nanocavity},\
  }\href {https://doi.org/10.1038/nature03119} {\bibfield  {journal} {\bibinfo
  {journal} {Nature}\ }\textbf {\bibinfo {volume} {432}},\ \bibinfo {pages}
  {200} (\bibinfo {year} {2004})}\BibitemShut {NoStop}%
\bibitem [{\citenamefont {Reithmaier}\ \emph {et~al.}(2004)\citenamefont
  {Reithmaier}, \citenamefont {S\k{e}k}, \citenamefont {L{\"o}ffler},
  \citenamefont {Hofmann}, \citenamefont {Kuhn}, \citenamefont {Reitzenstein},
  \citenamefont {Keldysh}, \citenamefont {Kulakovskii}, \citenamefont
  {Reinecke},\ and\ \citenamefont {Forchel}}]{reithmaier_strong_2004}%
  \BibitemOpen
  \bibfield  {author} {\bibinfo {author} {\bibfnamefont {J.~P.}\ \bibnamefont
  {Reithmaier}}, \bibinfo {author} {\bibfnamefont {G.}~\bibnamefont {S\k{e}k}},
  \bibinfo {author} {\bibfnamefont {A.}~\bibnamefont {L{\"o}ffler}}, \bibinfo
  {author} {\bibfnamefont {C.}~\bibnamefont {Hofmann}}, \bibinfo {author}
  {\bibfnamefont {S.}~\bibnamefont {Kuhn}}, \bibinfo {author} {\bibfnamefont
  {S.}~\bibnamefont {Reitzenstein}}, \bibinfo {author} {\bibfnamefont {L.~V.}\
  \bibnamefont {Keldysh}}, \bibinfo {author} {\bibfnamefont {V.~D.}\
  \bibnamefont {Kulakovskii}}, \bibinfo {author} {\bibfnamefont {T.~L.}\
  \bibnamefont {Reinecke}},\ and\ \bibinfo {author} {\bibfnamefont
  {A.}~\bibnamefont {Forchel}},\ }\bibfield  {title} {\bibinfo {title} {Strong
  coupling in a single quantum dot–semiconductor microcavity system},\ }\href
  {https://doi.org/10.1038/nature02969} {\bibfield  {journal} {\bibinfo
  {journal} {Nature}\ }\textbf {\bibinfo {volume} {432}},\ \bibinfo {pages}
  {197} (\bibinfo {year} {2004})}\BibitemShut {NoStop}%
\bibitem [{\citenamefont {Peter}\ \emph {et~al.}(2005)\citenamefont {Peter},
  \citenamefont {Senellart}, \citenamefont {Martrou}, \citenamefont
  {Lema\^{\i}tre}, \citenamefont {Hours}, \citenamefont {G\'erard},\ and\
  \citenamefont {Bloch}}]{PhysRevLett.95.067401}%
  \BibitemOpen
  \bibfield  {author} {\bibinfo {author} {\bibfnamefont {E.}~\bibnamefont
  {Peter}}, \bibinfo {author} {\bibfnamefont {P.}~\bibnamefont {Senellart}},
  \bibinfo {author} {\bibfnamefont {D.}~\bibnamefont {Martrou}}, \bibinfo
  {author} {\bibfnamefont {A.}~\bibnamefont {Lema\^{\i}tre}}, \bibinfo {author}
  {\bibfnamefont {J.}~\bibnamefont {Hours}}, \bibinfo {author} {\bibfnamefont
  {J.~M.}\ \bibnamefont {G\'erard}},\ and\ \bibinfo {author} {\bibfnamefont
  {J.}~\bibnamefont {Bloch}},\ }\bibfield  {title} {\bibinfo {title}
  {Exciton-photon strong-coupling regime for a single quantum dot embedded in a
  microcavity},\ }\href {https://doi.org/10.1103/PhysRevLett.95.067401}
  {\bibfield  {journal} {\bibinfo  {journal} {Phys. Rev. Lett.}\ }\textbf
  {\bibinfo {volume} {95}},\ \bibinfo {pages} {067401} (\bibinfo {year}
  {2005})}\BibitemShut {NoStop}%
\bibitem [{\citenamefont {De~Liberato}(2017)}]{de_liberato_virtual_2017}%
  \BibitemOpen
  \bibfield  {author} {\bibinfo {author} {\bibfnamefont {S.}~\bibnamefont
  {De~Liberato}},\ }\bibfield  {title} {{\selectlanguage {en}\bibinfo {title}
  {Virtual photons in the ground state of a dissipative system}},\ }\href
  {https://doi.org/10.1038/s41467-017-01504-5} {\bibfield  {journal} {\bibinfo
  {journal} {Nat Commun}\ }\textbf {\bibinfo {volume} {8}},\ \bibinfo {pages}
  {1465} (\bibinfo {year} {2017})}\BibitemShut {NoStop}%
\bibitem [{\citenamefont {Di~Stefano}\ \emph {et~al.}(2019)\citenamefont
  {Di~Stefano}, \citenamefont {Settineri}, \citenamefont {Macr{\`i}},
  \citenamefont {Garziano}, \citenamefont {Stassi}, \citenamefont {Savasta},\
  and\ \citenamefont {Nori}}]{di_stefano_resolution_2019}%
  \BibitemOpen
  \bibfield  {author} {\bibinfo {author} {\bibfnamefont {O.}~\bibnamefont
  {Di~Stefano}}, \bibinfo {author} {\bibfnamefont {A.}~\bibnamefont
  {Settineri}}, \bibinfo {author} {\bibfnamefont {V.}~\bibnamefont
  {Macr{\`i}}}, \bibinfo {author} {\bibfnamefont {L.}~\bibnamefont {Garziano}},
  \bibinfo {author} {\bibfnamefont {R.}~\bibnamefont {Stassi}}, \bibinfo
  {author} {\bibfnamefont {S.}~\bibnamefont {Savasta}},\ and\ \bibinfo {author}
  {\bibfnamefont {F.}~\bibnamefont {Nori}},\ }\bibfield  {title} {\bibinfo
  {title} {Resolution of gauge ambiguities in ultrastrong-coupling cavity
  quantum electrodynamics},\ }\href {https://doi.org/10.1038/s41567-019-0534-4}
  {\bibfield  {journal} {\bibinfo  {journal} {Nature Physics}\ }\textbf
  {\bibinfo {volume} {15}},\ \bibinfo {pages} {803} (\bibinfo {year}
  {2019})}\BibitemShut {NoStop}%
\bibitem [{\citenamefont {Taylor}\ \emph {et~al.}(2020)\citenamefont {Taylor},
  \citenamefont {Mandal}, \citenamefont {Zhou},\ and\ \citenamefont
  {Huo}}]{PhysRevLett.125.123602}%
  \BibitemOpen
  \bibfield  {author} {\bibinfo {author} {\bibfnamefont {M.~A.~D.}\
  \bibnamefont {Taylor}}, \bibinfo {author} {\bibfnamefont {A.}~\bibnamefont
  {Mandal}}, \bibinfo {author} {\bibfnamefont {W.}~\bibnamefont {Zhou}},\ and\
  \bibinfo {author} {\bibfnamefont {P.}~\bibnamefont {Huo}},\ }\bibfield
  {title} {\bibinfo {title} {Resolution of gauge ambiguities in molecular
  cavity quantum electrodynamics},\ }\href
  {https://doi.org/10.1103/PhysRevLett.125.123602} {\bibfield  {journal}
  {\bibinfo  {journal} {Phys. Rev. Lett.}\ }\textbf {\bibinfo {volume} {125}},\
  \bibinfo {pages} {123602} (\bibinfo {year} {2020})}\BibitemShut {NoStop}%
\bibitem [{\citenamefont {Gustin}\ \emph {et~al.}(2023)\citenamefont {Gustin},
  \citenamefont {Franke},\ and\ \citenamefont
  {Hughes}}]{gustin_gauge-invariant_2022}%
  \BibitemOpen
  \bibfield  {author} {\bibinfo {author} {\bibfnamefont {C.}~\bibnamefont
  {Gustin}}, \bibinfo {author} {\bibfnamefont {S.}~\bibnamefont {Franke}},\
  and\ \bibinfo {author} {\bibfnamefont {S.}~\bibnamefont {Hughes}},\
  }\bibfield  {title} {\bibinfo {title} {Gauge-invariant theory of truncated
  quantum light-matter interactions in arbitrary media},\ }\href
  {https://doi.org/10.1103/PhysRevA.107.013722} {\bibfield  {journal} {\bibinfo
   {journal} {Phys. Rev. A}\ }\textbf {\bibinfo {volume} {107}},\ \bibinfo
  {pages} {013722} (\bibinfo {year} {2023})}\BibitemShut {NoStop}%
\bibitem [{\citenamefont {Salmon}\ \emph {et~al.}(2022)\citenamefont {Salmon},
  \citenamefont {Gustin}, \citenamefont {Settineri}, \citenamefont {Stefano},
  \citenamefont {Zueco}, \citenamefont {Savasta}, \citenamefont {Nori},\ and\
  \citenamefont {Hughes}}]{salmon2021gauge}%
  \BibitemOpen
  \bibfield  {author} {\bibinfo {author} {\bibfnamefont {W.}~\bibnamefont
  {Salmon}}, \bibinfo {author} {\bibfnamefont {C.}~\bibnamefont {Gustin}},
  \bibinfo {author} {\bibfnamefont {A.}~\bibnamefont {Settineri}}, \bibinfo
  {author} {\bibfnamefont {O.~D.}\ \bibnamefont {Stefano}}, \bibinfo {author}
  {\bibfnamefont {D.}~\bibnamefont {Zueco}}, \bibinfo {author} {\bibfnamefont
  {S.}~\bibnamefont {Savasta}}, \bibinfo {author} {\bibfnamefont
  {F.}~\bibnamefont {Nori}},\ and\ \bibinfo {author} {\bibfnamefont
  {S.}~\bibnamefont {Hughes}},\ }\bibfield  {title} {\bibinfo {title}
  {Gauge-independent emission spectra and quantum correlations in the
  ultrastrong coupling regime of open system cavity-qed},\ }\href
  {https://doi.org/doi:10.1515/nanoph-2021-0718} {\bibfield  {journal}
  {\bibinfo  {journal} {Nanophotonics}\ }\textbf {\bibinfo {volume} {11}},\
  \bibinfo {pages} {1573} (\bibinfo {year} {2022})}\BibitemShut {NoStop}%
\bibitem [{\citenamefont {Mercurio}\ \emph {et~al.}(2022)\citenamefont
  {Mercurio}, \citenamefont {Macr\`{\i}}, \citenamefont {Gustin}, \citenamefont
  {Hughes}, \citenamefont {Savasta},\ and\ \citenamefont
  {Nori}}]{PhysRevResearch.4.023048}%
  \BibitemOpen
  \bibfield  {author} {\bibinfo {author} {\bibfnamefont {A.}~\bibnamefont
  {Mercurio}}, \bibinfo {author} {\bibfnamefont {V.}~\bibnamefont
  {Macr\`{\i}}}, \bibinfo {author} {\bibfnamefont {C.}~\bibnamefont {Gustin}},
  \bibinfo {author} {\bibfnamefont {S.}~\bibnamefont {Hughes}}, \bibinfo
  {author} {\bibfnamefont {S.}~\bibnamefont {Savasta}},\ and\ \bibinfo {author}
  {\bibfnamefont {F.}~\bibnamefont {Nori}},\ }\bibfield  {title} {\bibinfo
  {title} {Regimes of cavity {QED} under incoherent excitation: From weak to
  deep strong coupling},\ }\href
  {https://doi.org/10.1103/PhysRevResearch.4.023048} {\bibfield  {journal}
  {\bibinfo  {journal} {Phys. Rev. Research}\ }\textbf {\bibinfo {volume}
  {4}},\ \bibinfo {pages} {023048} (\bibinfo {year} {2022})}\BibitemShut
  {NoStop}%
\bibitem [{\citenamefont {Settineri}\ \emph {et~al.}(2021)\citenamefont
  {Settineri}, \citenamefont {Di~Stefano}, \citenamefont {Zueco}, \citenamefont
  {Hughes}, \citenamefont {Savasta},\ and\ \citenamefont
  {Nori}}]{Settineri2021Apr}%
  \BibitemOpen
  \bibfield  {author} {\bibinfo {author} {\bibfnamefont {A.}~\bibnamefont
  {Settineri}}, \bibinfo {author} {\bibfnamefont {O.}~\bibnamefont
  {Di~Stefano}}, \bibinfo {author} {\bibfnamefont {D.}~\bibnamefont {Zueco}},
  \bibinfo {author} {\bibfnamefont {S.}~\bibnamefont {Hughes}}, \bibinfo
  {author} {\bibfnamefont {S.}~\bibnamefont {Savasta}},\ and\ \bibinfo {author}
  {\bibfnamefont {F.}~\bibnamefont {Nori}},\ }\bibfield  {title} {\bibinfo
  {title} {Gauge freedom, quantum measurements, and time-dependent interactions
  in cavity {QED}},\ }\href {https://doi.org/10.1103/PhysRevResearch.3.023079}
  {\bibfield  {journal} {\bibinfo  {journal} {Phys. Rev. Research}\ }\textbf
  {\bibinfo {volume} {3}},\ \bibinfo {pages} {023079} (\bibinfo {year}
  {2021})}\BibitemShut {NoStop}%
\bibitem [{\citenamefont {Jaynes}\ and\ \citenamefont
  {Cummings}(1963)}]{jaynes_comparison_1963}%
  \BibitemOpen
  \bibfield  {author} {\bibinfo {author} {\bibfnamefont {E.}~\bibnamefont
  {Jaynes}}\ and\ \bibinfo {author} {\bibfnamefont {F.}~\bibnamefont
  {Cummings}},\ }\bibfield  {title} {\bibinfo {title} {Comparison of quantum
  and semiclassical radiation theories with application to the beam maser},\
  }\href {https://doi.org/10.1109/PROC.1963.1664} {\bibfield  {journal}
  {\bibinfo  {journal} {Proceedings of the IEEE}\ }\textbf {\bibinfo {volume}
  {51}},\ \bibinfo {pages} {89} (\bibinfo {year} {1963})}\BibitemShut {NoStop}%
\bibitem [{\citenamefont {Cummings}(2013)}]{cummings_reminiscing_2013}%
  \BibitemOpen
  \bibfield  {author} {\bibinfo {author} {\bibfnamefont {F.~W.}\ \bibnamefont
  {Cummings}},\ }\bibfield  {title} {{\selectlanguage {en}\bibinfo {title}
  {Reminiscing about thesis work with {E} {T} {Jaynes} at {Stanford} in the
  1950s}},\ }\href {https://doi.org/10.1088/0953-4075/46/22/220202} {\bibfield
  {journal} {\bibinfo  {journal} {J. Phys. B: At. Mol. Opt. Phys.}\ }\textbf
  {\bibinfo {volume} {46}},\ \bibinfo {pages} {220202} (\bibinfo {year}
  {2013})}\BibitemShut {NoStop}%
\bibitem [{\citenamefont {Braak}(2011)}]{braak_integrability_2011}%
  \BibitemOpen
  \bibfield  {author} {\bibinfo {author} {\bibfnamefont {D.}~\bibnamefont
  {Braak}},\ }\bibfield  {title} {\bibinfo {title} {Integrability of the {Rabi}
  model},\ }\href {https://doi.org/10.1103/PhysRevLett.107.100401} {\bibfield
  {journal} {\bibinfo  {journal} {Phys. Rev. Lett.}\ }\textbf {\bibinfo
  {volume} {107}},\ \bibinfo {pages} {100401} (\bibinfo {year}
  {2011})}\BibitemShut {NoStop}%
\bibitem [{\citenamefont {Stokes}\ and\ \citenamefont
  {Nazir}(2019)}]{adam_stokes_gauge_2019}%
  \BibitemOpen
  \bibfield  {author} {\bibinfo {author} {\bibfnamefont {A.}~\bibnamefont
  {Stokes}}\ and\ \bibinfo {author} {\bibfnamefont {A.}~\bibnamefont {Nazir}},\
  }\bibfield  {title} {\bibinfo {title} {Gauge ambiguities imply
  {Jaynes-Cummings} physics remains valid in ultrastrong coupling {QED}},\
  }\href {https://doi.org/https://doi.org/10.1038/s41467-018-08101-0}
  {\bibfield  {journal} {\bibinfo  {journal} {Nature communications}\ }\textbf
  {\bibinfo {volume} {10}},\ \bibinfo {pages} {499} (\bibinfo {year}
  {2019})}\BibitemShut {NoStop}%
\bibitem [{\citenamefont {Savasta}\ \emph {et~al.}(2021)\citenamefont
  {Savasta}, \citenamefont {Stefano},\ and\ \citenamefont
  {Nori}}]{savasta_thomasreichekuhn_2021}%
  \BibitemOpen
  \bibfield  {author} {\bibinfo {author} {\bibfnamefont {S.}~\bibnamefont
  {Savasta}}, \bibinfo {author} {\bibfnamefont {O.~D.}\ \bibnamefont
  {Stefano}},\ and\ \bibinfo {author} {\bibfnamefont {F.}~\bibnamefont
  {Nori}},\ }\bibfield  {title} {\bibinfo {title}
  {Thomas{\textendash}{Reiche}{\textendash}{Kuhn} ({TRK}) sum rule for
  interacting photons},\ }\href {https://doi.org/doi:10.1515/nanoph-2020-0433}
  {\bibfield  {journal} {\bibinfo  {journal} {Nanophotonics}\ }\textbf
  {\bibinfo {volume} {10}},\ \bibinfo {pages} {465} (\bibinfo {year}
  {2021})}\BibitemShut {NoStop}%
\bibitem [{\citenamefont {Starace}(1971)}]{starace_length_1971}%
  \BibitemOpen
  \bibfield  {author} {\bibinfo {author} {\bibfnamefont {A.~F.}\ \bibnamefont
  {Starace}},\ }\bibfield  {title} {\bibinfo {title} {Length and {Velocity}
  {Formulas} in {Approximate} {Oscillator}-{Strength} {Calculations}},\ }\href
  {https://doi.org/10.1103/PhysRevA.3.1242} {\bibfield  {journal} {\bibinfo
  {journal} {Phys. Rev. A}\ }\textbf {\bibinfo {volume} {3}},\ \bibinfo {pages}
  {1242} (\bibinfo {year} {1971})}\BibitemShut {NoStop}%
\bibitem [{\citenamefont {Carmichael}(2013)}]{carmichael_statistical_2013}%
  \BibitemOpen
  \bibfield  {author} {\bibinfo {author} {\bibfnamefont {H.~J.}\ \bibnamefont
  {Carmichael}},\ }\href@noop {} {\emph {\bibinfo {title} {Statistical
  {Methods} in {Quantum} {Optics} 1: {Master} {Equations} and {Fokker}-{Planck}
  {Equations}}}}\ (\bibinfo  {publisher} {Springer Science \& Business Media},\
  \bibinfo {year} {2013})\BibitemShut {NoStop}%
\bibitem [{\citenamefont {Carmichael}(1999)}]{carmichael_dissipation_1999}%
  \BibitemOpen
  \bibfield  {author} {\bibinfo {author} {\bibfnamefont {H.~J.}\ \bibnamefont
  {Carmichael}},\ }\bibfield  {title} {{\selectlanguage {en}\bibinfo {title}
  {Dissipation in {Quantum} {Mechanics}: {The} {Master} {Equation}
  {Approach}}},\ }in\ \href@noop {} {{\selectlanguage {en}\emph {\bibinfo
  {booktitle} {Statistical {Methods} in {Quantum} {Optics} 1: {Master}
  {Equations} and {Fokker}-{Planck} {Equations}}}}},\ \bibinfo {series and
  number} {Texts and {Monographs} in {Physics}},\ \bibinfo {editor} {edited by\
  \bibinfo {editor} {\bibfnamefont {H.~J.}\ \bibnamefont {Carmichael}}}\
  (\bibinfo  {publisher} {Springer},\ \bibinfo {address} {Berlin, Heidelberg},\
  \bibinfo {year} {1999})\ pp.\ \bibinfo {pages} {1--28}\BibitemShut {NoStop}%
\bibitem [{\citenamefont {Settineri}\ \emph {et~al.}(2018)\citenamefont
  {Settineri}, \citenamefont {Macr{\'i}}, \citenamefont {Ridolfo},
  \citenamefont {Di~Stefano}, \citenamefont {Kockum}, \citenamefont {Nori},\
  and\ \citenamefont {Savasta}}]{settineri_dissipation_2018}%
  \BibitemOpen
  \bibfield  {author} {\bibinfo {author} {\bibfnamefont {A.}~\bibnamefont
  {Settineri}}, \bibinfo {author} {\bibfnamefont {V.}~\bibnamefont
  {Macr{\'i}}}, \bibinfo {author} {\bibfnamefont {A.}~\bibnamefont {Ridolfo}},
  \bibinfo {author} {\bibfnamefont {O.}~\bibnamefont {Di~Stefano}}, \bibinfo
  {author} {\bibfnamefont {A.~F.}\ \bibnamefont {Kockum}}, \bibinfo {author}
  {\bibfnamefont {F.}~\bibnamefont {Nori}},\ and\ \bibinfo {author}
  {\bibfnamefont {S.}~\bibnamefont {Savasta}},\ }\bibfield  {title} {\bibinfo
  {title} {Dissipation and thermal noise in hybrid quantum systems in the
  ultrastrong-coupling regime},\ }\href
  {https://doi.org/10.1103/PhysRevA.98.053834} {\bibfield  {journal} {\bibinfo
  {journal} {Physical Review A}\ }\textbf {\bibinfo {volume} {98}},\ \bibinfo
  {pages} {053834} (\bibinfo {year} {2018})}\BibitemShut {NoStop}%
\bibitem [{\citenamefont {Ciuti}\ and\ \citenamefont
  {Carusotto}(2006)}]{ciuti_input-output_2006}%
  \BibitemOpen
  \bibfield  {author} {\bibinfo {author} {\bibfnamefont {C.}~\bibnamefont
  {Ciuti}}\ and\ \bibinfo {author} {\bibfnamefont {I.}~\bibnamefont
  {Carusotto}},\ }\bibfield  {title} {\bibinfo {title} {Input-output theory of
  cavities in the ultrastrong coupling regime: {The} case of time-independent
  cavity parameters},\ }\href {https://doi.org/10.1103/PhysRevA.74.033811}
  {\bibfield  {journal} {\bibinfo  {journal} {Phys. Rev. A}\ }\textbf {\bibinfo
  {volume} {74}},\ \bibinfo {pages} {033811} (\bibinfo {year}
  {2006})}\BibitemShut {NoStop}%
\bibitem [{\citenamefont {Beaudoin}\ \emph {et~al.}(2011)\citenamefont
  {Beaudoin}, \citenamefont {Gambetta},\ and\ \citenamefont
  {Blais}}]{beaudoin_dissipation_2011}%
  \BibitemOpen
  \bibfield  {author} {\bibinfo {author} {\bibfnamefont {F.}~\bibnamefont
  {Beaudoin}}, \bibinfo {author} {\bibfnamefont {J.~M.}\ \bibnamefont
  {Gambetta}},\ and\ \bibinfo {author} {\bibfnamefont {A.}~\bibnamefont
  {Blais}},\ }\bibfield  {title} {\bibinfo {title} {Dissipation and ultrastrong
  coupling in circuit {QED}},\ }\href
  {https://doi.org/10.1103/PhysRevA.84.043832} {\bibfield  {journal} {\bibinfo
  {journal} {Physical Review A}\ }\textbf {\bibinfo {volume} {84}},\ \bibinfo
  {pages} {043832} (\bibinfo {year} {2011})}\BibitemShut {NoStop}%
\bibitem [{\citenamefont
  {Le~Boit{\ifmmode\acute{e}\else\'{e}\fi}}(2020)}]{LeBoite2020Jul}%
  \BibitemOpen
  \bibfield  {author} {\bibinfo {author} {\bibfnamefont {A.}~\bibnamefont
  {Le~Boit{\ifmmode\acute{e}\else\'{e}\fi}}},\ }\bibfield  {title} {\bibinfo
  {title} {Theoretical methods for ultrastrong light{\textendash}matter
  interactions},\ }\href
  {https://doi.org/https://doi.org/10.1002/qute.201900140} {\bibfield
  {journal} {\bibinfo  {journal} {Adv. Quantum Technol.}\ }\textbf {\bibinfo
  {volume} {3}},\ \bibinfo {pages} {1900140} (\bibinfo {year}
  {2020})}\BibitemShut {NoStop}%
\bibitem [{\citenamefont {Hu}\ \emph {et~al.}(2015)\citenamefont {Hu},
  \citenamefont {Huang}, \citenamefont {Liao}, \citenamefont {Tian},\ and\
  \citenamefont {Goan}}]{PhysRevA.91.013812}%
  \BibitemOpen
  \bibfield  {author} {\bibinfo {author} {\bibfnamefont {D.}~\bibnamefont
  {Hu}}, \bibinfo {author} {\bibfnamefont {S.-Y.}\ \bibnamefont {Huang}},
  \bibinfo {author} {\bibfnamefont {J.-Q.}\ \bibnamefont {Liao}}, \bibinfo
  {author} {\bibfnamefont {L.}~\bibnamefont {Tian}},\ and\ \bibinfo {author}
  {\bibfnamefont {H.-S.}\ \bibnamefont {Goan}},\ }\bibfield  {title} {\bibinfo
  {title} {Quantum coherence in ultrastrong optomechanics},\ }\href
  {https://doi.org/10.1103/PhysRevA.91.013812} {\bibfield  {journal} {\bibinfo
  {journal} {Phys. Rev. A}\ }\textbf {\bibinfo {volume} {91}},\ \bibinfo
  {pages} {013812} (\bibinfo {year} {2015})}\BibitemShut {NoStop}%
\bibitem [{\citenamefont {Hughes}\ \emph {et~al.}(2021)\citenamefont {Hughes},
  \citenamefont {Settineri}, \citenamefont {Savasta},\ and\ \citenamefont
  {Nori}}]{PhysRevB.104.045431}%
  \BibitemOpen
  \bibfield  {author} {\bibinfo {author} {\bibfnamefont {S.}~\bibnamefont
  {Hughes}}, \bibinfo {author} {\bibfnamefont {A.}~\bibnamefont {Settineri}},
  \bibinfo {author} {\bibfnamefont {S.}~\bibnamefont {Savasta}},\ and\ \bibinfo
  {author} {\bibfnamefont {F.}~\bibnamefont {Nori}},\ }\bibfield  {title}
  {\bibinfo {title} {Resonant raman scattering of single molecules under
  simultaneous strong cavity coupling and ultrastrong optomechanical coupling
  in plasmonic resonators: Phonon-dressed polaritons},\ }\href
  {https://doi.org/10.1103/PhysRevB.104.045431} {\bibfield  {journal} {\bibinfo
   {journal} {Phys. Rev. B}\ }\textbf {\bibinfo {volume} {104}},\ \bibinfo
  {pages} {045431} (\bibinfo {year} {2021})}\BibitemShut {NoStop}%
\bibitem [{\citenamefont {Wilson-Rae}\ and\ \citenamefont
  {Imamo\ifmmode~\breve{g}\else \u{g}\fi{}lu}(2002)}]{PhysRevB.65.235311}%
  \BibitemOpen
  \bibfield  {author} {\bibinfo {author} {\bibfnamefont {I.}~\bibnamefont
  {Wilson-Rae}}\ and\ \bibinfo {author} {\bibfnamefont {A.}~\bibnamefont
  {Imamo\ifmmode~\breve{g}\else \u{g}\fi{}lu}},\ }\bibfield  {title} {\bibinfo
  {title} {Quantum dot cavity-qed in the presence of strong electron-phonon
  interactions},\ }\href {https://doi.org/10.1103/PhysRevB.65.235311}
  {\bibfield  {journal} {\bibinfo  {journal} {Phys. Rev. B}\ }\textbf {\bibinfo
  {volume} {65}},\ \bibinfo {pages} {235311} (\bibinfo {year}
  {2002})}\BibitemShut {NoStop}%
\bibitem [{\citenamefont {Roy-Choudhury}\ and\ \citenamefont
  {Hughes}(2015)}]{PhysRevB.92.205406}%
  \BibitemOpen
  \bibfield  {author} {\bibinfo {author} {\bibfnamefont {K.}~\bibnamefont
  {Roy-Choudhury}}\ and\ \bibinfo {author} {\bibfnamefont {S.}~\bibnamefont
  {Hughes}},\ }\bibfield  {title} {\bibinfo {title} {Quantum theory of the
  emission spectrum from quantum dots coupled to structured photonic reservoirs
  and acoustic phonons},\ }\href {https://doi.org/10.1103/PhysRevB.92.205406}
  {\bibfield  {journal} {\bibinfo  {journal} {Phys. Rev. B}\ }\textbf {\bibinfo
  {volume} {92}},\ \bibinfo {pages} {205406} (\bibinfo {year}
  {2015})}\BibitemShut {NoStop}%
\bibitem [{\citenamefont {Neuman}\ and\ \citenamefont
  {Aizpurua}(2018)}]{Neuman2018}%
  \BibitemOpen
  \bibfield  {author} {\bibinfo {author} {\bibfnamefont {T.}~\bibnamefont
  {Neuman}}\ and\ \bibinfo {author} {\bibfnamefont {J.}~\bibnamefont
  {Aizpurua}},\ }\bibfield  {title} {\bibinfo {title} {Origin of the asymmetric
  light emission from molecular exciton{\textendash}polaritons},\ }\href
  {https://doi.org/10.1364/optica.5.001247} {\bibfield  {journal} {\bibinfo
  {journal} {Optica}\ }\textbf {\bibinfo {volume} {5}},\ \bibinfo {pages}
  {1247} (\bibinfo {year} {2018})}\BibitemShut {NoStop}%
\bibitem [{\citenamefont {Salmon}(2021)}]{salmon_master_2021}%
  \BibitemOpen
  \bibfield  {author} {\bibinfo {author} {\bibfnamefont {W.}~\bibnamefont
  {Salmon}},\ }\href@noop {} {\bibinfo {title} {Master equations for computing
  gauge-invariant observables in the ultrastrong coupling regime of
  cavity-{QED}}} (\bibinfo {year} {2021}),\ \bibinfo {note} {{M.Sc. Thesis},
  {Queen's University}, {Canada}}\BibitemShut {NoStop}%
\bibitem [{\citenamefont {Lax}(1963)}]{lax_formal_1963}%
  \BibitemOpen
  \bibfield  {author} {\bibinfo {author} {\bibfnamefont {M.}~\bibnamefont
  {Lax}},\ }\bibfield  {title} {\bibinfo {title} {Formal {Theory} of {Quantum}
  {Fluctuations} from a {Driven} {State}},\ }\href
  {https://doi.org/10.1103/PhysRev.129.2342} {\bibfield  {journal} {\bibinfo
  {journal} {Phys. Rev.}\ }\textbf {\bibinfo {volume} {129}},\ \bibinfo {pages}
  {2342} (\bibinfo {year} {1963})}\BibitemShut {NoStop}%
\bibitem [{\citenamefont {Dicke}(1954)}]{dicke_coherence_1954}%
  \BibitemOpen
  \bibfield  {author} {\bibinfo {author} {\bibfnamefont {R.~H.}\ \bibnamefont
  {Dicke}},\ }\bibfield  {title} {\bibinfo {title} {Coherence in {Spontaneous}
  {Radiation} {Processes}},\ }\href {https://doi.org/10.1103/PhysRev.93.99}
  {\bibfield  {journal} {\bibinfo  {journal} {Phys. Rev.}\ }\textbf {\bibinfo
  {volume} {93}},\ \bibinfo {pages} {99} (\bibinfo {year} {1954})}\BibitemShut
  {NoStop}%
\bibitem [{\citenamefont {Hepp}\ and\ \citenamefont
  {Lieb}(1973)}]{hepp_superradiant_1973}%
  \BibitemOpen
  \bibfield  {author} {\bibinfo {author} {\bibfnamefont {K.}~\bibnamefont
  {Hepp}}\ and\ \bibinfo {author} {\bibfnamefont {E.~H.}\ \bibnamefont
  {Lieb}},\ }\bibfield  {title} {{\selectlanguage {en}\bibinfo {title} {On the
  superradiant phase transition for molecules in a quantized radiation field:
  the dicke maser model}},\ }\href
  {https://doi.org/10.1016/0003-4916(73)90039-0} {\bibfield  {journal}
  {\bibinfo  {journal} {Annals of Physics}\ }\textbf {\bibinfo {volume} {76}},\
  \bibinfo {pages} {360} (\bibinfo {year} {1973})}\BibitemShut {NoStop}%
\bibitem [{\citenamefont {Wang}\ and\ \citenamefont
  {Hioe}(1973)}]{wang_phase_1973}%
  \BibitemOpen
  \bibfield  {author} {\bibinfo {author} {\bibfnamefont {Y.~K.}\ \bibnamefont
  {Wang}}\ and\ \bibinfo {author} {\bibfnamefont {F.~T.}\ \bibnamefont
  {Hioe}},\ }\bibfield  {title} {\bibinfo {title} {Phase {Transition} in the
  {Dicke} {Model} of {Superradiance}},\ }\href
  {https://doi.org/10.1103/PhysRevA.7.831} {\bibfield  {journal} {\bibinfo
  {journal} {Phys. Rev. A}\ }\textbf {\bibinfo {volume} {7}},\ \bibinfo {pages}
  {831} (\bibinfo {year} {1973})}\BibitemShut {NoStop}%
\bibitem [{\citenamefont {Emary}\ and\ \citenamefont
  {Brandes}(2003)}]{emary_quantum_2003}%
  \BibitemOpen
  \bibfield  {author} {\bibinfo {author} {\bibfnamefont {C.}~\bibnamefont
  {Emary}}\ and\ \bibinfo {author} {\bibfnamefont {T.}~\bibnamefont
  {Brandes}},\ }\bibfield  {title} {\bibinfo {title} {Quantum {Chaos}
  {Triggered} by {Precursors} of a {Quantum} {Phase} {Transition}: {The}
  {Dicke} {Model}},\ }\href {https://doi.org/10.1103/PhysRevLett.90.044101}
  {\bibfield  {journal} {\bibinfo  {journal} {Phys. Rev. Lett.}\ }\textbf
  {\bibinfo {volume} {90}},\ \bibinfo {pages} {044101} (\bibinfo {year}
  {2003})}\BibitemShut {NoStop}%
\bibitem [{\citenamefont {Baumann}\ \emph {et~al.}(2010)\citenamefont
  {Baumann}, \citenamefont {Guerlin}, \citenamefont {Brennecke},\ and\
  \citenamefont {Esslinger}}]{baumann_dicke_2010}%
  \BibitemOpen
  \bibfield  {author} {\bibinfo {author} {\bibfnamefont {K.}~\bibnamefont
  {Baumann}}, \bibinfo {author} {\bibfnamefont {C.}~\bibnamefont {Guerlin}},
  \bibinfo {author} {\bibfnamefont {F.}~\bibnamefont {Brennecke}},\ and\
  \bibinfo {author} {\bibfnamefont {T.}~\bibnamefont {Esslinger}},\ }\bibfield
  {title} {\bibinfo {title} {Dicke quantum phase transition with a superfluid
  gas in an optical cavity},\ }\href {https://doi.org/10.1038/nature09009}
  {\bibfield  {journal} {\bibinfo  {journal} {Nature}\ }\textbf {\bibinfo
  {volume} {464}},\ \bibinfo {pages} {1301} (\bibinfo {year}
  {2010})}\BibitemShut {NoStop}%
\bibitem [{\citenamefont {Garraway}(2011)}]{garraway_dicke_2011}%
  \BibitemOpen
  \bibfield  {author} {\bibinfo {author} {\bibfnamefont {B.~M.}\ \bibnamefont
  {Garraway}},\ }\bibfield  {title} {\bibinfo {title} {The {Dicke} model in
  quantum optics: {Dicke} model revisited},\ }\href
  {https://doi.org/10.1098/rsta.2010.0333} {\bibfield  {journal} {\bibinfo
  {journal} {Philosophical Transactions of the Royal Society A: Mathematical,
  Physical and Engineering Sciences}\ }\textbf {\bibinfo {volume} {369}},\
  \bibinfo {pages} {1137} (\bibinfo {year} {2011})}\BibitemShut {NoStop}%
\bibitem [{\citenamefont {Baden}\ \emph {et~al.}(2014)\citenamefont {Baden},
  \citenamefont {Arnold}, \citenamefont {Grimsmo}, \citenamefont {Parkins},\
  and\ \citenamefont {Barrett}}]{baden_realization_2014}%
  \BibitemOpen
  \bibfield  {author} {\bibinfo {author} {\bibfnamefont {M.~P.}\ \bibnamefont
  {Baden}}, \bibinfo {author} {\bibfnamefont {K.~J.}\ \bibnamefont {Arnold}},
  \bibinfo {author} {\bibfnamefont {A.~L.}\ \bibnamefont {Grimsmo}}, \bibinfo
  {author} {\bibfnamefont {S.}~\bibnamefont {Parkins}},\ and\ \bibinfo {author}
  {\bibfnamefont {M.~D.}\ \bibnamefont {Barrett}},\ }\bibfield  {title}
  {\bibinfo {title} {Realization of the {Dicke} {Model} {Using}
  {Cavity}-{Assisted} {Raman} {Transitions}},\ }\href
  {https://doi.org/10.1103/PhysRevLett.113.020408} {\bibfield  {journal}
  {\bibinfo  {journal} {Phys. Rev. Lett.}\ }\textbf {\bibinfo {volume} {113}},\
  \bibinfo {pages} {020408} (\bibinfo {year} {2014})}\BibitemShut {NoStop}%
\bibitem [{\citenamefont {Bastarrachea-Magnani}\ \emph
  {et~al.}(2015)\citenamefont {Bastarrachea-Magnani}, \citenamefont {López-del
  Carpio}, \citenamefont {Lerma-Hernández},\ and\ \citenamefont
  {Hirsch}}]{bastarrachea-magnani_chaos_2015}%
  \BibitemOpen
  \bibfield  {author} {\bibinfo {author} {\bibfnamefont {M.~A.}\ \bibnamefont
  {Bastarrachea-Magnani}}, \bibinfo {author} {\bibfnamefont {B.}~\bibnamefont
  {López-del Carpio}}, \bibinfo {author} {\bibfnamefont {S.}~\bibnamefont
  {Lerma-Hernández}},\ and\ \bibinfo {author} {\bibfnamefont {J.~G.}\
  \bibnamefont {Hirsch}},\ }\bibfield  {title} {{\selectlanguage {en}\bibinfo
  {title} {Chaos in the {Dicke} model: quantum and semiclassical analysis}},\
  }\href {https://doi.org/10.1088/0031-8949/90/6/068015} {\bibfield  {journal}
  {\bibinfo  {journal} {Phys. Scr.}\ }\textbf {\bibinfo {volume} {90}},\
  \bibinfo {pages} {068015} (\bibinfo {year} {2015})}\BibitemShut {NoStop}%
\bibitem [{\citenamefont {Klinder}\ \emph {et~al.}(2015)\citenamefont
  {Klinder}, \citenamefont {Keßler}, \citenamefont {Wolke}, \citenamefont
  {Mathey},\ and\ \citenamefont {Hemmerich}}]{klinder_dynamical_2015}%
  \BibitemOpen
  \bibfield  {author} {\bibinfo {author} {\bibfnamefont {J.}~\bibnamefont
  {Klinder}}, \bibinfo {author} {\bibfnamefont {H.}~\bibnamefont {Keßler}},
  \bibinfo {author} {\bibfnamefont {M.}~\bibnamefont {Wolke}}, \bibinfo
  {author} {\bibfnamefont {L.}~\bibnamefont {Mathey}},\ and\ \bibinfo {author}
  {\bibfnamefont {A.}~\bibnamefont {Hemmerich}},\ }\bibfield  {title}
  {{\selectlanguage {EN}\bibinfo {title} {Dynamical phase transition in the
  open {Dicke} model}},\ }\href {https://doi.org/10.1073/pnas.1417132112}
  {\bibfield  {journal} {\bibinfo  {journal} {Proceedings of the National
  Academy of Sciences}\ }\textbf {\bibinfo {volume} {112}},\ \bibinfo {pages}
  {3290} (\bibinfo {year} {2015})}\BibitemShut {NoStop}%
\bibitem [{\citenamefont {Larson}\ and\ \citenamefont
  {Irish}(2017)}]{larson_remarks_2017}%
  \BibitemOpen
  \bibfield  {author} {\bibinfo {author} {\bibfnamefont {J.}~\bibnamefont
  {Larson}}\ and\ \bibinfo {author} {\bibfnamefont {E.~K.}\ \bibnamefont
  {Irish}},\ }\bibfield  {title} {{\selectlanguage {en}\bibinfo {title} {Some
  remarks on ‘superradiant’ phase transitions in light-matter systems}},\
  }\href {https://doi.org/10.1088/1751-8121/aa65dc} {\bibfield  {journal}
  {\bibinfo  {journal} {J. Phys. A: Math. Theor.}\ }\textbf {\bibinfo {volume}
  {50}},\ \bibinfo {pages} {174002} (\bibinfo {year} {2017})}\BibitemShut
  {NoStop}%
\bibitem [{\citenamefont {Kirton}\ \emph {et~al.}(2019)\citenamefont {Kirton},
  \citenamefont {Roses}, \citenamefont {Keeling},\ and\ \citenamefont
  {Dalla~Torre}}]{kirton_introduction_2019}%
  \BibitemOpen
  \bibfield  {author} {\bibinfo {author} {\bibfnamefont {P.}~\bibnamefont
  {Kirton}}, \bibinfo {author} {\bibfnamefont {M.~M.}\ \bibnamefont {Roses}},
  \bibinfo {author} {\bibfnamefont {J.}~\bibnamefont {Keeling}},\ and\ \bibinfo
  {author} {\bibfnamefont {E.~G.}\ \bibnamefont {Dalla~Torre}},\ }\bibfield
  {title} {{\selectlanguage {en}\bibinfo {title} {Introduction to the {Dicke}
  {Model}: {From} {Equilibrium} to {Nonequilibrium}, and {Vice} {Versa}}},\
  }\href {https://doi.org/10.1002/qute.201800043} {\bibfield  {journal}
  {\bibinfo  {journal} {Advanced Quantum Technologies}\ }\textbf {\bibinfo
  {volume} {2}},\ \bibinfo {pages} {1800043} (\bibinfo {year}
  {2019})}\BibitemShut {NoStop}%
\bibitem [{\citenamefont {Lambert}\ \emph {et~al.}(2016)\citenamefont
  {Lambert}, \citenamefont {Matsuzaki}, \citenamefont {Kakuyanagi},
  \citenamefont {Ishida}, \citenamefont {Saito},\ and\ \citenamefont
  {Nori}}]{lambert_superradiance_2016}%
  \BibitemOpen
  \bibfield  {author} {\bibinfo {author} {\bibfnamefont {N.}~\bibnamefont
  {Lambert}}, \bibinfo {author} {\bibfnamefont {Y.}~\bibnamefont {Matsuzaki}},
  \bibinfo {author} {\bibfnamefont {K.}~\bibnamefont {Kakuyanagi}}, \bibinfo
  {author} {\bibfnamefont {N.}~\bibnamefont {Ishida}}, \bibinfo {author}
  {\bibfnamefont {S.}~\bibnamefont {Saito}},\ and\ \bibinfo {author}
  {\bibfnamefont {F.}~\bibnamefont {Nori}},\ }\bibfield  {title} {\bibinfo
  {title} {Superradiance with an ensemble of superconducting flux qubits},\
  }\href {https://doi.org/10.1103/PhysRevB.94.224510} {\bibfield  {journal}
  {\bibinfo  {journal} {Phys. Rev. B}\ }\textbf {\bibinfo {volume} {94}},\
  \bibinfo {pages} {224510} (\bibinfo {year} {2016})}\BibitemShut {NoStop}%
\bibitem [{\citenamefont {Shammah}\ \emph {et~al.}(2017)\citenamefont
  {Shammah}, \citenamefont {Lambert}, \citenamefont {Nori},\ and\ \citenamefont
  {De~Liberato}}]{shammah_superradiance_2017}%
  \BibitemOpen
  \bibfield  {author} {\bibinfo {author} {\bibfnamefont {N.}~\bibnamefont
  {Shammah}}, \bibinfo {author} {\bibfnamefont {N.}~\bibnamefont {Lambert}},
  \bibinfo {author} {\bibfnamefont {F.}~\bibnamefont {Nori}},\ and\ \bibinfo
  {author} {\bibfnamefont {S.}~\bibnamefont {De~Liberato}},\ }\bibfield
  {title} {\bibinfo {title} {Superradiance with local phase-breaking effects},\
  }\href {https://doi.org/10.1103/PhysRevA.96.023863} {\bibfield  {journal}
  {\bibinfo  {journal} {Phys. Rev. A}\ }\textbf {\bibinfo {volume} {96}},\
  \bibinfo {pages} {023863} (\bibinfo {year} {2017})}\BibitemShut {NoStop}%
\bibitem [{\citenamefont {Shammah}\ \emph {et~al.}(2018)\citenamefont
  {Shammah}, \citenamefont {Ahmed}, \citenamefont {Lambert}, \citenamefont
  {De~Liberato},\ and\ \citenamefont {Nori}}]{shammah_open_2018}%
  \BibitemOpen
  \bibfield  {author} {\bibinfo {author} {\bibfnamefont {N.}~\bibnamefont
  {Shammah}}, \bibinfo {author} {\bibfnamefont {S.}~\bibnamefont {Ahmed}},
  \bibinfo {author} {\bibfnamefont {N.}~\bibnamefont {Lambert}}, \bibinfo
  {author} {\bibfnamefont {S.}~\bibnamefont {De~Liberato}},\ and\ \bibinfo
  {author} {\bibfnamefont {F.}~\bibnamefont {Nori}},\ }\bibfield  {title}
  {\bibinfo {title} {Open quantum systems with local and collective incoherent
  processes: Efficient numerical simulations using permutational invariance},\
  }\href {https://doi.org/10.1103/PhysRevA.98.063815} {\bibfield  {journal}
  {\bibinfo  {journal} {Phys. Rev. A}\ }\textbf {\bibinfo {volume} {98}},\
  \bibinfo {pages} {063815} (\bibinfo {year} {2018})}\BibitemShut {NoStop}%
\bibitem [{\citenamefont {Lolli}\ \emph {et~al.}(2015)\citenamefont {Lolli},
  \citenamefont {Baksic}, \citenamefont {Nagy}, \citenamefont {Manucharyan},\
  and\ \citenamefont {Ciuti}}]{lolli_ancillary_2015}%
  \BibitemOpen
  \bibfield  {author} {\bibinfo {author} {\bibfnamefont {J.}~\bibnamefont
  {Lolli}}, \bibinfo {author} {\bibfnamefont {A.}~\bibnamefont {Baksic}},
  \bibinfo {author} {\bibfnamefont {D.}~\bibnamefont {Nagy}}, \bibinfo {author}
  {\bibfnamefont {V.~E.}\ \bibnamefont {Manucharyan}},\ and\ \bibinfo {author}
  {\bibfnamefont {C.}~\bibnamefont {Ciuti}},\ }\bibfield  {title} {\bibinfo
  {title} {Ancillary {Qubit} {Spectroscopy} of {Vacua} in {Cavity} and
  {Circuit} {Quantum} {Electrodynamics}},\ }\href
  {https://doi.org/10.1103/PhysRevLett.114.183601} {\bibfield  {journal}
  {\bibinfo  {journal} {Phys. Rev. Lett.}\ }\textbf {\bibinfo {volume} {114}},\
  \bibinfo {pages} {183601} (\bibinfo {year} {2015})}\BibitemShut {NoStop}%
\bibitem [{\citenamefont {Jaako}\ \emph {et~al.}(2016)\citenamefont {Jaako},
  \citenamefont {Xiang}, \citenamefont {Garcia-Ripoll},\ and\ \citenamefont
  {Rabl}}]{jaako_ultrastrong-coupling_2016}%
  \BibitemOpen
  \bibfield  {author} {\bibinfo {author} {\bibfnamefont {T.}~\bibnamefont
  {Jaako}}, \bibinfo {author} {\bibfnamefont {Z.-L.}\ \bibnamefont {Xiang}},
  \bibinfo {author} {\bibfnamefont {J.~J.}\ \bibnamefont {Garcia-Ripoll}},\
  and\ \bibinfo {author} {\bibfnamefont {P.}~\bibnamefont {Rabl}},\ }\bibfield
  {title} {\bibinfo {title} {Ultrastrong-coupling phenomena beyond the {Dicke}
  model},\ }\href {https://doi.org/10.1103/PhysRevA.94.033850} {\bibfield
  {journal} {\bibinfo  {journal} {Physical Review A}\ }\textbf {\bibinfo
  {volume} {94}},\ \bibinfo {pages} {033850} (\bibinfo {year}
  {2016})}\BibitemShut {NoStop}%
\bibitem [{\citenamefont {Dimer}\ \emph
  {et~al.}(2007{\natexlab{a}})\citenamefont {Dimer}, \citenamefont {Estienne},
  \citenamefont {Parkins},\ and\ \citenamefont
  {Carmichael}}]{dimer_proposed_2007}%
  \BibitemOpen
  \bibfield  {author} {\bibinfo {author} {\bibfnamefont {F.}~\bibnamefont
  {Dimer}}, \bibinfo {author} {\bibfnamefont {B.}~\bibnamefont {Estienne}},
  \bibinfo {author} {\bibfnamefont {A.~S.}\ \bibnamefont {Parkins}},\ and\
  \bibinfo {author} {\bibfnamefont {H.~J.}\ \bibnamefont {Carmichael}},\
  }\bibfield  {title} {\bibinfo {title} {Proposed realization of the
  {Dicke}-model quantum phase transition in an optical cavity {QED} system},\
  }\href {https://doi.org/10.1103/PhysRevA.75.013804} {\bibfield  {journal}
  {\bibinfo  {journal} {Phys. Rev. A}\ }\textbf {\bibinfo {volume} {75}},\
  \bibinfo {pages} {013804} (\bibinfo {year} {2007}{\natexlab{a}})}\BibitemShut
  {NoStop}%
\bibitem [{\citenamefont {Garbe}\ \emph
  {et~al.}(2017{\natexlab{a}})\citenamefont {Garbe}, \citenamefont {Egusquiza},
  \citenamefont {Solano}, \citenamefont {Ciuti}, \citenamefont {Coudreau},
  \citenamefont {Milman},\ and\ \citenamefont
  {Felicetti}}]{garbe_superradiant_2017}%
  \BibitemOpen
  \bibfield  {author} {\bibinfo {author} {\bibfnamefont {L.}~\bibnamefont
  {Garbe}}, \bibinfo {author} {\bibfnamefont {I.~L.}\ \bibnamefont
  {Egusquiza}}, \bibinfo {author} {\bibfnamefont {E.}~\bibnamefont {Solano}},
  \bibinfo {author} {\bibfnamefont {C.}~\bibnamefont {Ciuti}}, \bibinfo
  {author} {\bibfnamefont {T.}~\bibnamefont {Coudreau}}, \bibinfo {author}
  {\bibfnamefont {P.}~\bibnamefont {Milman}},\ and\ \bibinfo {author}
  {\bibfnamefont {S.}~\bibnamefont {Felicetti}},\ }\bibfield  {title} {\bibinfo
  {title} {Superradiant phase transition in the ultrastrong-coupling regime of
  the two-photon {Dicke} model},\ }\href
  {https://doi.org/10.1103/PhysRevA.95.053854} {\bibfield  {journal} {\bibinfo
  {journal} {Phys. Rev. A}\ }\textbf {\bibinfo {volume} {95}},\ \bibinfo
  {pages} {053854} (\bibinfo {year} {2017}{\natexlab{a}})}\BibitemShut
  {NoStop}%
\bibitem [{\citenamefont {Chen}\ and\ \citenamefont
  {Zhang}(2018{\natexlab{a}})}]{chen_finite-size_2018}%
  \BibitemOpen
  \bibfield  {author} {\bibinfo {author} {\bibfnamefont {X.-Y.}\ \bibnamefont
  {Chen}}\ and\ \bibinfo {author} {\bibfnamefont {Y.-Y.}\ \bibnamefont
  {Zhang}},\ }\bibfield  {title} {\bibinfo {title} {Finite-size scaling
  analysis in the two-photon {Dicke} model},\ }\href
  {https://doi.org/10.1103/PhysRevA.97.053821} {\bibfield  {journal} {\bibinfo
  {journal} {Phys. Rev. A}\ }\textbf {\bibinfo {volume} {97}},\ \bibinfo
  {pages} {053821} (\bibinfo {year} {2018}{\natexlab{a}})}\BibitemShut
  {NoStop}%
\bibitem [{\citenamefont {Garziano}\ \emph {et~al.}(2020)\citenamefont
  {Garziano}, \citenamefont {Settineri}, \citenamefont {Di~Stefano},
  \citenamefont {Savasta},\ and\ \citenamefont {Nori}}]{garziano_gauge_2020}%
  \BibitemOpen
  \bibfield  {author} {\bibinfo {author} {\bibfnamefont {L.}~\bibnamefont
  {Garziano}}, \bibinfo {author} {\bibfnamefont {A.}~\bibnamefont {Settineri}},
  \bibinfo {author} {\bibfnamefont {O.}~\bibnamefont {Di~Stefano}}, \bibinfo
  {author} {\bibfnamefont {S.}~\bibnamefont {Savasta}},\ and\ \bibinfo {author}
  {\bibfnamefont {F.}~\bibnamefont {Nori}},\ }\bibfield  {title} {\bibinfo
  {title} {Gauge invariance of the {Dicke} and {Hopfield} models},\ }\href
  {https://doi.org/10.1103/PhysRevA.102.023718} {\bibfield  {journal} {\bibinfo
   {journal} {Phys. Rev. A}\ }\textbf {\bibinfo {volume} {102}},\ \bibinfo
  {pages} {023718} (\bibinfo {year} {2020})}\BibitemShut {NoStop}%
\bibitem [{\citenamefont {Bhaseen}\ \emph {et~al.}(2012)\citenamefont
  {Bhaseen}, \citenamefont {Mayoh}, \citenamefont {Simons},\ and\ \citenamefont
  {Keeling}}]{bhaseen_dynamics_2012}%
  \BibitemOpen
  \bibfield  {author} {\bibinfo {author} {\bibfnamefont {M.~J.}\ \bibnamefont
  {Bhaseen}}, \bibinfo {author} {\bibfnamefont {J.}~\bibnamefont {Mayoh}},
  \bibinfo {author} {\bibfnamefont {B.~D.}\ \bibnamefont {Simons}},\ and\
  \bibinfo {author} {\bibfnamefont {J.}~\bibnamefont {Keeling}},\ }\bibfield
  {title} {\bibinfo {title} {Dynamics of nonequilibrium {Dicke} models},\
  }\href {https://doi.org/10.1103/PhysRevA.85.013817} {\bibfield  {journal}
  {\bibinfo  {journal} {Phys. Rev. A}\ }\textbf {\bibinfo {volume} {85}},\
  \bibinfo {pages} {013817} (\bibinfo {year} {2012})}\BibitemShut {NoStop}%
\bibitem [{\citenamefont {Aedo}\ and\ \citenamefont
  {Lamata}(2018)}]{aedo_analog_2018}%
  \BibitemOpen
  \bibfield  {author} {\bibinfo {author} {\bibfnamefont {I.}~\bibnamefont
  {Aedo}}\ and\ \bibinfo {author} {\bibfnamefont {L.}~\bibnamefont {Lamata}},\
  }\bibfield  {title} {\bibinfo {title} {Analog quantum simulation of
  generalized {Dicke} models in trapped ions},\ }\href
  {https://doi.org/10.1103/PhysRevA.97.042317} {\bibfield  {journal} {\bibinfo
  {journal} {Phys. Rev. A}\ }\textbf {\bibinfo {volume} {97}},\ \bibinfo
  {pages} {042317} (\bibinfo {year} {2018})}\BibitemShut {NoStop}%
\bibitem [{\citenamefont {De~Bernardis}\ \emph
  {et~al.}(2018{\natexlab{b}})\citenamefont {De~Bernardis}, \citenamefont
  {Jaako},\ and\ \citenamefont {Rabl}}]{de_bernardis_cavity_2018}%
  \BibitemOpen
  \bibfield  {author} {\bibinfo {author} {\bibfnamefont {D.}~\bibnamefont
  {De~Bernardis}}, \bibinfo {author} {\bibfnamefont {T.}~\bibnamefont
  {Jaako}},\ and\ \bibinfo {author} {\bibfnamefont {P.}~\bibnamefont {Rabl}},\
  }\bibfield  {title} {\bibinfo {title} {Cavity quantum electrodynamics in the
  nonperturbative regime},\ }\href {https://doi.org/10.1103/PhysRevA.97.043820}
  {\bibfield  {journal} {\bibinfo  {journal} {Phys. Rev. A}\ }\textbf {\bibinfo
  {volume} {97}},\ \bibinfo {pages} {043820} (\bibinfo {year}
  {2018}{\natexlab{b}})}\BibitemShut {NoStop}%
\bibitem [{\citenamefont {Stokes}\ and\ \citenamefont
  {Nazir}(2020)}]{stokes_uniqueness_2020}%
  \BibitemOpen
  \bibfield  {author} {\bibinfo {author} {\bibfnamefont {A.}~\bibnamefont
  {Stokes}}\ and\ \bibinfo {author} {\bibfnamefont {A.}~\bibnamefont {Nazir}},\
  }\bibfield  {title} {\bibinfo {title} {Uniqueness of the phase transition in
  many-dipole cavity quantum electrodynamical systems},\ }\href
  {https://doi.org/10.1103/PhysRevLett.125.143603} {\bibfield  {journal}
  {\bibinfo  {journal} {Phys. Rev. Lett.}\ }\textbf {\bibinfo {volume} {125}},\
  \bibinfo {pages} {143603} (\bibinfo {year} {2020})}\BibitemShut {NoStop}%
\bibitem [{\citenamefont {Reitzenstein}\ \emph {et~al.}(2010)\citenamefont
  {Reitzenstein}, \citenamefont {B\"ockler}, \citenamefont {L\"offler},
  \citenamefont {H\"ofling}, \citenamefont {Worschech}, \citenamefont
  {Forchel}, \citenamefont {Yao},\ and\ \citenamefont
  {Hughes}}]{reitzenstein_polarization-dependent_2010}%
  \BibitemOpen
  \bibfield  {author} {\bibinfo {author} {\bibfnamefont {S.}~\bibnamefont
  {Reitzenstein}}, \bibinfo {author} {\bibfnamefont {C.}~\bibnamefont
  {B\"ockler}}, \bibinfo {author} {\bibfnamefont {A.}~\bibnamefont
  {L\"offler}}, \bibinfo {author} {\bibfnamefont {S.}~\bibnamefont
  {H\"ofling}}, \bibinfo {author} {\bibfnamefont {L.}~\bibnamefont
  {Worschech}}, \bibinfo {author} {\bibfnamefont {A.}~\bibnamefont {Forchel}},
  \bibinfo {author} {\bibfnamefont {P.}~\bibnamefont {Yao}},\ and\ \bibinfo
  {author} {\bibfnamefont {S.}~\bibnamefont {Hughes}},\ }\bibfield  {title}
  {\bibinfo {title} {Polarization-dependent strong coupling in elliptical
  high-$q$ micropillar cavities},\ }\href
  {https://doi.org/10.1103/PhysRevB.82.235313} {\bibfield  {journal} {\bibinfo
  {journal} {Phys. Rev. B}\ }\textbf {\bibinfo {volume} {82}},\ \bibinfo
  {pages} {235313} (\bibinfo {year} {2010})}\BibitemShut {NoStop}%
\bibitem [{\citenamefont {Kim}\ \emph {et~al.}(2011)\citenamefont {Kim},
  \citenamefont {Sridharan}, \citenamefont {Shen}, \citenamefont {Solomon},\
  and\ \citenamefont {Waks}}]{kim_strong_2011}%
  \BibitemOpen
  \bibfield  {author} {\bibinfo {author} {\bibfnamefont {H.}~\bibnamefont
  {Kim}}, \bibinfo {author} {\bibfnamefont {D.}~\bibnamefont {Sridharan}},
  \bibinfo {author} {\bibfnamefont {T.~C.}\ \bibnamefont {Shen}}, \bibinfo
  {author} {\bibfnamefont {G.~S.}\ \bibnamefont {Solomon}},\ and\ \bibinfo
  {author} {\bibfnamefont {E.}~\bibnamefont {Waks}},\ }\bibfield  {title}
  {{\selectlanguage {EN}\bibinfo {title} {Strong coupling between two quantum
  dots and a photonic crystal cavity using magnetic field tuning}},\ }\href
  {https://doi.org/10.1364/OE.19.002589} {\bibfield  {journal} {\bibinfo
  {journal} {Opt. Express}\ }\textbf {\bibinfo {volume} {19}},\ \bibinfo
  {pages} {2589} (\bibinfo {year} {2011})}\BibitemShut {NoStop}%
\bibitem [{\citenamefont {Majumdar}\ \emph {et~al.}(2012)\citenamefont
  {Majumdar}, \citenamefont {Bajcsy}, \citenamefont {Rundquist}, \citenamefont
  {Kim},\ and\ \citenamefont {Vu\ifmmode \check{c}\else
  \v{c}\fi{}kovi\ifmmode~\acute{c}\else
  \'{c}\fi{}}}]{majumdar_phonon-mediated_2012}%
  \BibitemOpen
  \bibfield  {author} {\bibinfo {author} {\bibfnamefont {A.}~\bibnamefont
  {Majumdar}}, \bibinfo {author} {\bibfnamefont {M.}~\bibnamefont {Bajcsy}},
  \bibinfo {author} {\bibfnamefont {A.}~\bibnamefont {Rundquist}}, \bibinfo
  {author} {\bibfnamefont {E.}~\bibnamefont {Kim}},\ and\ \bibinfo {author}
  {\bibfnamefont {J.}~\bibnamefont {Vu\ifmmode \check{c}\else
  \v{c}\fi{}kovi\ifmmode~\acute{c}\else \'{c}\fi{}}},\ }\bibfield  {title}
  {\bibinfo {title} {Phonon-mediated coupling between quantum dots through an
  off-resonant microcavity},\ }\href
  {https://doi.org/10.1103/PhysRevB.85.195301} {\bibfield  {journal} {\bibinfo
  {journal} {Phys. Rev. B}\ }\textbf {\bibinfo {volume} {85}},\ \bibinfo
  {pages} {195301} (\bibinfo {year} {2012})}\BibitemShut {NoStop}%
\bibitem [{\citenamefont {Maragkou}\ \emph {et~al.}(2013)\citenamefont
  {Maragkou}, \citenamefont {Sánchez-Muñoz}, \citenamefont {Lazić},
  \citenamefont {Chernysheva}, \citenamefont {van~der Meulen}, \citenamefont
  {González-Tudela}, \citenamefont {Tejedor}, \citenamefont {Martínez},
  \citenamefont {Prieto}, \citenamefont {Postigo},\ and\ \citenamefont
  {Calleja}}]{maragkou_bichromatic_2013}%
  \BibitemOpen
  \bibfield  {author} {\bibinfo {author} {\bibfnamefont {M.}~\bibnamefont
  {Maragkou}}, \bibinfo {author} {\bibfnamefont {C.}~\bibnamefont
  {Sánchez-Muñoz}}, \bibinfo {author} {\bibfnamefont {S.}~\bibnamefont
  {Lazić}}, \bibinfo {author} {\bibfnamefont {E.}~\bibnamefont {Chernysheva}},
  \bibinfo {author} {\bibfnamefont {H.~P.}\ \bibnamefont {van~der Meulen}},
  \bibinfo {author} {\bibfnamefont {A.}~\bibnamefont {González-Tudela}},
  \bibinfo {author} {\bibfnamefont {C.}~\bibnamefont {Tejedor}}, \bibinfo
  {author} {\bibfnamefont {L.~J.}\ \bibnamefont {Martínez}}, \bibinfo {author}
  {\bibfnamefont {I.}~\bibnamefont {Prieto}}, \bibinfo {author} {\bibfnamefont
  {P.~A.}\ \bibnamefont {Postigo}},\ and\ \bibinfo {author} {\bibfnamefont
  {J.~M.}\ \bibnamefont {Calleja}},\ }\bibfield  {title} {\bibinfo {title}
  {Bichromatic dressing of a quantum dot detected by a remote second quantum
  dot},\ }\href {https://doi.org/10.1103/PhysRevB.88.075309} {\bibfield
  {journal} {\bibinfo  {journal} {Phys. Rev. B}\ }\textbf {\bibinfo {volume}
  {88}},\ \bibinfo {pages} {075309} (\bibinfo {year} {2013})}\BibitemShut
  {NoStop}%
\bibitem [{\citenamefont {Bourassa}\ \emph {et~al.}(2009)\citenamefont
  {Bourassa}, \citenamefont {Gambetta}, \citenamefont {Abdumalikov},
  \citenamefont {Astafiev}, \citenamefont {Nakamura},\ and\ \citenamefont
  {Blais}}]{bourassa_ultrastrong_2009}%
  \BibitemOpen
  \bibfield  {author} {\bibinfo {author} {\bibfnamefont {J.}~\bibnamefont
  {Bourassa}}, \bibinfo {author} {\bibfnamefont {J.~M.}\ \bibnamefont
  {Gambetta}}, \bibinfo {author} {\bibfnamefont {A.~A.}\ \bibnamefont
  {Abdumalikov}}, \bibinfo {author} {\bibfnamefont {O.}~\bibnamefont
  {Astafiev}}, \bibinfo {author} {\bibfnamefont {Y.}~\bibnamefont {Nakamura}},\
  and\ \bibinfo {author} {\bibfnamefont {A.}~\bibnamefont {Blais}},\ }\bibfield
   {title} {\bibinfo {title} {Ultrastrong coupling regime of cavity qed with
  phase-biased flux qubits},\ }\href
  {https://doi.org/10.1103/PhysRevA.80.032109} {\bibfield  {journal} {\bibinfo
  {journal} {Phys. Rev. A}\ }\textbf {\bibinfo {volume} {80}},\ \bibinfo
  {pages} {032109} (\bibinfo {year} {2009})}\BibitemShut {NoStop}%
\bibitem [{\citenamefont {Yahiaoui}\ \emph {et~al.}(2022)\citenamefont
  {Yahiaoui}, \citenamefont {Chase}, \citenamefont {Kyaw}, \citenamefont {Tay},
  \citenamefont {Baydin}, \citenamefont {Noe}, \citenamefont {Song},
  \citenamefont {Kono}, \citenamefont {Agrawal}, \citenamefont {Bamba},\ and\
  \citenamefont {Searles}}]{Yahiaoui2022}%
  \BibitemOpen
  \bibfield  {author} {\bibinfo {author} {\bibfnamefont {R.}~\bibnamefont
  {Yahiaoui}}, \bibinfo {author} {\bibfnamefont {Z.~A.}\ \bibnamefont {Chase}},
  \bibinfo {author} {\bibfnamefont {C.}~\bibnamefont {Kyaw}}, \bibinfo {author}
  {\bibfnamefont {F.}~\bibnamefont {Tay}}, \bibinfo {author} {\bibfnamefont
  {A.}~\bibnamefont {Baydin}}, \bibinfo {author} {\bibfnamefont {G.~T.}\
  \bibnamefont {Noe}}, \bibinfo {author} {\bibfnamefont {J.}~\bibnamefont
  {Song}}, \bibinfo {author} {\bibfnamefont {J.}~\bibnamefont {Kono}}, \bibinfo
  {author} {\bibfnamefont {A.}~\bibnamefont {Agrawal}}, \bibinfo {author}
  {\bibfnamefont {M.}~\bibnamefont {Bamba}},\ and\ \bibinfo {author}
  {\bibfnamefont {T.~A.}\ \bibnamefont {Searles}},\ }\bibfield  {title}
  {\bibinfo {title} {Dicke-cooperativity-assisted ultrastrong coupling
  enhancement in terahertz metasurfaces},\ }\href
  {https://doi.org/10.1021/acs.nanolett.2c01892} {\bibfield  {journal}
  {\bibinfo  {journal} {Nano Letters}\ }\textbf {\bibinfo {volume} {22}},\
  \bibinfo {pages} {9788} (\bibinfo {year} {2022})}\BibitemShut {NoStop}%
\bibitem [{\citenamefont {Sánchez-Barquilla}\ \emph
  {et~al.}(2022)\citenamefont {Sánchez-Barquilla}, \citenamefont
  {Fernández-Domínguez}, \citenamefont {Feist},\ and\ \citenamefont
  {García-Vidal}}]{sanchez-barquilla_theoretical_2022}%
  \BibitemOpen
  \bibfield  {author} {\bibinfo {author} {\bibfnamefont {M.}~\bibnamefont
  {Sánchez-Barquilla}}, \bibinfo {author} {\bibfnamefont {A.~I.}\ \bibnamefont
  {Fernández-Domínguez}}, \bibinfo {author} {\bibfnamefont {J.}~\bibnamefont
  {Feist}},\ and\ \bibinfo {author} {\bibfnamefont {F.~J.}\ \bibnamefont
  {García-Vidal}},\ }\bibfield  {title} {\bibinfo {title} {A theoretical
  perspective on molecular polaritonics},\ }\href
  {https://doi.org/10.1021/acsphotonics.2c00048} {\bibfield  {journal}
  {\bibinfo  {journal} {ACS Photonics}\ }\textbf {\bibinfo {volume} {9}},\
  \bibinfo {pages} {1830} (\bibinfo {year} {2022})}\BibitemShut {NoStop}%
\bibitem [{\citenamefont {Benz}\ \emph {et~al.}(2016)\citenamefont {Benz},
  \citenamefont {Schmidt}, \citenamefont {Dreismann}, \citenamefont
  {Chikkaraddy}, \citenamefont {Zhang}, \citenamefont {Demetriadou},
  \citenamefont {Carnegie}, \citenamefont {Ohadi}, \citenamefont {de~Nijs},
  \citenamefont {Esteban}, \citenamefont {Aizpurua},\ and\ \citenamefont
  {Baumberg}}]{benz_single-molecule_2016}%
  \BibitemOpen
  \bibfield  {author} {\bibinfo {author} {\bibfnamefont {F.}~\bibnamefont
  {Benz}}, \bibinfo {author} {\bibfnamefont {M.~K.}\ \bibnamefont {Schmidt}},
  \bibinfo {author} {\bibfnamefont {A.}~\bibnamefont {Dreismann}}, \bibinfo
  {author} {\bibfnamefont {R.}~\bibnamefont {Chikkaraddy}}, \bibinfo {author}
  {\bibfnamefont {Y.}~\bibnamefont {Zhang}}, \bibinfo {author} {\bibfnamefont
  {A.}~\bibnamefont {Demetriadou}}, \bibinfo {author} {\bibfnamefont
  {C.}~\bibnamefont {Carnegie}}, \bibinfo {author} {\bibfnamefont
  {H.}~\bibnamefont {Ohadi}}, \bibinfo {author} {\bibfnamefont
  {B.}~\bibnamefont {de~Nijs}}, \bibinfo {author} {\bibfnamefont
  {R.}~\bibnamefont {Esteban}}, \bibinfo {author} {\bibfnamefont
  {J.}~\bibnamefont {Aizpurua}},\ and\ \bibinfo {author} {\bibfnamefont
  {J.~J.}\ \bibnamefont {Baumberg}},\ }\bibfield  {title} {\bibinfo {title}
  {Single-molecule optomechanics in ``picocavities''},\ }\href
  {https://doi.org/https://www.science.org/doi/10.1126/sciadv.abp9285}
  {\bibfield  {journal} {\bibinfo  {journal} {Science}\ }\textbf {\bibinfo
  {volume} {354}},\ \bibinfo {pages} {726} (\bibinfo {year}
  {2016})}\BibitemShut {NoStop}%
\bibitem [{\citenamefont {Chikkaraddy}\ \emph {et~al.}(2016)\citenamefont
  {Chikkaraddy}, \citenamefont {de~Nijs}, \citenamefont {Benz}, \citenamefont
  {Barrow}, \citenamefont {Scherman}, \citenamefont {Rosta}, \citenamefont
  {Demetriadou}, \citenamefont {Fox}, \citenamefont {Hess},\ and\ \citenamefont
  {Baumberg}}]{chikkaraddy_single-molecule_2016}%
  \BibitemOpen
  \bibfield  {author} {\bibinfo {author} {\bibfnamefont {R.}~\bibnamefont
  {Chikkaraddy}}, \bibinfo {author} {\bibfnamefont {B.}~\bibnamefont
  {de~Nijs}}, \bibinfo {author} {\bibfnamefont {F.}~\bibnamefont {Benz}},
  \bibinfo {author} {\bibfnamefont {S.~J.}\ \bibnamefont {Barrow}}, \bibinfo
  {author} {\bibfnamefont {O.~A.}\ \bibnamefont {Scherman}}, \bibinfo {author}
  {\bibfnamefont {E.}~\bibnamefont {Rosta}}, \bibinfo {author} {\bibfnamefont
  {A.}~\bibnamefont {Demetriadou}}, \bibinfo {author} {\bibfnamefont
  {P.}~\bibnamefont {Fox}}, \bibinfo {author} {\bibfnamefont {O.}~\bibnamefont
  {Hess}},\ and\ \bibinfo {author} {\bibfnamefont {J.~J.}\ \bibnamefont
  {Baumberg}},\ }\bibfield  {title} {\bibinfo {title} {Single-molecule strong
  coupling at room temperature in plasmonic nanocavities},\ }\href
  {https://doi.org/10.1038/nature17974} {\bibfield  {journal} {\bibinfo
  {journal} {Nature}\ }\textbf {\bibinfo {volume} {535}},\ \bibinfo {pages}
  {127} (\bibinfo {year} {2016})}\BibitemShut {NoStop}%
\bibitem [{\citenamefont {Al-Ani}\ \emph {et~al.}(2022)\citenamefont {Al-Ani},
  \citenamefont {As’ham}, \citenamefont {Klochan}, \citenamefont {Hattori},
  \citenamefont {Huang},\ and\ \citenamefont
  {Miroshnichenko}}]{al-ani_recent_2022}%
  \BibitemOpen
  \bibfield  {author} {\bibinfo {author} {\bibfnamefont {I.~A.~M.}\
  \bibnamefont {Al-Ani}}, \bibinfo {author} {\bibfnamefont {K.}~\bibnamefont
  {As’ham}}, \bibinfo {author} {\bibfnamefont {O.}~\bibnamefont {Klochan}},
  \bibinfo {author} {\bibfnamefont {H.~T.}\ \bibnamefont {Hattori}}, \bibinfo
  {author} {\bibfnamefont {L.}~\bibnamefont {Huang}},\ and\ \bibinfo {author}
  {\bibfnamefont {A.~E.}\ \bibnamefont {Miroshnichenko}},\ }\bibfield  {title}
  {\bibinfo {title} {Recent advances on strong light-matter coupling in
  atomically thin {TMDC} semiconductor materials},\ }\href
  {https://doi.org/10.1088/2040-8986/ac5cd7} {\bibfield  {journal} {\bibinfo
  {journal} {Journal of Optics}\ }\textbf {\bibinfo {volume} {24}},\ \bibinfo
  {pages} {053001} (\bibinfo {year} {2022})}\BibitemShut {NoStop}%
\bibitem [{\citenamefont {Albrechtsen}\ \emph {et~al.}(2022)\citenamefont
  {Albrechtsen}, \citenamefont {Vosoughi~Lahijani}, \citenamefont
  {Christiansen}, \citenamefont {Nguyen}, \citenamefont {Casses}, \citenamefont
  {Hansen}, \citenamefont {Stenger}, \citenamefont {Sigmund}, \citenamefont
  {Jansen}, \citenamefont {Mørk},\ and\ \citenamefont
  {Stobbe}}]{albrechtsen_nanometer-scale_2022}%
  \BibitemOpen
  \bibfield  {author} {\bibinfo {author} {\bibfnamefont {M.}~\bibnamefont
  {Albrechtsen}}, \bibinfo {author} {\bibfnamefont {B.}~\bibnamefont
  {Vosoughi~Lahijani}}, \bibinfo {author} {\bibfnamefont {R.~E.}\ \bibnamefont
  {Christiansen}}, \bibinfo {author} {\bibfnamefont {V.~T.~H.}\ \bibnamefont
  {Nguyen}}, \bibinfo {author} {\bibfnamefont {L.~N.}\ \bibnamefont {Casses}},
  \bibinfo {author} {\bibfnamefont {S.~E.}\ \bibnamefont {Hansen}}, \bibinfo
  {author} {\bibfnamefont {N.}~\bibnamefont {Stenger}}, \bibinfo {author}
  {\bibfnamefont {O.}~\bibnamefont {Sigmund}}, \bibinfo {author} {\bibfnamefont
  {H.}~\bibnamefont {Jansen}}, \bibinfo {author} {\bibfnamefont
  {J.}~\bibnamefont {Mørk}},\ and\ \bibinfo {author} {\bibfnamefont
  {S.}~\bibnamefont {Stobbe}},\ }\bibfield  {title} {\bibinfo {title}
  {Nanometer-scale photon confinement in topology-optimized dielectric
  cavities},\ }\href
  {https://doi.org/https://doi.org/10.1038/s41467-022-33874-w} {\bibfield
  {journal} {\bibinfo  {journal} {Nat Commun}\ }\textbf {\bibinfo {volume}
  {13}},\ \bibinfo {pages} {6281} (\bibinfo {year} {2022})}\BibitemShut
  {NoStop}%
\bibitem [{\citenamefont {Cui}\ and\ \citenamefont
  {Raymer}(2006)}]{cui_emission_2006}%
  \BibitemOpen
  \bibfield  {author} {\bibinfo {author} {\bibfnamefont {G.}~\bibnamefont
  {Cui}}\ and\ \bibinfo {author} {\bibfnamefont {M.~G.}\ \bibnamefont
  {Raymer}},\ }\bibfield  {title} {\bibinfo {title} {Emission spectra and
  quantum efficiency of single-photon sources in the {cavity-QED}
  strong-coupling regime},\ }\href {https://doi.org/10.1103/PhysRevA.73.053807}
  {\bibfield  {journal} {\bibinfo  {journal} {Phys. Rev. A}\ }\textbf {\bibinfo
  {volume} {73}},\ \bibinfo {pages} {053807} (\bibinfo {year}
  {2006})}\BibitemShut {NoStop}%
\bibitem [{\citenamefont {del Valle}\ \emph {et~al.}(2012)\citenamefont {del
  Valle}, \citenamefont {Gonzalez-Tudela}, \citenamefont {Laussy},
  \citenamefont {Tejedor},\ and\ \citenamefont
  {Hartmann}}]{del_valle_theory_2012}%
  \BibitemOpen
  \bibfield  {author} {\bibinfo {author} {\bibfnamefont {E.}~\bibnamefont {del
  Valle}}, \bibinfo {author} {\bibfnamefont {A.}~\bibnamefont
  {Gonzalez-Tudela}}, \bibinfo {author} {\bibfnamefont {F.~P.}\ \bibnamefont
  {Laussy}}, \bibinfo {author} {\bibfnamefont {C.}~\bibnamefont {Tejedor}},\
  and\ \bibinfo {author} {\bibfnamefont {M.~J.}\ \bibnamefont {Hartmann}},\
  }\bibfield  {title} {\bibinfo {title} {Theory of {Frequency}-{Filtered} and
  {Time}-{Resolved} {N}-{Photon} {Correlations}},\ }\href
  {https://doi.org/10.1103/PhysRevLett.109.183601} {\bibfield  {journal}
  {\bibinfo  {journal} {Physical Review Letters}\ }\textbf {\bibinfo {volume}
  {109}},\ \bibinfo {pages} {183601} (\bibinfo {year} {2012})}\BibitemShut
  {NoStop}%
\bibitem [{\citenamefont {Johansson}\ \emph {et~al.}(2012)\citenamefont
  {Johansson}, \citenamefont {Nation},\ and\ \citenamefont
  {Nori}}]{johansson2012qutip}%
  \BibitemOpen
  \bibfield  {author} {\bibinfo {author} {\bibfnamefont {J.~R.}\ \bibnamefont
  {Johansson}}, \bibinfo {author} {\bibfnamefont {P.~D.}\ \bibnamefont
  {Nation}},\ and\ \bibinfo {author} {\bibfnamefont {F.}~\bibnamefont {Nori}},\
  }\bibfield  {title} {\bibinfo {title} {Qutip: An open-source python framework
  for the dynamics of open quantum systems},\ }\href
  {https://doi.org/https://doi.org/10.1016/j.cpc.2012.02.021} {\bibfield
  {journal} {\bibinfo  {journal} {Computer Physics Communications}\ }\textbf
  {\bibinfo {volume} {183}},\ \bibinfo {pages} {1760} (\bibinfo {year}
  {2012})}\BibitemShut {NoStop}%
\bibitem [{\citenamefont {Johansson}\ \emph {et~al.}(2013)\citenamefont
  {Johansson}, \citenamefont {Nation},\ and\ \citenamefont
  {Nori}}]{johansson_qutip_2013}%
  \BibitemOpen
  \bibfield  {author} {\bibinfo {author} {\bibfnamefont {J.~R.}\ \bibnamefont
  {Johansson}}, \bibinfo {author} {\bibfnamefont {P.~D.}\ \bibnamefont
  {Nation}},\ and\ \bibinfo {author} {\bibfnamefont {F.}~\bibnamefont {Nori}},\
  }\bibfield  {title} {\bibinfo {title} {{QuTiP} 2: {A} {Python} framework for
  the dynamics of open quantum systems},\ }\href
  {https://doi.org/10.1016/j.cpc.2012.11.019} {\bibfield  {journal} {\bibinfo
  {journal} {Computer Physics Communications}\ }\textbf {\bibinfo {volume}
  {184}},\ \bibinfo {pages} {1234} (\bibinfo {year} {2013})}\BibitemShut
  {NoStop}%
\bibitem [{\citenamefont {Macr\`{\i}}\ \emph {et~al.}(2022)\citenamefont
  {Macr\`{\i}}, \citenamefont {Mercurio}, \citenamefont {Nori}, \citenamefont
  {Savasta},\ and\ \citenamefont {S\'anchez Mu\~noz}}]{macri_spontaneous_2022}%
  \BibitemOpen
  \bibfield  {author} {\bibinfo {author} {\bibfnamefont {V.}~\bibnamefont
  {Macr\`{\i}}}, \bibinfo {author} {\bibfnamefont {A.}~\bibnamefont
  {Mercurio}}, \bibinfo {author} {\bibfnamefont {F.}~\bibnamefont {Nori}},
  \bibinfo {author} {\bibfnamefont {S.}~\bibnamefont {Savasta}},\ and\ \bibinfo
  {author} {\bibfnamefont {C.}~\bibnamefont {S\'anchez Mu\~noz}},\ }\bibfield
  {title} {\bibinfo {title} {Spontaneous scattering of {Raman} photons from
  cavity-{QED} systems in the ultrastrong coupling regime},\ }\href
  {https://doi.org/10.1103/PhysRevLett.129.273602} {\bibfield  {journal}
  {\bibinfo  {journal} {Phys. Rev. Lett.}\ }\textbf {\bibinfo {volume} {129}},\
  \bibinfo {pages} {273602} (\bibinfo {year} {2022})}\BibitemShut {NoStop}%
\bibitem [{\citenamefont {Dimer}\ \emph
  {et~al.}(2007{\natexlab{b}})\citenamefont {Dimer}, \citenamefont {Estienne},
  \citenamefont {Parkins},\ and\ \citenamefont
  {Carmichael}}]{PhysRevA.75.013804}%
  \BibitemOpen
  \bibfield  {author} {\bibinfo {author} {\bibfnamefont {F.}~\bibnamefont
  {Dimer}}, \bibinfo {author} {\bibfnamefont {B.}~\bibnamefont {Estienne}},
  \bibinfo {author} {\bibfnamefont {A.~S.}\ \bibnamefont {Parkins}},\ and\
  \bibinfo {author} {\bibfnamefont {H.~J.}\ \bibnamefont {Carmichael}},\
  }\bibfield  {title} {\bibinfo {title} {Proposed realization of the
  {Dicke}-model quantum phase transition in an optical cavity {QED} system},\
  }\href {https://doi.org/10.1103/PhysRevA.75.013804} {\bibfield  {journal}
  {\bibinfo  {journal} {Phys. Rev. A}\ }\textbf {\bibinfo {volume} {75}},\
  \bibinfo {pages} {013804} (\bibinfo {year} {2007}{\natexlab{b}})}\BibitemShut
  {NoStop}%
\bibitem [{\citenamefont {Garbe}\ \emph
  {et~al.}(2017{\natexlab{b}})\citenamefont {Garbe}, \citenamefont {Egusquiza},
  \citenamefont {Solano}, \citenamefont {Ciuti}, \citenamefont {Coudreau},
  \citenamefont {Milman},\ and\ \citenamefont
  {Felicetti}}]{PhysRevA.95.053854}%
  \BibitemOpen
  \bibfield  {author} {\bibinfo {author} {\bibfnamefont {L.}~\bibnamefont
  {Garbe}}, \bibinfo {author} {\bibfnamefont {I.~L.}\ \bibnamefont
  {Egusquiza}}, \bibinfo {author} {\bibfnamefont {E.}~\bibnamefont {Solano}},
  \bibinfo {author} {\bibfnamefont {C.}~\bibnamefont {Ciuti}}, \bibinfo
  {author} {\bibfnamefont {T.}~\bibnamefont {Coudreau}}, \bibinfo {author}
  {\bibfnamefont {P.}~\bibnamefont {Milman}},\ and\ \bibinfo {author}
  {\bibfnamefont {S.}~\bibnamefont {Felicetti}},\ }\bibfield  {title} {\bibinfo
  {title} {Superradiant phase transition in the ultrastrong-coupling regime of
  the two-photon {Dicke} model},\ }\href
  {https://doi.org/10.1103/PhysRevA.95.053854} {\bibfield  {journal} {\bibinfo
  {journal} {Phys. Rev. A}\ }\textbf {\bibinfo {volume} {95}},\ \bibinfo
  {pages} {053854} (\bibinfo {year} {2017}{\natexlab{b}})}\BibitemShut
  {NoStop}%
\bibitem [{\citenamefont {Chen}\ and\ \citenamefont
  {Zhang}(2018{\natexlab{b}})}]{PhysRevA.97.053821}%
  \BibitemOpen
  \bibfield  {author} {\bibinfo {author} {\bibfnamefont {X.-Y.}\ \bibnamefont
  {Chen}}\ and\ \bibinfo {author} {\bibfnamefont {Y.-Y.}\ \bibnamefont
  {Zhang}},\ }\bibfield  {title} {\bibinfo {title} {Finite-size scaling
  analysis in the two-photon {Dicke} model},\ }\href
  {https://doi.org/10.1103/PhysRevA.97.053821} {\bibfield  {journal} {\bibinfo
  {journal} {Phys. Rev. A}\ }\textbf {\bibinfo {volume} {97}},\ \bibinfo
  {pages} {053821} (\bibinfo {year} {2018}{\natexlab{b}})}\BibitemShut
  {NoStop}%
\bibitem [{\citenamefont {Johansson}\ \emph {et~al.}(2006)\citenamefont
  {Johansson}, \citenamefont {Saito}, \citenamefont {Meno}, \citenamefont
  {Nakano}, \citenamefont {Ueda}, \citenamefont {Semba},\ and\ \citenamefont
  {Takayanagi}}]{PhysRevLett.96.127006}%
  \BibitemOpen
  \bibfield  {author} {\bibinfo {author} {\bibfnamefont {J.}~\bibnamefont
  {Johansson}}, \bibinfo {author} {\bibfnamefont {S.}~\bibnamefont {Saito}},
  \bibinfo {author} {\bibfnamefont {T.}~\bibnamefont {Meno}}, \bibinfo {author}
  {\bibfnamefont {H.}~\bibnamefont {Nakano}}, \bibinfo {author} {\bibfnamefont
  {M.}~\bibnamefont {Ueda}}, \bibinfo {author} {\bibfnamefont {K.}~\bibnamefont
  {Semba}},\ and\ \bibinfo {author} {\bibfnamefont {H.}~\bibnamefont
  {Takayanagi}},\ }\bibfield  {title} {\bibinfo {title} {Vacuum {Rabi}
  oscillations in a macroscopic superconducting qubit {$LC$} oscillator
  system},\ }\href {https://doi.org/10.1103/PhysRevLett.96.127006} {\bibfield
  {journal} {\bibinfo  {journal} {Phys. Rev. Lett.}\ }\textbf {\bibinfo
  {volume} {96}},\ \bibinfo {pages} {127006} (\bibinfo {year}
  {2006})}\BibitemShut {NoStop}%
\bibitem [{\citenamefont {Hughes}\ and\ \citenamefont
  {Agarwal}(2017)}]{hughes_anisotropy-induced_2017}%
  \BibitemOpen
  \bibfield  {author} {\bibinfo {author} {\bibfnamefont {S.}~\bibnamefont
  {Hughes}}\ and\ \bibinfo {author} {\bibfnamefont {G.~S.}\ \bibnamefont
  {Agarwal}},\ }\bibfield  {title} {\bibinfo {title} {Anisotropy-{Induced}
  {Quantum} {Interference} and {Population} {Trapping} between {Orthogonal}
  {Quantum} {Dot} {Exciton} {States} in {Semiconductor} {Cavity} {Systems}},\
  }\href {https://doi.org/10.1103/PhysRevLett.118.063601} {\bibfield  {journal}
  {\bibinfo  {journal} {Phys. Rev. Lett.}\ }\textbf {\bibinfo {volume} {118}},\
  \bibinfo {pages} {063601} (\bibinfo {year} {2017})}\BibitemShut {NoStop}%
\end{thebibliography}%

\end{document}